\documentclass[11pt,letterpaper]{article}

\usepackage{amsmath,amssymb,amsfonts,amsthm,bbm}
\usepackage{mathtools}   % for \norm
\usepackage{stmaryrd} % for \llbracket     1 \rrbracket
\usepackage{enumitem}   % for customized enumerate

\usepackage[titletoc,title]{appendix}  % Appendix A
\usepackage{booktabs}  % For \toprule, \midrule, \bottomrule
\usepackage{multirow} % For \multirow command
\usepackage{siunitx}   % For proper number alignment
\usepackage[normalem]{ulem}
\usepackage[table,xcdraw,dvipsnames]{xcolor}   % for \textcolor, \colorlet, ...
\usepackage{hyperref}
\newcommand\myshade{85}
\colorlet{mylinkcolor}{YellowOrange}
\colorlet{mycitecolor}{Aquamarine}
\colorlet{myurlcolor}{violet}
\hypersetup{
  linkcolor  = mylinkcolor!\myshade!black,
  citecolor  = mycitecolor!\myshade!black,
  urlcolor   = myurlcolor!\myshade!black,
  colorlinks = true,
}
\usepackage{graphicx}
\usepackage{caption}
\usepackage{subcaption}
\usepackage[top=1in, bottom=1in, left=1in, right=1in]{geometry}
\usepackage{algorithm,algorithmicx}
\usepackage{algpseudocode}

\newcommand{\bfm}[1]{\ensuremath{\boldsymbol{#1}}} % bm

\def\ba{\bfm a}   \def\bA{\bfm A}  
\def\bb{\bfm b}   \def\bB{\bfm B}  
     
   \def\bD{\bfm D}  
   \def\bE{\bfm E}  \def\EE{\mathbb{E}}
  \def\bF{\bfm F}  
\def\bg{\bfm g}   \def\bG{\bfm G}  
   \def\bH{\bfm H}  
   \def\bI{\bfm I}

   \def\bM{\bfm M}  
     
     \def\OO{\mathbb{O}}
   \def\bP{\bfm P}  \def\PP{\mathbb{P}}
   \def\bQ{\bfm Q}  
   \def\bR{\bfm R}  \def\RR{\mathbb{R}}

\def\bu{\bfm u}   \def\bU{\bfm U}  
\def\bv{\bfm v}   \def\bV{\bfm V}  
   \def\bW{\bfm W}  
   \def\bX{\bfm X}  
   \def\bY{\bfm Y}

\def\calA{{\cal  A}} 
\def\calB{{\cal  B}} 
 
 \def\cD{{\cal  D}}
 
\def\calF{{\cal  F}} \def\cF{{\cal  F}}

 \def\cN{{\cal  N}}

\def\calT{{\cal  T}}

 \def\cX{{\cal  X}}

\newcommand{\bfsym}[1]{\ensuremath{\boldsymbol{#1}}}

              \def\bSigma{\bfsym \Sigma}
         \def\bLambda {\bfsym {\Lambda}}

          \def\bPhi{\bfsym {\Phi}}

\DeclareMathOperator{\Cov}{Cov}

\DeclareMathOperator{\rank}{rank}

\newtheorem{lemma}{Lemma}[section]
\newtheorem{theorem}{Theorem}[section]
\newtheorem{proposition}{Proposition}[section]

\theoremstyle{definition}
\newtheorem{remark}{Remark}
\newtheorem{assumption}{Assumption}

\usepackage[authoryear,sort]{natbib}  %[numbers]

\usepackage{graphicx}
\graphicspath{{figs/}}

\usepackage{comment}

\renewcommand{\hat}{\widehat}
\renewcommand{\tilde}{\widetilde}

\newcommand{\cano}[1]{\widetilde{\mathbf{#1}}}

\newcommand{\bfv}{\mathbf{f}}

\title{\bf Modewise Additive Factor Model for Matrix Time Series}

\author{Elynn Chen$^\sharp$ \hspace{6ex}
Yuefeng Han$^\dag$ \hspace{6ex}
Jiayu Li$^\natural$ \hspace{6ex} 
Ke Xu$^\diamond$\thanks{\scriptsize{
Correspondence to E. Chen (E-mail: elynn.chen@stern.nyu.edu) and Y. Han (E-mail: yuefeng.han@nd.edu).
Coauthors are ordered alphabetically. E.~Chen is supported in part by NSF Grants DMS-2412577; Y.~Han is supported in part by Grants DMS-2412578.}} \\ \normalsize
\medskip
$^{\sharp,\natural}$New York University \hspace{6ex}
$^{\dag,\diamond}$University of Notre Dame
}
    
\date{}

\begin{document}
%-------
%
%  title
%

\maketitle

\begin{abstract}
We introduce a Modewise Additive Factor Model (MAFM) for matrix-valued time series that captures row-specific and column-specific latent effects through an additive structure, offering greater flexibility than multiplicative frameworks such as Tucker and CP factor models. In MAFM, each observation decomposes into a row-factor component, a column-factor component, and noise, allowing distinct sources of variation along different modes to be modeled separately. We develop a computationally efficient two-stage estimation procedure: Modewise Inner-product Eigendecomposition (MINE) for initialization, followed by Complement-Projected Alternating Subspace Estimation (COMPAS) for iterative refinement. The key methodological innovation is that orthogonal complement projections completely eliminate cross-modal interference when estimating each loading space. We establish convergence rates for the estimated factor loading matrices under proper conditions. We further derive asymptotic distributions for the loading matrix estimators and develop consistent covariance estimators, yielding a data-driven inference framework that enables confidence interval construction and hypothesis testing. As a technical contribution of independent interest, we establish matrix Bernstein inequalities for quadratic forms of dependent matrix time series. Numerical experiments on synthetic and real data demonstrate the advantages of the proposed method over existing approaches.
\end{abstract}

%-------
%
%  main text
%
%\begin{singlespace}
%\tableofcontents
%\end{singlespace}

\section{Introduction}  \label{sec:intro}

Factor analysis is a fundamental tool for understanding common dependence among high-dimensional observations. Classical vector factor models provide a parsimonious representation of co-movements through a small number of latent factors, treating each observation as a long vector driven by shared components. These models have been rigorously developed for vector-valued time series in statistics, economics and finance, with seminal works \citep{connor1986performance, bai2002determining, bai2003inferential, forni2004generalized, forni2005generalized,lam2011estimation, lam2012factor,fan2016projected} establishing consistency and inferential theory as both sample size and dimension grow.
However, when observations possess natural matrix or tensor structure, such as panels of units and variables, spatiotemporal measurements, or import-export volumes across product categories, vector factor models fail to capture important multi-way dependencies \citep{wang2019factor,chen2023statistical}. Vectorizing a matrix or tensor severs inherent row and column relationships, discarding structural information and often yielding less efficient and less interpretable estimates.

This limitation has motivated substantial recent interest in matrix- and tensor-valued factor models. By preserving multi-way structure, these models capture within- and across-mode dependencies that would be lost through flattening, often yielding more interpretable factors and improved estimation efficiency. A prominent class of such models employs a multiplicative architecture, where a latent factor tensor interacts with loading matrices along each mode. For a matrix time series $\{\bX_t\}_{t=1}^n$, a prototypical example is the Tucker factor model \citep{wang2019factor}:
\begin{align*}
\bX_t = \bA \bF_t \bB^\top + \bE_t,    
\end{align*}
where $\bF_t\in\RR^{r_1 \times r_2}$ is a matrix of latent factors and $\bA\in\RR^{d_1\times r_1},\bB\in\RR^{d_2\times r_2}$ are row and column loading matrices. Entry-wise, $X_{t,ij} = \ba_i^\top \bF_t \bb_j + E_{t,ij}$, so the latent factors act globally across all rows and columns. 
Recent research has extended this model and its estimation to more refined methodologies, including iterative refinement, formal inference, and more realistic matrix time-series settings such as time-varying and threshold models
\citep{chen2020constrained,chen2023statistical,yu2022projected,yu2024dynamic,chen2025factor,chen2024time,liu2022identification}.
Further, this Tucker factor model framework extends naturally to tensor time series \citep{chen2022factor, han2022rank,han2024tensor, babii2025tensor, chen2024semi,chen2024rank}, while a related but more constrained variant is the CP factor model \citep{chang2023modelling, chang2024identification, chen2026estimation, han2024cp,chen2025diffusion, bolivar2025threshold}, which assumes a superdiagonal latent factor tensor.

Despite their advantages over vector factor models, these multiplicative frameworks share a fundamental structural limitation: co-movement across the entire data array is driven by a single global factor $\bF_t$. Row-specific and column-specific effects are not independent sources of variation but rather different projections of the same underlying factors. For instance, consider a retail chain analyzing sales data across stores (rows) and product categories (columns). Tucker and CP factor models implicitly assume that store-level dynamics (e.g., local competition, management quality) and product-level dynamics (e.g., supply chain disruptions, marketing campaigns) are manifestations of the same core factors, an assumption that may be overly restrictive in practice.

To overcome this limitation, \cite{lam2025matrix} introduced a main-effect Tucker factor model with spectral decomposition-based estimation, while \cite{yuan2023twoway} proposed a two-way dynamic factor model using quasi-maximum likelihood estimation. In this paper, we introduce and analyze a Modewise Additive Factor Model (MAFM) for matrix time series that separately captures row-specific and column-specific latent effects. Each observation $\bX_t \in \RR^{d_1 \times d_2}$ decomposes into a \emph{row-factor component}, a \emph{column-factor component}, and noise: 
\begin{equation}\label{eqn: factor model}
\bX_t = \bF_t \bA^\top + \bB \bG_t^\top + \bE_t,
\end{equation}
where $\bF_t \in \RR^{d_1 \times r_1}$ and $\bG_t \in \RR^{d_2 \times r_2}$ are latent factor matrices for rows and columns, $\bA \in \RR^{d_2 \times r_1}$ and $\bB \in \RR^{d_1 \times r_2}$ are the corresponding loading matrices, and $\bE_t \in \RR^{d_1 \times d_2}$ is a noise matrix. Write $\bF_t = (\bfv_{t,1},\bfv_{t,2},\dots,\bfv_{t,d_1})^\top$ with $\bfv_{t,i}\in\RR^{r_1}$ being the latent factor for row $i$, and $\bG_t = (\bg_{t,1},\bg_{t,2},\dots,\bg_{t,d_2})^\top$ with $\bg_{t,j}\in\RR^{r_2}$ being the latent factor for column $j$. 
Each entry of $\bX_t$ decomposes as
\begin{align*}
X_{t,ij} = \bfv_{t,i}^\top \ba_j + \bb_i^\top \bg_{t,j} + E_{t,ij},    
\end{align*}
where the first term captures row-specific effects and the second captures column-specific effects. By disentangling these mode-specific sources of variation, our model offers greater flexibility than multiplicative frameworks.

To illustrate, revisit the retail example. Let $\bX_t$ contain sales figures with rows indexing stores and columns indexing product categories. The factor $\bfv_{t, i}$ captures store-specific drivers (e.g., location, management) at time $t$, while $\bg_{t,j}$ captures product-specific drivers (e.g., seasonality, branding). The loading $\ba_j \in \RR^{r_1}$ reflects how store factors influence product $j$, and $\bb_i \in \RR^{r_2}$ reflects how product factors influence store $i$. The additive structure $\bfv_{t,i}^\top \ba_j + \bb_i^\top \bg_{t,j}$ naturally separates these distinct sources of variation, offering a more realistic representation when store-level and product-level dynamics are driven by fundamentally different factors.

In this paper, we develop a computationally efficient two-stage procedure for MAFM. The first stage, \emph{Modewise Inner-product Eigendecomposition} (MINE), computes eigendecompositions of the modewise sample covariance matrices of $\bX_t$ to obtain initial estimates. While the signal from one mode induces structural bias when estimating the other, this bias has weaker signal strength than the target component and can be controlled. The second stage, \emph{COMplement-Projected Alternating Subspace Estimation} (COMPAS), iteratively refines these estimates through orthogonal complement projections, a key methodological innovation. Unlike the orthogonal projections in Tucker and CP factor models \citep{han2024tensor, han2024cp}, our orthogonal complement projections completely eliminate cross-modal interference: when estimating the loading matrix for one mode, projecting onto the orthogonal complement perfectly removes the signal from the other mode. Our theoretical analysis provides details of these improvements.

In the theoretical analysis, we establish statistical upper bounds on the estimation errors of the factor loading matrices for the proposed algorithms. MINE yields good initial estimates despite bias from cross-modal contamination. COMPAS eliminates this bias through iterative refinement. Notably, unlike orthogonal projection-based algorithms in Tucker and CP factor models, COMPAS does not amplify the signal-to-noise ratio through projection; nevertheless, it achieves comparable convergence rates (see Remark~\ref{rmk:comparison}). We also prove that using only a partial complement is statistically suboptimal, justifying the use of full orthogonal complement projections in COMPAS.

Beyond estimation, we establish asymptotic normality of the estimated factor loading matrices and develop consistent estimators for variance-covariance components under proper conditions, yielding a data-driven statistical inference framework which is an aspect largely underexplored in the factor model literature. These results enable valid confidence interval construction and hypothesis testing, tools that are essential for empirical applications. Our inferential approach, based on spectral representations of the estimators, differs fundamentally from techniques used in vector factor models \citep{bai2003inferential} and Tucker factor models \citep{chen2023statistical, yu2022projected}. Our approach can be extended naturally to other matrix and tensor factor model problems, such as inference of factor loading matrices using lagged autocovariance matrices \citep{han2024tensor, chang2023modelling}. Additionally, we derive sharp tail probability inequalities for quadratic forms of matrix time series under exponential-type tail conditions and temporal dependence, which may be of independent interest for the analysis of matrix and tensor time series models.

\subsection{Notation}

We use lowercase letters (e.g., $a, b, c$) for scalars, bold lowercase letters (e.g., $\ba, \bb$) for vectors, and bold uppercase letters (e.g., $\bA, \bB$) for matrices. For two sequences of real numbers $\{a_n\}$ and $\{b_n\}$, write $a_n = O(b_n)$ if $|a_n| \le C|b_n|$ for some constant $C > 0$ and all sufficiently large $n$, and $a_n = o(b_n)$ if $a_n/b_n \to 0$. Write $a_n \lesssim b_n$ (resp. $a_n \gtrsim b_n$) if $a_n \le Cb_n$ (resp. $a_n \ge Cb_n$) for some $C > 0$, and $a_n \asymp b_n$ if both $a_n \lesssim b_n$ and $a_n \gtrsim b_n$ hold. The notation $O_{\mathbb{P}}(\cdot)$ and $o_{\mathbb{P}}(\cdot)$ is used in their standard probabilistic senses. Write $a \wedge b = \min\{a, b\}$ and $a \vee b = \max\{a, b\}$. %We use $C, C_1, c, c_1, \ldots$ to denote generic constants, whose actual values may vary from line to line.

For a matrix $\bM$, $\Vec(\bM)$ denotes the vectorization obtained by stacking columns. The $p \times p$ identity matrix is $\bI_p$. The Stiefel manifold $\OO^{p \times r} = \{\bU \in \RR^{p \times r} : \bU^{\top}\bU = \bI_{r}\}$ is the set of $p\times r$ matrices with orthonormal columns. For any $\bU \in \OO^{p\times r}$, its orthogonal complement is denoted $\bU_\perp \in \OO^{p \times (p-r)}$. For symmetric matrices $\bA, \bB \in \RR^{p \times p}$, write $\bA \preceq \bB$ (resp. $\bA \prec \bB$) if $\bB - \bA$ is positive semidefinite (resp. positive definite), and similarly for $\succeq$ and $\succ$.

For a matrix $\bA \in \RR^{p_1 \times p_2}$ of rank $r$, let its singular value decomposition (SVD) be $\bA = \bU_1 \bLambda \bU_2^{\top}$, where $\bU_1 \in \OO^{p_1 \times r}$ and $\bU_2 \in \OO^{p_2 \times r}$ are the left and right singular matrices, and $\bLambda = {\rm diag}(\sigma_1(\bA), \ldots, \sigma_r(\bA))$ contains the singular values in descending order: $\sigma_1(\bA) \ge \cdots \ge \sigma_r(\bA) > 0$. We write $\sigma_{\max}(\bA) = \sigma_1(\bA)$ and $\sigma_{\min}(\bA) = \sigma_r(\bA)$. The projection matrices onto the left and right singular subspaces are $\bP_{\bA} = \bU_1\bU_1^{\top}$ and $\bP_{\bA^{\top}} = \bU_2\bU_2^{\top}$. The spectral and Frobenius norms are denoted as $\|\bA\|_2=\|\bA\|=\sigma_1(\bA)$ and $\|\bA\|_{\mathrm{F}}$, respectively.
For two orthonormal matrices $\bU, \widehat \bU \in \OO^{p \times r}$, let $\sigma_1\ge \sigma_2 \ge \cdots \ge \sigma_r\ge 0$ be the singular values of $\bU^\top \widehat \bU$. A natural measure of distance between their column spaces is
$\|\widehat \bU\widehat \bU^\top - \bU\bU^\top\|_2=\sqrt{1-\sigma_r^2},$
which equals the sine of the largest principal angle between the column spaces.

\subsection{Organization}
The remainder of the paper is organized as follows. Section~\ref{sec:model} discusses identifiability of the modewise additive factor model. Section~\ref{sec:estimation} presents the two-stage estimation procedure: MINE initialization and COMPAS refinement. Section~\ref{sec:theory} establishes statistical convergence rates of our estimators, asymptotic normality of factor loading matrices, and consistent estimation of variance-covariance components. 
Section~\ref{sec:ineq} develops sharp Bernstein-type inequalities for matrix time series. Section~\ref{sec:simul} reports simulation results examining estimation accuracy and inferential validity under various settings. Section~\ref{sec:real} applies the proposed method to real-world matrix time series data. Section \ref{sec:discussion} concludes the paper. Additional simulations, all proofs and technical lemmas are deferred to the Appendix.

\section{Identifiability of Modewise Additive Factor Model} \label{sec:model}

Identifiability is a foundational prerequisite for the estimation and interpretation of factor models. Like existing vector and Tucker factor models, our Modewise Additive Factor Model (MAFM) in \eqref{eqn: factor model} is invariant under the transformation
$(\bA\bH_{\bA}, \bB\bH_{\bB}, \bF_t\bH_{\bA}^{-1}, \bG_t\bH_{\bB}^{-1})$ for any invertible matrices $\bH_{\bA} \in \RR^{r_1 \times r_1}$ and $\bH_{\bB} \in \RR^{r_2 \times r_2}$. Nevertheless, the column spaces of $\bA$ and $\bB$, i.e., the factor loading spaces, are uniquely defined. The primary objective of this work is to identify and consistently estimate these subspaces.

To resolve the rotational and scaling ambiguities, we introduce a canonical representation of the model. Let $\bA = \bU_{\bA}\bLambda_{\bA}\bW_{\bA}^\top$ and $\bB = \bU_{\bB}\bLambda_{\bB}\bW_{\bB}^\top$ denote the SVD of the loading matrices. Absorbing the right singular matrices and singular values into the factors yields the canonical form:
\begin{equation}
\label{eqn: cano factor model}
\bX_t = \cano F_t \bU_{\bA}^\top + \bU_{\bB} \cano G_t^\top + \bE_t:=\bM_t+\bE_t,
\end{equation}
where $\cano F_t = \bF_t \bW_{\bA}\bLambda_{\bA} \in \RR^{d_1 \times r_1}$ and $\cano G_t = \bG_t \bW_{\bB}\bLambda_{\bB} \in \RR^{d_2 \times r_2}$. This representation forms the basis of our estimation procedure.

To ensure that the loading spaces spanned by $\bU_{\bA}$ and $\bU_{\bB}$ are uniquely recoverable from the observed data $\{\bX_t\}_{t=1}^n$, we impose the following mild structural assumptions:

(\textbf{C1}) The factor processes $\{\bF_t\}_{t=1}^n$ and $\{\bG_t\}_{t=1}^n$ are weakly stationary with mean zero and mutually independent.

(\textbf{C2}) For any nonzero vectors $\bu \in \RR^{d_1}$ and $\bv \in \RR^{d_2}$, the covariance matrices of $\bA\bF_t^\top\bu$ (row-factor part) and $\bB\bG_t^\top\bv$ (column-factor part) have ranks $r_1$ and $r_2$, respectively. 

Condition (C1) ensures that the row and column factors are uncorrelated with mean zero, allowing their contributions to the signal covariance to be separated. The weak stationarity further guarantees a consistent covariance structure across all time points, providing a well-defined population target for estimation. Specifically, the covariance tensor of the signal component $\bM_t$ admits the additive decomposition
\begin{align}\label{eqn: covariance tensor}
\text{Cov}(\bM_t) 
:= & \EE((\cano\bF_t\bU_{\bA}^{\top} + \bU_{\bB}\cano\bG_t^{\top}) \otimes (\cano\bF_t\bU_{\bA}^{\top} + \bU_{\bB}\cano\bG_t^{\top})) \notag\\
= & \EE((\cano\bF_t \bU_{\bA}^{\top}) \otimes (\cano\bF_t\bU_{\bA}^{\top})) + \EE((\bU_{\bB}\cano\bG_t^{\top}) \otimes (\bU_{\bB}\cano\bG_t^{\top})) \notag\\
= &\EE(\cano\bF_t \otimes \cano\bF_t) \times_2 \bU_{\bA} \times_4 \bU_{\bA} + \EE(\cano\bG_t \otimes \cano\bG_t) \times_1 \bU_{\bB} \times_3 \bU_{\bB} ,    
\end{align}
where $\otimes$ denotes the tensor outer product, and the mode-$k$ product $\cX\times_k \bA$ between $\mathcal{X}\in\mathbb{R}^{d_1\times\cdots\times d_M}$ and $\bA\in\mathbb{R}^{\tilde d_k\times d_k}$ produces a tensor of size $d_1\times\cdots\times d_{k-1}\times\tilde d_k\times d_{k+1}\times\cdots\times d_M$ with entries  
$(\cX\times_k \bA)_{i_1\cdots i_{k-1} j i_{k+1}\cdots i_M} = \sum_{i_k=1}^{d_k}\mathcal{X}_{i_1\cdots i_k\cdots i_M}A_{j i_k}$. This decomposition, a direct consequence of the additive model structure and the uncorrelatedness in (C1), is fundamental to separating the two factor loading spaces. Condition (C2) ensures that $r_1$ and $r_2$ are the intrinsic dimensions of the row-factor and column-factor components, which is crucial for identifiability.

\begin{proposition}\label{prop: identifiability}
Suppose the matrix time series $\bX_t \in \RR^{d_1 \times d_2}$ follows the modewise additive factor model \eqref{eqn: factor model} with $\bF_t$ and $\bG_t$ satisfying Conditions (C1)--(C2). Then, the loading spaces spanned by \(\bU_{\bB}\) and \(\bU_{\bA}\), respectively, are identifiable up to orthogonal rotations. Specifically, for any $\bU_{\bB}^{\prime} \in \OO^{d_1 \times r_2}$, if $\|\bU_{\bB}^{\prime}\bU_{\bB}^{\prime \top} - \bU_{\bB}\bU_{\bB}^{\top}\| \neq 0$, then we have $\PP\left(\bU_{\bB\perp}^{\prime}\bU_{\bB\perp}^{\prime \top}\bM_t = \mathbf{0}_{d_1 \times d_2} \right) = 0$. Similarly, for any $\bU_{\bA}^{\prime} \in \OO^{d_2 \times r_1}$, if $\|\bU_{\bA}^{\prime}\bU_{\bA}^{\prime \top} - \bU_{\bA}\bU_{\bA}^{\top}\| \neq 0$, then we have $\PP\left(\bM_t\bU_{\bA\perp}^{\prime}\bU_{\bA\perp}^{\prime \top} = \mathbf{0}_{d_1 \times d_2} \right) = 0$.
\end{proposition}

Proposition \ref{prop: identifiability} establishes the identifiability of the factor loading spaces.

%In essence, this proposition guarantees that the true factor loading spaces are the only subspaces that can fully capture the signal variation in their respective modes. Any attempt to characterize the signal with an incorrect subspace will leave a non-zero residual signal with probability one, making the true subspaces empirically discoverable.

\begin{remark}
\label{remark:1}
\cite{yuan2023twoway} consider a similar factor model for high-dimensional matrix time series but impose substantially stronger assumptions for model identification. In addition to uncorrelatedness between row and column factors (C1), their framework requires: (i) mutual uncorrelatedness across rows of each factor matrix, i.e., $\text{Cov}(\bfv_{t, i}, \bfv_{t,j}) = 0$ and $\text{Cov}(\bg_{t, i}, \bg_{t,j}) = 0 $ for all $ i \neq j$; (ii) identical covariance and autocovariance structures across rows of each factor matrix, i.e., $\text{Cov}(\bfv_{t, i}, \bfv_{t', i})$ and $\text{Cov}(\bg_{t,j}, \bg_{t',j} )$ are constant across $i,j$ for all $t, t' \in [n]$; and (iii) diagonal covariance matrices $\text{Cov}(\bfv_{t, i})$ and $\text{Cov}(\bg_{t,j})$. These assumptions are restrictive and atypical in both time series analysis and factor modeling. In contrast, our model identification relies on considerably milder conditions. As we demonstrate below, our estimation procedure also operates under standard regularity conditions that are broadly applicable.

\end{remark}

\section{Estimation}
\label{sec:estimation}

In this section, we develop estimators for the factor loading matrices $\bU_{\bA}$ and $\bU_{\bB}$. The additive structure of the model poses a unique statistical challenge: when estimating the loading matrix $\bU_{\bA}$, the column-factor signal $\bB\bG_t^\top$ acts as structured contamination in $\bX_t^\top \bX_t$, and vice versa. This cross-modal interference can introduce substantial bias into standard spectral decomposition based estimators. To address this issue, we propose a computationally efficient two-stage spectral procedure that decouples the row and column factors. Our approach first extracts initial estimates and then iteratively refines them through orthogonal complement projections.

\noindent\textbf{Initialization.} We construct initial estimators for the factor loading matrices via a two-step PCA. Specifically, 
%$\widehat{\bU}_{\bA}^{(0)} = \text{Eigen}_{r_1} (n^{-1} \sum_{t=1}^n \bX_t^\top \bX_t ) $ and $\widehat{\bU}_{\bB}^{(0)} = \text{Eigen}_{r_2} (n^{-1} \sum_{t=1}^n \bX_t \bX_t^\top )$, 
\begin{align*}
\widehat{\bU}_{\bA}^{(0)} = \text{Eigen}_{r_1} \left(\frac1n \sum_{t=1}^n \bX_t^\top \bX_t \right) \quad \text{and}\quad \widehat{\bU}_{\bB}^{(0)} = \text{Eigen}_{r_2} \left(\frac1n \sum_{t=1}^n \bX_t \bX_t^\top \right)   , 
\end{align*}
where Eigen$_m$ stands for the matrix composed of the top $m$ eigenvectors corresponding to the largest $m$ eigenvalues. This procedure, detailed in Algorithm~\ref{alg: MINE} and named Modewise INner-product Eigendecomposition (MINE), provides initializations for our iterative refinement.

The rationale follows from the modewise covariance structure of the observations. Assuming $\cano\bF_t,\cano\bG_t,\bE_t$ are mutually independent, the population column-wise covariance of $\bX_t$ is
\begin{align*}
\mathbb{E}[\bX_t^\top \bX_t] = \mathbb{E}[\bU_{\bA}\cano\bF_t^\top \cano\bF_t\bU_{\bA}^\top] + \mathbb{E}[\cano\bG_t\bU_{\bB}^{\top} \bU_{\bB}\cano\bG_t^\top] + \mathbb{E}[\bE_t^\top\bE_t].    
\end{align*}
The dominant low-rank component is the first term, $\bU_{\bA}\EE[\cano\bF_t^\top \cano\bF_t]\bU_{\bA}^\top$, which has rank $r_1$. Although the second term $\EE[\cano\bG_t\bU_{\bB}^{\top} \bU_{\bB}\cano\bG_t^\top]$ also contributes spiked eigenvalues, these grow at a slower rate. Consequently, the top $r_1$ eigenvectors of this modewise covariance primarily capture the loading matrix $\bU_{\bA}$. A symmetric argument applies to the loading matrix $\bU_{\bB}$.

\begin{algorithm}[ht]
\caption{Modewise INner-product Eigendecomposition (MINE)}\label{alg: MINE}
\begin{algorithmic}[1]
\Require Matrix time series $\{\bX_t\}_{t=1}^n \in \mathbb{R}^{d_1 \times d_2}$, target rank $(r_1, r_2)$. 

\State Compute the loading matrices composed of the first $r_1$ or $r_2$ eigenvectors corresponding to the largest $r_1$ or $r_2$ eigenvalues
\begin{align*}
\widehat{\bU}_{\bA}^{(0)} = \text{Eigen}_{r_1} \left( \frac{1}{n} \sum_{t=1}^n \bX_t^\top \bX_t \right) \quad \text{and} \quad \widehat{\bU}_{\bB}^{(0)} = \text{Eigen}_{r_2} \left( \frac{1}{n} \sum_{t=1}^n \bX_t \bX_t^\top \right)  .
\end{align*}
%\Return $\widehat{\bU}_{\bA}^{(0)}$, $\widehat{\bU}_{\bB}^{(0)}$

\Ensure Initialization $\widehat{\bU}_{\bA}^{(0)}$, $\widehat{\bU}_{\bB}^{(0)}$.
\end{algorithmic}
\end{algorithm}

\noindent\textbf{Iterative Refinement.} After obtaining initial estimates via MINE (Algorithm~\ref{alg: MINE}), we apply an orthogonal complement projection algorithm to refine them. The key idea is to isolate the signal from one mode by annihilating the signal from the other.
It is motivated by the following observation. 
Let $\bU_{\bA\perp},\bU_{\bB\perp}$ be the orthogonal complements of $\bU_{\bA},\bU_{\bB}$, respectively.
Multiplying $\bX_t$ by the orthogonal complement $\bU_{\bA\perp}$, model \eqref{eqn: cano factor model} yields
\begin{align}\label{eq:mafm-ideal}
\bX_t \bU_{\bA\perp} =  \bU_{\bB} \widetilde{\bG}_t^\top \bU_{\bA\perp} + \bE_t \bU_{\bA\perp} \in \RR^{d_1\times(d_2-r_1)}.    
\end{align}
The orthogonal complement projection eliminates the row-factor component $\widetilde{\bF}_t \bU_{\bA}^\top$, while preserving the column-factor component. Under proper conditions on the projected noise matrix $\bE_t \bU_{\bA\perp}$, estimation of the loading matrix $\bU_{\bB}$ based on $\bX_t \bU_{\bA\perp}$ can be made significantly more accurate. A symmetric argument applies to estimating $\bU_{\bA}$ via multiplying $\bX_t^\top$ by $\bU_{\bB\perp}$. We note that since $r_1\ll d_2 $ and $ r_2\ll d_1$, unlike orthogonal projections in Tucker and CP factor models \citep{han2024tensor,han2024cp}, the orthogonal complement projection does not reduce the noise variation. 

In practice, we do not know $\bU_{\bA\perp},\bU_{\bB\perp}$. Analogous to back-fitting algorithms, we iteratively estimate the loading matrices $\bU_{\bB}$ and $\bU_{\bA}$ at iteration $i$ based on $\bX_t \widehat\bU_{\bA\perp}^{(i-1)} ,\bX_t^\top \widehat\bU_{\bB\perp}^{(i)}$,
% \begin{align*}
% &\bX_t \widehat\bU_{\bA\perp}^{(i-1)} ,\quad \bX_t^\top \widehat\bU_{\bA\perp}^{(i)} ,
% \end{align*}
using the estimate $\widehat\bU_{\bA\perp}^{(i-1)}$ obtained in the previous iteration and the estimate $\widehat\bU_{\bB\perp}^{(i)}$ obtained in the current iteration. As we shall show in the next section, such an iterative procedure leads to a much improved statistical rate in the high dimensional matrix factor model scenarios, as if $\bU_{\bA\perp},\bU_{\bB\perp}$ are known and we indeed observe $\bX_t \bU_{\bA\perp},\bX_t^\top \bU_{\bB\perp}$ that follows model \eqref{eq:mafm-ideal}.

The complete procedure is presented in Algorithm~\ref{alg: COMPAS}, named COMplement-Projected Alternating Subspace Estimation (COMPAS). Although projection onto any subset of columns from $\bU_{\bA\perp}$ (or $\bU_{\bB\perp}$) eliminates the corresponding row-factor component (or column-factor component), our theoretical analysis (Theorem~\ref{thm: partial_iteration}) establishes that projection onto the full orthogonal complement yields optimal estimation efficiency. Hence, the use of the complete orthogonal complement is a critical design choice.

\begin{algorithm}[ht]
\caption{COMplement-Projected Alternating Subspace Estimation (COMPAS)}
\label{alg: COMPAS}
\begin{algorithmic}[1]
\Require Matrix time series $\{\bX_t\}_{t=1}^n \in \mathbb{R}^{d_1 \times d_2}$, target rank $(r_1, r_2)$, initialization $\widehat{\bU}_{\bA}^{(0)}, \widehat{\bU}_{\bB}^{(0)}$, the tolerance parameter $\epsilon>0$, maximal number of iterations $T_0$. 

\State Set $i=0$.
\Repeat 
\State Let $i=i+1$

\State Compute $\widehat{\bU}_{\bB}^{(i)}, \widehat{\bU}_{\bA}^{(i)}$:
\begin{align*}
\widehat{\bU}_{\bB}^{(i)} = \text{Eigen}_{r_2} \left(\frac{1}{n}\sum_{t=1}^n \bX_t \widehat{\bU}_{\bA\perp}^{(i-1)} \widehat{\bU}_{\bA\perp}^{(i-1)\top} \bX_t^\top \right), \     \widehat{\bU}_{\bA}^{(i)} =\text{Eigen}_{r_1} \left(\frac{1}{n} \sum_{t=1}^n \bX_t^\top \widehat{\bU}_{\bB\perp}^{(i)} \widehat{\bU}_{\bB\perp}^{(i)\top} \bX_t \right).
\end{align*}

\Until $i=T_0$ or $\|\widehat{\bU}_{\bB}^{(i)}\widehat{\bU}_{\bB}^{(i)\top}-\widehat{\bU}_{\bB}^{(i-1)}\widehat{\bU}_{\bB}^{(i-1)\top}\|\le\epsilon$ and $\|\widehat{\bU}_{\bA}^{(i)}\widehat{\bU}_{\bA}^{(i)\top}-\widehat{\bU}_{\bA}^{(i-1)}\widehat{\bU}_{\bA}^{(i-1)\top}\|\le\epsilon$.

\Ensure Estimates $\widehat{\bU}_{\bA}=\widehat{\bU}_{\bA}^{(i)}, \widehat{\bU}_{\bB}=\widehat{\bU}_{\bB}^{(i)}$.
\end{algorithmic}
\end{algorithm}

\noindent\textbf{Factor Estimation.} Once the loading matrices have been estimated via COMPAS (Algorithm~\ref{alg: COMPAS}), the final step is to recover the latent factors. We construct estimators by projecting the observed data onto the estimated signal subspaces, thereby isolating the contribution of each mode. Let $\widehat{\bU}_{\bA}$ and $\widehat{\bU}_{\bB}$ be the final estimates of the loading matrices from COMPAS, with $\widehat{\bU}_{\bA\perp}$ and $\widehat{\bU}_{\bB\perp}$ being their respective orthogonal complements. The estimators for the canonical factors $\cano F_t$ and $\cano G_t$ are
\begin{align}\label{eq:factors}
\hat \bF_t 
=  \hat \bU_{\bB\perp}\hat \bU_{\bB\perp}^\top \bX_t \hat \bU_{\bA}, \quad \hat \bG_t^\top 
= \hat \bU_{\bB}^\top \bX_t,
\end{align}
or equivalently, $\hat \bF_t = \bX_t \hat \bU_{\bA}, \hat \bG_t^\top = \hat \bU_{\bB}^\top \bX_t \hat \bU_{\bA \perp} \hat \bU_{\bA \perp}^\top.$
With the estimated loading matrices and factors, we construct the fitted value $\widehat{\bX}_t$ for the observation at time $t$ by removing the component orthogonal to both factor loading spaces:
%. One way to define the fitted value is by removing the component of the data that is orthogonal to both estimated signal spaces:
\begin{align*}
\widehat{\bX}_t = \bX_t - \widehat{\bU}_{\bB\perp}\widehat{\bU}_{\bB\perp}^\top \bX_t \widehat{\bU}_{\bA\perp}\widehat{\bU}_{\bA\perp}^\top.    
\end{align*}

\section{Theoretical Properties}
\label{sec:theory}

In this section, we shall investigate the statistical properties of the proposed algorithms described in the last section. Our theories provide consistency guarantees and characterize statistical error rates for estimating the factor loading matrices $\bU_{\bA}$ and $\bU_{\bB}$, under proper regularity conditions. We also establish the asymptotic distributions for each row of the estimated factor loading matrices.

\subsection{Convergence Rates for Factor Loading Estimation}

To present theoretical properties of the proposed procedures, we impose the following assumptions.

\begin{assumption}[Noise]
\label{assum: error term}
The noise processes $\{\bE_t\}_{t=1}^n$ are independent Gaussian matrices, conditioning on the factor processes $\{\bF_t\}_{t=1}^n$ and $ \{\bG_t\}_{t=1}^n$. Moreover, there exist constants $0 < \underline{\sigma} \le \overline{\sigma}$ such that for all $\bu \in \RR^{d_1d_2}$,
\[
\underline{\sigma}^2 \| \bu \|_{2}^2 \le \EE\big( \bu^\top \text{vec}(\bE_t) \big)^2 \le \overline{\sigma}^2 \| \bu \|_{2}^2.
\]    
Denote $\bSigma_{\bE} = \Cov(\Vec(\bE_t))$ and $\bSigma_{\bE^{\top}} = \Cov(\Vec(\bE_t^{\top}))$. It follows immediately that $\underline{\sigma}^2 \bI_{d_1d_2} \preceq \bSigma_{\bE}, \bSigma_{\bE^{\top}} \preceq \overline{\sigma}^2 \bI_{d_1d_2}$.
\end{assumption}

Assumption \ref{assum: error term} parallels noise conditions in \cite{lam2011estimation,lam2012factor, han2024tensor, han2024cp, chen2022factor}. It accommodates general dependence patterns among individual time series while maintaining analytical tractability. The Gaussian assumption, imposed for technical convenience, ensures fast convergence rates in our analysis. In principle, this could be relaxed to sub-Gaussian or heavier-tailed distributions. However, such generalizations would substantially complicate the formulas, conditions, and statistical outcomes in our time series framework without yielding additional insights. To maintain focus on the core contributions, we adopt the Gaussian error assumption, which simplifies exposition without compromising the fundamental methodology or findings.

\begin{assumption}[Mixing conditions of factors]
\label{assum: alpha-mixing}
The factor processes $\{\bF_t\}_{t=1}^n$ and $\{\bG_t\}_{t=1}^n$ are independent, weakly stationary, and strong $\alpha$-mixing with mixing coefficients $\alpha_{ F}(h)$ and $\alpha_{ G}(h)$ satisfying
\begin{align*}
\alpha_{ F}(h) \le \exp\left({-c_{ F} h^{\theta_1}}\right), \quad \alpha_{ G}(h) \le \exp\left({-c_{ G} h^{\theta_1}}\right),    
\end{align*}
for some constants $c_{ F}, c_{ G} > 0 $ and $\theta_1 >0$, where
\begin{align*}
\alpha_{ F}(h) &:= \sup_t \big\{|\PP(A \cap B) - \PP(A)\PP(B)| : A \in \cF(\bF_s, s \le t), B \in \cF(\bF_s, s \geq t + h) \big\}, \\
\alpha_{ G}(h) &:= \sup_t \big\{ |\PP(A \cap B) - \PP(A)\PP(B)| : A \in \cF(\bG_s, s \le t), B \in \cF(\bG_s, s \geq t + h) \big\}.    
\end{align*}

\end{assumption}

Assumption~\ref{assum: alpha-mixing} accommodates a broad class of time series models, including causal ARMA processes with continuously distributed innovations \citep{tsay2010analysis,tsay2018nonlinear,fan2003nonlinear}. 
For notational simplicity, we impose a common $\theta_1$ for both factor processes, though different values could be accommodated without altering the proofs. The independence between factor processes $\bF_t$ and $\bG_t$ is a structural assumption that ensures separation: row factors drive across-column variation, while column factors drive across-row variation, without overlap.

\begin{assumption}[Independent rows of the factors]
\label{assum: independent factor rows}  
% 1. [Independence between factors] For any $t$ and measurable sets $\bA \subset \RR^{d_1 \times r_1}$ and $\bB \subset \RR^{d_2 \times r_2}$, we have:
% \[
% \PP\left(\bF_t \in \bA, \bG_t \in \bB\right) = \PP\left(\bF_t \in \bA\right)\PP\left(\bG_t \in \bB\right).
% \]
%2. [Independent rows of factors]  
(i) For $\bF_t = (\bfv_{t,1}, \bfv_{t,2}, \dots, \bfv_{t,d_1})^\top$, the rows $\bfv_{t, i} \in \RR^{r_1}$ are independent across $i$, with mean zero. Similarly, for $\bG_t = (\bg_{t,1}, \bg_{t,2}, \dots, \bg_{t,d_2})^\top$, the rows $\bg_{t,j}\in \RR^{r_2}$ are independent across $j$, with mean zero.

%3. [Covariance matrices of factor rows] 
(ii) The covariance matrices $\Cov(\bfv_{t, i}) = \bSigma_{\bfv, i}$ and $\Cov(\bg_{t, j}) = \bSigma_{\bg, j}$ satisfy $\underline{\sigma}_{ F}^2 \bI_{r_1} \preceq \bSigma_{\bfv, i} \preceq \overline{\sigma}_{F}^2 \bI_{r_1}$ and $\underline{\sigma}_{G}^2 \bI_{r_2} \preceq \bSigma_{\bg, j} \preceq \overline{\sigma}_{G}^2 \bI_{r_2}$.
\end{assumption}

Assumption~\ref{assum: independent factor rows} imposes key structural properties on the latent factors. Part (i) is technically important: the independence of rows within each factor matrix enables us to leverage concentration inequalities for independent exponential-tail random variables when analyzing the cross-sectional dimension of factor rows. Part (ii) is a standard condition ensuring that the eigenvalues of the covariance of each factor row are uniformly bounded above and below. As detailed in the remark below, this assumption translates directly into bounds on the eigenvalues of key population covariance matrices, thereby quantifying the signal strength of the factor processes.

\begin{remark}\label{remark:signal_strength}
Let $\bP_{F} \in \RR^{d_1 \times d_1}$ and $\bP_{G} \in \RR^{d_2 \times d_2}$ be deterministic orthogonal projection matrices. Assumption~\ref{assum: independent factor rows} implies:

\noindent (i): $\underline{\sigma}_{ F}^2 \rank(\bP_{F}) \lesssim \sigma_{r_1}(\EE[\bF_t^\top \bP_{F} \bF_t]) \le \|\EE[\bF_t^\top \bP_{F} \bF_t]\|_2 \lesssim \overline{\sigma}_{F}^2 \rank(\bP_{F})\;$, and $\underline{\sigma}_{G}^2 \rank(\bP_{G}) \lesssim \sigma_{r_2}(\EE[\bG_t^\top \bP_{G} \bG_t]) \le \|\EE[\bG_t^\top \bP_{G} \bG_t]\|_2 \lesssim \overline{\sigma}_{G}^2 \rank(\bP_{G})$.

\noindent (ii): $\overline{\sigma}_{F}^2 r_1 \lesssim \sigma_{r_1}\left(\EE[\bF_t \bF_t^\top]\right) \le \left\|\EE[\bF_t \bF_t^\top\right\| \lesssim \overline{\sigma}_{F}^2 r_1$, and $\overline{\sigma}_{G}^2 r_2 \lesssim \sigma_{r_2}\left(\EE[\bG_t \bG_t^\top]\right) \le \left\|\EE[\bG_t \bG_t^\top]\right\|_2 \lesssim \overline{\sigma}_{G}^2 r_2 $.

To see (i), note that under Assumption~\ref{assum: independent factor rows},
\begin{equation*}
\begin{split}
& \EE\bigl(\bF_t^{\top}\bP_{F}\bF_t\bigr)
\;=\;\sum_{i=1}^{d_1}[\bP_{F}]_{i, i}\EE\bigl(\bfv_{t,\,i}\,\bfv_{t,\,i}^{\top}\bigr)
\;=\sum_{i=1}^{d_1}[\bP_{F}]_{i, i}\,\bSigma_{\bfv, i}, \\
\quad
& \EE\bigl(\bG_t^{\top}\bP_{G}\bG_t\bigr)
\;=\;\sum_{j=1}^{d_2}[\bP_{G}]_{j, j}\EE\bigl(\bg_{t,\,j}\,\bg_{t,\,j}^{\top}\bigr)
\;=\sum_{j=1}^{d_2}[\bP_{G}]_{j, j}\,\bSigma_{\bg, j}.
\end{split}
\end{equation*}
Since $\bP_{F}$ and $\bP_{G}$ are projection matrices, the bounds in Assumption~\ref{assum: independent factor rows}(ii) yield part (i).
For (ii), observe that $\EE(\bF_t\bF_t^{\top}) = [\EE(\bfv_{t, i}^\top \bfv_{t, j})]_{i, j}$.
By the independence and zero-mean conditions, \(\EE(\bfv_{t, i}^\top \bfv_{t, j}) = \EE(\bfv_{t, i}^\top \bfv_{t, i})\mathbf{1}(i =j)\), so
\[
\EE(\bF_t\bF_t^{\top}) 
= 
\operatorname{diag}\!\big(\EE[\bfv_{t, 1}^\top \bfv_{t, 1}], \dots, \EE[\bfv_{t, d_1}^\top \bfv_{t, d_1}]\big).
\]
Since $\EE(\bfv_{t, i}^{\top}\bfv_{t, i})= \operatorname{tr}(\bSigma_{\bfv, i})$, where $\bSigma_{\bfv, i} \in \RR^{r_1 \times r_1}$, part (ii) follows.

Since the factor matrices $\bF_t$ and $\bG_t$ have growing row dimensions, the signal strength of the factor processes plays a role analogous to loading strength in typical Tucker factor models, providing a foundation for our theoretical analysis.
\end{remark}

\begin{assumption}[Tail behavior of factor rows]
\label{assum: factor tail bound}
For any unit vectors $\bu \in \RR^{d_1}$ and $\bv \in \RR^{d_2}$,
\begin{align*}
&\max_{i=1, \cdots, d_1} \mathbb{P} \left( \left|\bu^\top \bfv_{t, i} \right| \geq \overline{\sigma}_{F} \cdot x \right) \le C_1 \exp \{-c_1 x^{\theta_2} \},    \\
%\end{align*}
%and
%\begin{align*}
&\max_{j=1, \cdots, d_2} \mathbb{P} \left( \left|\bv^\top \bg_{t, j} \right| \geq \overline{\sigma}_{G} \cdot x \right) \le C_2 \exp \{-c_2 x^{\theta_2} \},    
\end{align*}
where $C_1, C_2, c_1, c_2$ are positive constants and $0 < \theta_2 \le 2$.
\end{assumption}

Assumption~\ref{assum: factor tail bound} requires that the tail probabilities of factor rows decay exponentially, with $\theta_2=2$ corresponding to sub-Gaussian distributions. Combining Assumptions~\ref{assum: alpha-mixing} and \ref{assum: factor tail bound}, we develop novel matrix Bernstein type inequalities (e.g., Lemma \ref{lem: concentration GtWG FtWF main text}) for temporally dependent random matrices with exponential tails, extending classical results for i.i.d. random matrices. These concentration bounds may be of independent interest.

\begin{assumption}[Factor loading strength]
\label{assum: loading matrices scaling}
For some constants $\delta_0, \delta_1$ with $0 \le \delta_0 \le \delta_1 < 1$, assume that
\begin{align*}
d_2^{(1-\delta_1)/2} \lesssim \sigma_{r_1}(\bA) \le \|\bA\|_2 \lesssim d_2^{(1-\delta_0)/2},\quad\text{and}\quad d_1^{(1-\delta_1)/2} \lesssim \sigma_{r_2}(\bB) \le \|\bB\|_2 \lesssim d_1^{(1-\delta_0)/2}.    
\end{align*}

\end{assumption}

Assumption~\ref{assum: loading matrices scaling} is similar to the signal strength condition in \cite{lam2012factor,han2024tensor,chen2026estimation}, and aligns with the pervasive condition on factor loadings \citep{stock2002forecasting, bai2003inferential}. It plays a key role in identifying the common factors
and idiosyncratic noises in \eqref{eqn: cano factor model}. The indices $\delta_0,\delta_1$ measure factor loading strength, or the rate of signal strength growth as the dimension $d_k$ increases: $\delta_0$ characterizes the strength of the strongest factors, while $\delta_1$ corresponds to the weakest. When $\delta_0=\delta_1=0$, the factors are called strong factors; otherwise, the factors are called weak factors. For notational simplicity, we adopt common values of $\delta_0$ and $\delta_1$ across both dimensions, $d_1$ and $d_2$, although different values can be accommodated without affecting the analysis. The gap $\delta_1-\delta_0$ further characterizes the heterogeneity in loading strengths.

\begin{remark}[Variance Components]
For clarity, the variance components $\overline{\sigma}_{F}$, $\underline{\sigma}_{ F}$, $\overline{\sigma}_{G}$, $\underline{\sigma}_{G}$, $\overline{\sigma}$, $\underline{\sigma}$ are treated as fixed constants and absorbed into generic constants in the theorem statements. This allows the convergence rates to emphasize the roles of sample size $n$ and dimensions $(d_1,d_2)$ in determining estimation accuracy. Appendix~B provides full versions of all theoretical results with explicit dependence on variance components.
\end{remark}

Under these assumptions, we now derive non-asymptotic error bounds for our two-stage estimation procedure. The first result establishes the accuracy of the initial factor loading estimator $\widehat{\bU}_{\bA}^{(0)}$ and $\widehat{\bU}_{\bB}^{(0)}$ from Algorithm~\ref{alg: MINE}.

\begin{theorem}
\label{thm: initialization}
Suppose Assumptions~\ref{assum: error term}-\ref{assum: loading matrices scaling} hold. Let $1/\varphi = 1/\theta_1 + 2/\theta_2$. In an event with probability at least $1- \exp(-c_1n^{\varphi}d_1^{\varphi/2}) - \exp(-c_1d_2)$, the following error bound holds for the estimation of the loading matrix $\bU_{\bA}$ using Algorithm \ref{alg: MINE},
\begin{align}
\label{eqn: initial error bound UA}
\big\| \widehat{\bU}_{\bA}^{(0)} \widehat{\bU}_{\bA}^{(0)\top} - \bU_{\bA} \bU_{\bA}^\top \big\| 
\le & \frac{C_1}{d_2^{1-\delta_1}} + 
\frac{C_1 d_2^{\delta_1-\frac{\delta_{0}}{2}}}{\sqrt{nd_1}} + C_1  \left(\frac{d_2^{1/\varphi}d_1^{(1-\delta_0)/2}r_2 \left(d_1^{(1-\delta_0)/2} + d_2^{(1-\delta_0)/2}\right)}{nd_2^{1-\delta_1}d_1}\right),
\end{align}
where $c_1,C_1>0$ are some constants. Similarly, in an event with probability at least $1- \exp(-c_2n^{\varphi}d_2^{\varphi/2})-\exp(- c_2d_1)$, the following error bound holds for the estimation of the loading matrix $\bU_{\bB}$ using Algorithm \ref{alg: MINE},
\begin{align}
\label{eqn: initial error bound UB}
\big\| \widehat{\bU}_{\bB}^{(0)} \widehat{\bU}_{\bB}^{(0)\top} -  \bU_{\bB} \bU_{\bB}^\top \big\| \le & \frac{C_2}{d_1^{1-\delta_1}} + 
\frac{C_2 d_1^{\delta_1-\frac{\delta_{0}}{2}}}{\sqrt{nd_2}} + C_2 \left(\frac{d_1^{1/\varphi}d_2^{(1-\delta_0)/2}r_1 \left(d_2^{(1-\delta_0)/2} + d_1^{(1-\delta_0)/2}\right)}{nd_1^{1-\delta_1}d_2}\right),  
\end{align}
where $c_2,C_2>0$ are some constants.
Here, $C_1, C_2 $ depend on $\overline{\sigma}_{F}, \underline{\sigma}_{ F}, \overline{\sigma}_{G}, \underline{\sigma}_{G}, \overline{\sigma}$ and $\underline{\sigma}$.

\end{theorem}

The error bounds in Theorem~\ref{thm: initialization} comprise three distinct components. The first term corresponds to the average of the noise-by-noise cross-products. If all factors are strong factors with $\delta_0=\delta_1=0$, this term reduces to order $1/d_2$ or $1/d_1$, consistent with non-iterative Tucker factor models \citep{chen2023statistical}. This rate illustrates the ``blessing of dimensionality'' \citep{barigozzi2023quasi}: larger cross-sectional dimensions help average out noise. The second term arises from the average of the signal-by-noise cross-products. Under strong factors, the order of rates $(nd_1)^{-1/2}$ and $(nd_2)^{-1/2}$ also matches Tucker factor models \citep{han2024tensor}. The third term captures cross-modal interference, reflecting the interaction between row and column factors as well as contamination from the irrelevant factor mode.

\begin{remark}[Sample size requirement]
Consider the strong factor model with $\delta_0=\delta_1=0$, and comparable dimensions with $d_1 \asymp d_2$. When $\theta_1 = 1$ (standard exponential mixing decay) and $\theta_2=2$ (sub-Gaussian tails), the resulting exponent is $\varphi=1/2$. In this case, achieving the desired error bounds requires $n \gtrsim \max\{d_1, d_2\}$, which is a relatively modest requirement. When $\theta_1 \to \infty$ so that the factor processes approach independence, and $\theta_2=2$, we have $\varphi \to 1$. In that scenario, the sample size requirement becomes $n\gtrsim 1$, independent of dimensions $d_1,d_2$, which is an attractive property for high-dimensional settings.
\end{remark}

Next, we analyze the statistical performance of the iterative refinement (Algorithm \ref{alg: COMPAS}). The algorithm eliminates cross-modal interference by projecting onto the orthogonal complement of the estimated factor loading spaces. For instance, when estimating $\bU_{\bB}$, projection via $\bU_{\bA\perp}$ removes the row-factor component $\widetilde{\bF}_t \bU_{\bA}^\top$ in \eqref{eqn: cano factor model}. Ideally (i.e., when the true complement projection matrices $\bU_{\bA\perp},\bU_{\bB\perp}$ are used), this would eliminate the third terms in \eqref{eqn: initial error bound UA} and \eqref{eqn: initial error bound UB}, reducing the error bounds to
\begin{align}\label{eqn:final bdd}
\varepsilon_{\bA}^{\text{(main)}} = \frac{1 }{ d_2^{1-\delta_1}} + 
\frac{ d_2^{\delta_1-\frac{\delta_{0}}{2}}}{ \sqrt{n d_1}}, \quad  \varepsilon_{\bB}^{\text{(main)}} = \frac{1 }{ d_1^{1-\delta_1}} + 
\frac{ d_1^{\delta_1-\frac{\delta_{0}}{2}}}
{ \sqrt{n d_2}}     .    
\end{align}
In addition, define the contraction rate of each iteration,
\begin{align}
\eta_{\bA} &= \frac{d_1^{(1-\delta_0)/2}r_2  \left((d_2+d_1r_2)^{1/\varphi} d_2^{(1-\delta_0)/2} + d_2^{1/\varphi} d_1^{(1-\delta_0)/2}\right)}{ nd_2^{1-\delta_1}d_1},  \label{eqn: contraction rate UA} \\
\eta_{\bB} &= \frac{d_2^{(1-\delta_0)/2}r_1  \left((d_1+d_2r_1)^{1/\varphi} d_1^{(1-\delta_0)/2} +  d_1^{1/\varphi} d_2^{(1-\delta_0)/2}\right)}{  nd_1^{1-\delta_1}d_2}. \label{eqn: contraction rate UB}
\end{align}
The following theorem provides conditions under which the iterative refinement achieves the ideal rates.

\begin{theorem}
\label{thm: iteration}
Suppose Assumptions~\ref{assum: error term}-\ref{assum: loading matrices scaling} hold. Let $1/\varphi = 1/\theta_1 + 2/\theta_2$. Let $\varepsilon_{\bA}^{(0)} = \| \widehat{\bU}_{\bA}^{(0)} \widehat{\bU}_{\bA}^{(0)\top}  - \bU_{\bA} \bU_{\bA}^\top \|$ and $\varepsilon_{\bB}^{(0)} = \|  \widehat{\bU}_{\bB}^{(0)} \widehat{\bU}_{\bB}^{(0)\top}  - \bU_{\bB} \bU_{\bB}^\top \|$ be the error of initial estimators such that $\varepsilon_{\bA}^{(0)}, \varepsilon_{\bB}^{(0)} < 1$. Let $\widehat \bU_{\bA}, \widehat \bU_{\bB}$ be the output of Algorithm~\ref{alg: COMPAS}. 
Then, the following statements hold for certain numerical constants $c_1,c_2$ depending on $\overline{\sigma}_{F}, \underline{\sigma}_{ F},  \overline{\sigma}_{G}, \underline{\sigma}_{G},  \overline{\sigma}, \underline{\sigma}$ only: when $c_1\eta_{\bA}\le \rho_1 <1 $ and $c_2\eta_{\bB}\le \rho_2 <1$, then after at most 
$\tau = \lfloor \max\{\log(\varepsilon_{\bA}^{(0)}/\varepsilon_{\bA}^{\text{(main)}}),$ $ \log(\varepsilon_{\bB}^{(0)}/\varepsilon_{\bB}^{\text{(main)}})\} / \log(1/(\rho_1 \rho_2)) \rfloor$
iterations, the following upper bounds hold simultaneously in an event with probability at least $1- \exp(-cn^{\varphi}(d_1 \wedge d_2)^{\varphi/2})- \exp(-c(d_1 \wedge d_2)) $:
\begin{align}
\| \widehat{\bU}_{\bA} \widehat{\bU}_{\bA}^\top - \bU_{\bA} \bU_{\bA}^\top \|
\le & C_1 \varepsilon_{\bA}^{\text{(main)}} + C_1\eta_{\bA} \varepsilon_{\bB}^{\text{(main)}} ,  \label{eqn: refined error bound UA}\\
\|  \widehat{\bU}_{\bB} \widehat{\bU}_{\bB}^\top  - \bU_{\bB} \bU_{\bB}^\top \|
\le & C_2 \varepsilon_{\bB}^{\text{(main)}} + C_2\eta_{\bB} \varepsilon_{\bA}^{\text{(main)}} , \label{eqn: refined error bound UB}
\end{align}
where $c>0$, $C_1,C_2>0$ are constants depending on  $\overline{\sigma}_{F}, \underline{\sigma}_{ F}, \overline{\sigma}_{G}, \underline{\sigma}_{G}, \overline{\sigma}$, $\underline{\sigma}$, and $\varepsilon_{\bA}^{\text{(main)}}$, $ \varepsilon_{\bB}^{\text{(main)}}$ are defined in \eqref{eqn:final bdd}.
%Here, $C_1, C_2$ depend on the variance components, $\overline{\sigma}_{F}, \underline{\sigma}_{ F}, \overline{\sigma}_{G}, \underline{\sigma}_{G}, \overline{\sigma}$ and $\underline{\sigma}$.
\end{theorem}

The key insight in analyzing COMPAS (Algorithm~\ref{alg: COMPAS}) is that each iteration achieves error contraction, with contraction rates $\rho_1$ and $\rho_2$ for estimating $\bU_{\bA}$ and $\bU_{\bB}$, respectively. Unlike Tucker and CP factor models \citep{han2024tensor, han2024cp}, COMPAS does not reduce noise variation; instead, it eliminates only the third term in the initialization bounds \eqref{eqn: initial error bound UA} and \eqref{eqn: initial error bound UB}, which represents cross-modal interference. The contraction factors $\eta_{\bA}$ and $\eta_{\bB}$ govern the vanishing rate of this term, rather than affecting the oracle accuracy of the loading space estimators as in \cite{han2024tensor}. This reflects the role of alternating orthogonal complement projection: each mode benefits from refined estimates of the other, so that the error $\| \widehat{\bU}_{\bA} \widehat{\bU}_{\bA}^\top - \bU_{\bA} \bU_{\bA}^\top \|$ includes the leading error term of $\widehat{\bU}_{\bB}$ scaled by $\eta_{\bA}$, and vice versa.

The first two error terms from initialization persist in the refined bound. The first term corresponds to the average of the noise-by-noise cross-products, while the second captures the average of the signal-by-noise cross-products. This decomposition clarifies how factor loading strength, noise variance, sample size, and dimensionality jointly determine estimation accuracy, paralleling the error structure in Tucker and CP factor models.

\begin{remark}[Comparison with other matrix factor models] \label{rmk:comparison}
Consider $d_1\asymp d_2$ with fixed ranks $r_1,r_2$. Using covariance tensors (rather than auto-covariance tensors), CP factor models \citep{han2024cp,chen2026estimation} achieve loading vectors error bounds of order $O(d_1^{-2+2\delta_1}+n^{-1/2}d_1^{-1/2+2\delta_1})$, while Tucker factor models \citep{han2024tensor} achieve loading space error bounds $O(d_1^{-2+2\delta_1}+n^{-1/2}d_1^{-1/2-\delta_0+2\delta_1})$. Compared with the rates of our MAFM $O(d_1^{-1+\delta_1}+n^{-1/2}d_1^{-1/2-\delta_0/2+\delta_1})$ from Theorem \ref{thm: iteration}, the first noise-by-noise error term in our bounds is always worse than in CP and Tucker factor models. For the second signal-by-noise error term, CP factor models are always worse than ours, while the comparison with Tucker factor models depends on the sign of $\delta_0/2-\delta_1$. Thus, convergence rates alone do not distinguish among these model classes. Although our initialization resembles the TIPUP initialization in \cite{chen2022factor, han2024tensor}, the distinct model structures and projection designs can be exploited for model selection.

\cite{yuan2023twoway} propose a similar two-way dynamic factor model restricted to strong factors ($\delta_0=\delta_1=0$). Their error bounds yield convergence rate $O_{\PP}(d_1^{-1}+d_1^{-1/2} n^{-1/2})$, matching ours under the strong factor setting. However, their results are primarily asymptotic, whereas our analysis employs weaker conditions and provides non-asymptotic guarantees that extend to weak factor settings.

\end{remark}

A key design choice in Algorithm~\ref{alg: COMPAS} is the use of the full orthogonal complement (e.g., $\bU_{\bA\perp} \in \RR^{d_2 \times (d_2-r_1)}$) in the projection step. The matrix $\bU_{\bA\perp}$ spans all $d_2-r_1$ directions orthogonal to $\bU_{\bA} \in \RR^{d_2 \times r_1}$, thereby leveraging the maximum information unrelated to $\bU_{\bA}$ when estimating $\bU_{\bB}$. One might consider a computationally cheaper variant that projects onto a smaller subspace of the complement. Specifically, suppose that we replace $\bU_{\bA\perp} \in \OO^{d_2 \times (d_2-r_1)}$ and $\bU_{\bB\perp} \in \OO^{d_1 \times (d_1-r_2)}$ with partial complements $\bV_{\bA} \in \OO^{d_2 \times s_2}$ and $\bV_{\bB} \in \OO^{d_1 \times s_1}$, respectively, where $s_2<d_2-r_1$ and $s_1<d_1-r_2$. Let $\cano \bV_{\bA} \in \OO^{d_2 \times s_2}$ and $\cano \bV_{\bB} \in \OO^{d_1 \times s_1}$ be the corresponding estimates of $\bV_{\bA}$ and $\bV_{\bB}$ from Algorithm~\ref{alg: COMPAS}, and let $\cano \bU_{\bA}$ and $\cano \bU_{\bB}$ denote the resulting loading matrix estimates. 
Theorem~\ref{thm: partial_iteration} formally confirms that using fewer than the full complement directions incurs a larger error bound, reflecting a loss of statistical efficiency.

\begin{theorem}[Partial complement projection] 
\label{thm: partial_iteration}
Suppose Assumptions \ref{assum: error term}-\ref{assum: loading matrices scaling} hold. Let $1/\varphi = 1/\theta_1 + 2/\theta_2$. 
%Let $\eta_{\bA}$ and $\eta_{\bB}$ be the same as defined in Equation~\eqref{eqn: contraction rate UA} and \eqref{eqn: contraction rate UB}, respectively. after a certain number of iterations,
Then, the following statements hold for certain numerical constants $c_1,c_2$ depending on $\overline{\sigma}_{F}, \underline{\sigma}_{ F},  \overline{\sigma}_{G}, \underline{\sigma}_{G}, \overline{\sigma}, \underline{\sigma}$ only: when $c_1 \eta_{\bA}\le \rho <1 $ and $c_2 \eta_{\bB}\le \rho <1$, then the following upper bounds hold simultaneously in an event with probability at least $1 - \exp(-cn^{\varphi}(d_1 \wedge d_2)^{\varphi/2})- \exp(-c(d_1 \wedge d_2))$:
\begin{align*}
\| \cano{\bU}_{\bA} \cano{\bU}_{\bA}^{\top} - \bU_{\bA} \bU_{\bA}^\top \|
\le & C_1\left(\frac{1}{d_2^{1-\delta_1}} + 
\frac{d_2^{\delta_1-\frac{\delta_{0}}{2}}}
{\sqrt{n s_1}}\right) + C_1 \left(\frac{d_1}{s_1} \cdot \eta_{\bA}\right)  \left(\frac{1}{d_1^{1-\delta_1}} + 
\frac{d_1^{\delta_1-\frac{\delta_{0}}{2}}}
{\sqrt{n s_2}}\right), \\
\| \cano{\bU}_{\bB} \cano{\bU}_{\bB}^{\top} - \bU_{\bB} \bU_{\bB}^\top \|
\le & C_2\left(\frac{1}{d_1^{1-\delta_1}} + 
\frac{d_1^{\delta_1-\frac{\delta_{0}}{2}}}
{\sqrt{n s_2}}\right) + C_2 \left(\frac{d_2}{s_2} \cdot \eta_{\bB}\right)  \left(\frac{1}{d_2^{1-\delta_1}} + 
\frac{d_2^{\delta_1-\frac{\delta_{0}}{2}}}
{\sqrt{n s_1}}\right).
\end{align*}
where $c>0$, and $C_1,C_2>0$ are constants depending on  $\overline{\sigma}_{F}, \underline{\sigma}_{ F}, \overline{\sigma}_{G}, \underline{\sigma}_{G}, \overline{\sigma}$, $\underline{\sigma}$.
%Here, $C_1, C_2$ depend on the variance components, $\overline{\sigma}_{F}, \underline{\sigma}_{ F}, \overline{\sigma}_{G}, \underline{\sigma}_{G}, \overline{\sigma}$ and $\underline{\sigma}$.
\end{theorem}

Compared with Theorem~\ref{thm: iteration}, which uses the full complement, Theorem~\ref{thm: partial_iteration} shows that restricting to $s_1 < d_1-r_2$ and $s_2 < d_2-r_1$ directions inflates the signal-by-noise terms by replacing $\sqrt{n d_1}$ and $\sqrt{n d_2}$ with $\sqrt{n s_1}$ and $\sqrt{n s_2}$. This efficiency loss has a clear mechanistic explanation. The update for $\bU_{\bB}$ relies on the row-wise sample covariance of the projected data $\bX_t \bU_{\bA\perp}$. The signal strength in this projection, which is essential for separating signal from noise, is determined by the smallest nonzero eigenvalue of the signal covariance matrix. As established in Remark~\ref{remark:signal_strength}, this eigenvalue is proportional to the rank of the projection matrix. Using the full complement projection ensures maximum signal variation for updating each mode, leading to tighter error contraction. In contrast, a partial complement of rank $s_2 < d_2-r_1$ discards signal information, inflating the final estimation error. While the algorithm remains valid with partial complement projections, the convergence rates and error bounds are strictly worse.

\subsection{Statistical Inference for Factor Loadings}

In factor models, each row of the loading matrix represents how a particular unit (e.g. a country or an asset) loads onto the latent factors. For example, in finance, asset pricing models posit that each stock’s return depends linearly on a set of common factors, with the factor loadings capturing the asset’s risk exposures or beta coefficients, to systematic market factors \citep{ROSS1976341, connor1986performance,lettau2022high,giglio2022factor,conlon2025asset}. In macroeconomics, a loading row may measure a country's sensitivity to global business cycle shocks \citep{bai2003inferential,houssa2008monetary,forni2021policy}. Inference on these row-specific loadings is crucial for testing hypotheses about unit heterogeneity (e.g., which countries significantly load on a global factor) or about the stability of the factor structure over time (e.g., whether certain assets cease to load on a market factor).

In this section, we establish the asymptotic distributions of the estimated factor loading matrices. Let 
\begin{equation}
\label{eqn: epsilon bar A bar B}
\overline{\varepsilon}_{\bA} = \varepsilon_{\bA}^{\text{(main)}} + \eta_{\bA} \cdot \varepsilon_{\bB}^{\text{(main)}}, \text{and} \quad 
\overline{\varepsilon}_{\bB} = \varepsilon_{\bB}^{\text{(main)}} + \eta_{\bB} \cdot \varepsilon_{\bA}^{\text{(main)}}
\end{equation}
denote the error bounds for estimating loading matrices $\bU_{\bA}$ and $\bU_{\bB}$ after applying Algorithms~\ref{alg: MINE} and \ref{alg: COMPAS}, as established in \eqref{eqn: refined error bound UA} and \eqref{eqn: refined error bound UB}. By Theorem~\ref{thm: iteration}, $\| \widehat{\bU}_{\bA}\widehat{\bU}_{\bA}^{\top} - \bU_{\bA}\bU_{\bA}^{\top} \| \le \overline{\varepsilon}_{\bA}$ and $\| \widehat{\bU}_{\bB}\widehat{\bU}_{\bB}^{\top} - \bU_{\bB}\bU_{\bB}^{\top} \| \le \overline{\varepsilon}_{\bB}$ with probability at least $1- \exp(-cn^{\varphi}(d_1 \wedge d_2)^{\varphi/2}) - \exp(-c (d_1 \wedge d_2))$.
Recall the noise covariance matrices $\bSigma_{\bE} = \Cov\left(\Vec\left(\bE_t\right)\right)\in \RR^{d_1d_2 \times d_1d_2}$ and $\bSigma_{\bE^{\top}} = \Cov\left(\Vec\left(\bE_t^{\top}\right)\right) \in \RR^{d_1d_2 \times d_1d_2}$. Let
\begin{equation} \label{eqn: sample projected factor covariance}
\bSigma_{\bF} = \EE \cano \bF_t^{\top} \bU_{\bB\perp} \bU_{\bB\perp}^{\top} \cano \bF_t, 
\quad \bSigma_{\bG} = \EE \cano \bG_t^{\top} \bU_{\bA\perp} \bU_{\bA\perp}^{\top} \cano \bG_t.
\end{equation}
be the population row-wise covariance matrices of the projected factors $\bU_{\bB\perp}\bU_{\bB\perp}^{\top}\cano \bF_t$ and $\bU_{\bA\perp}\bU_{\bA\perp}^{\top}\cano \bG_t $, respectively. 
%The following theorem shows the asymptotic normality of each row of the estimated factor loading matrices.

\begin{remark}
The population row-wise covariance matrices of the projected factors in \eqref{eqn: sample projected factor covariance} involve orthogonal complement projections. For example, when estimating $\bU_{\bA}$, we project $\bM_t$ in \eqref{eqn: cano factor model} onto $\bU_{\bB\perp}\bU_{\bB\perp}^{\top}$ to remove the contribution from the column-factor component. 
\end{remark}

\begin{theorem}\label{thm: asymptotic normality}
Suppose the conditions of Theorem~\ref{thm: iteration} hold. %Let $\widehat{\bU}_{\bA}$ and $\widehat{\bU}_{\bB}$ be the output of sequentially applying Algorithms~\ref{alg: MINE} and \ref{alg: COMPAS}.
Let $\calF_{r_k}$ denote the Borel $\sigma$-algebra on $\RR^{r_k}$ for $k=1,2$, and let $e_i$ denote the $i$-th standard basis vector. Then, there exist rotation matrices $\bR_{\bA} = \min_{\bR \in \OO^{r_1 \times r_1}}\|\widehat{\bU}_{\bA} - \bU_{\bA}\bR\|$ and $\bR_{\bB} = \min_{\bR \in \OO^{r_2 \times r_2}}\|\widehat{\bU}_{\bB} - \bU_{\bB}\bR\| \in \OO^{r_2 \times r_2}$, such that for all $i \in [d_2]$ and $j \in [d_1]$,
\begin{align}
\sup_{\calA \in \calF_{r_1}}\left|\PP\left(\sqrt{n} \bR_{\bA}^{\top}\bSigma_{\bA,i}^{-\frac{1}{2}} \bSigma_{\bF}\bR_{\bA}\left(\widehat{\bU}_{\bA}^{\top}- \bR_{\bA}^{\top}\bU_{\bA}^{\top}\right)e_i \in \calA\right) - \PP(\mathcal{N}(0, \bI_{r_1}) \in \calA)\right| 
\le & C_1\Omega_{\bA} ,     \label{eqn: asymp normality of UAei} \\
\sup_{\calB \in \calF_{r_2}}\left|\PP\left(\sqrt{n} \bR_{\bB}^{\top}\bSigma_{\bB,j}^{-\frac{1}{2}}  \bSigma_{\bG}\bR_{\bB}\left(\widehat{\bU}_{\bB}^{\top}- \bR_{\bB}^{\top}\bU_{\bB}^{\top}\right)e_j \in \calB\right) - \PP(\mathcal{N}(0, \bI_{r_2}) \in \calB)\right|
\le & C_2\Omega_{\bB},  \label{eqn: asymp normality of UBej}
\end{align}
where $C_1,C_2>0$ are constants depending on $\overline{\sigma}_{F}, \underline{\sigma}_{ F}, \overline{\sigma}_{G}, \underline{\sigma}_{G}, \overline{\sigma}$, $\underline{\sigma}$, the covariance matrices of the signal-by-noise cross-products are
\begin{align}
\bSigma_{\bA,i} &= \EE\left[\left(e_i^{\top}\bU_{\bA\perp}\bU_{\bA\perp}^{\top} \otimes\cano \bF_t^{\top} \bU_{\bB \perp}\bU_{\bB \perp}^{\top}\right) \bSigma_{\bE} \left(\bU_{\bA\perp}\bU_{\bA\perp}^{\top}e_i \otimes \bU_{\bB \perp} \bU_{\bB \perp}^{\top} \cano \bF_t\right)\right],    \label{eqn: Sigma_A}\\
\bSigma_{\bB,j} &=\EE\left[\left(e_j^{\top}\bU_{\bB\perp}\bU_{\bB\perp}^{\top} \otimes  \cano \bG_t^{\top} \bU_{\bA\perp}\bU_{\bA\perp}^{\top}\right) \bSigma_{\bE^{\top}} \left(\bU_{\bB\perp}\bU_{\bB\perp}^{\top}e_j \otimes \bU_{\bA\perp} \bU_{\bA\perp}^{\top} \cano \bG_t \right)\right],  \label{eqn: Sigma_B}
\end{align}
and the error bounds are
\begin{align*}
\Omega_{\bA}
= & d_2^{\delta_1-\delta_0}r_1\sqrt{\frac{1}{nd_1}} + d_2^{(\delta_1-\delta_0)/2} \sqrt{r_1\log(nd_1)} \overline{\varepsilon}_{\bB} + d_2^{(\delta_1-\delta_0)/2} \cdot \sqrt{n} d_2^{(1-\delta_0)/2}d_1^{1/2} (\overline{\varepsilon}_{\bA}^2 + \overline{\varepsilon}_{\bA}\overline{\varepsilon}_{\bB} ), \\    
\Omega_{\bB}
= & d_1^{\delta_1-\delta_0}r_2 \sqrt{\frac{1}{nd_2}} + d_1^{(\delta_1-\delta_0)/2}  \sqrt{r_2\log(nd_2)} \overline{\varepsilon}_{\bA} + d_1^{(\delta_1-\delta_0)/2} \cdot \sqrt{n} d_1^{(1-\delta_0)/2}d_2^{1/2}  (\overline{\varepsilon}_{\bB}^2 + \overline{\varepsilon}_{\bA}\overline{\varepsilon}_{\bB} ).
\end{align*}
%Here, $C_1, C_2$ depending on the variance components, $\overline{\sigma}_{F}, \underline{\sigma}_{ F}, \overline{\sigma}_{G}, \underline{\sigma}_{G}, \overline{\sigma}$ and $\underline{\sigma}$.
\end{theorem}

Theorem \ref{thm: asymptotic normality} establishes the asymptotic normality of each row of the estimated factor loading matrices, with means aligned to the true values via rotations $\bR_{\bA},\bR_{\bB}$. The asymptotic variances involve the column-wise covariance matrices of the projected factors ($\bSigma_{\bF}$ and $\bSigma_{\bG}$), and the covariance matrices of the signal-by-noise cross-products ($\bSigma_{\bA,i}$ and $\bSigma_{\bB,j}$), analogous to statistical inference in conventional factor models. However, our inferential framework relies on a fine-grained analysis of the covariance matrix of the projected data. Specifically, we employ spectral representation tools from \cite{xia2021normal} to achieve a second-order accurate characterization of matrix SVD, enabling precise control of the leading term and sharp bounds on higher-order terms. This differs fundamentally from techniques used in vector factor models \citep{bai2003inferential} and Tucker factor models \citep{chen2023statistical, yu2022projected}, which adopt the approach of \cite{bai2003inferential}. Our framework offers two advantages: first, we establish explicit convergence rates for asymptotic normality, which previous methods cannot provide; second, the technique in \cite{bai2003inferential} relies on a manually designed rotation matrix for error decomposition, which does not extend to our setting due to the modewise additive structure and high dimensionality of the factor processes. Our approach can be  generalized to inference in other matrix and tensor factor models, such as CP factor models \citep{chang2023modelling,chang2024identification}.

The convergence rates of asymptotic normality $\Omega_{\bA}$ and $\Omega_{\bB}$ comprise three terms. Taking $\Omega_{\bA}$ as an example: (i) the first term $r_1\sqrt{1/(nd_1)}d_2^{\delta_1-\delta_0}$ is the Berry Esseen type bound for signal-by-noise cross-product of the first-order term; (ii) the second term, $ d_2^{(\delta_1-\delta_0)/2} \sqrt{r_1\log(nd_1)} \overline{\varepsilon}_{\bB}$, arises from estimation error in $\bU_{\bB}$ propagating to the first-order term; (iii) the third term $d_2^{(\delta_1-\delta_0)/2} \cdot \sqrt{n} d_2^{(1-\delta_0)/2}d_1^{1/2} (\overline{\varepsilon}_{\bA}^2 + \overline{\varepsilon}_{\bA}\overline{\varepsilon}_{\bB} )$ reflects the noise-by-noise crossproducts of the first order term, while the normal approximation is derived based on the first-order signal-by-noise cross-products induced by the perturbation $\widehat{\bU}_{\bA} - \bU_{\bA}\bR_{\bA}$.

\begin{remark}
Assume $d_1 \asymp d_2$ with fixed ranks $r_1,r_2$. Up to logarithmic factors, the error bounds satisfy
$
\Omega_{\bA} \asymp \Omega_{\bB} 
%%\asymp & d^{\delta_1-\delta_0} \sqrt{\frac{1}{n d}} + \left(\frac{1}{d^{1-\delta_1}}+\frac{d^{\delta_1-\frac{\delta_0}{2}}}{\sqrt{n d}}\right) d^{(\delta_1-\delta_0)/2} + \left(\frac{1}{d^{1-\delta_1}}+\frac{d^{\delta_1-\frac{\delta_0}{2}}}{\sqrt{n d}}\right)^2 \sqrt{n}d^{1 + \delta_1/2 - \delta_0} \\
%%d^{\frac{3}{2} \delta_1-\frac{\delta_0}{2}-1}+\frac{d^{\frac{3}{2} \delta_1-\delta_0-1 / 2}}{\sqrt{n}}+\sqrt{n} d^{\frac{5}{2} \delta_1-\delta_0-1}+\frac{d^{\frac{5}{2} \delta_1-2 \delta_0}}{\sqrt{n}}
\asymp  d_1^{3\delta_1/2-\delta_0/2-1} + n^{-1/2} d_1^{3\delta_1/2-\delta_0-1/2} +n^{-1/2} d_1^{5\delta_1/2 -2 \delta_0} +n^{1/2} d_1^{ 5\delta_1/2-\delta_0-1} .
$
If further assume $\delta_1=\delta_0$ (equal factor loading strength), this simplifies to
$\Omega_{\bA} \asymp \Omega_{\bB} \asymp  d_1^{\delta_0-1} + n^{-1/2} d_1^{\delta_0/2 } +n^{1/2} d_1^{ 3\delta_0/2-1}$.
\end{remark}

\begin{remark}
In the strong factor model case with $\delta_0 = \delta_1 = 0$ and $d_1 \asymp d_2$, we have $\Omega_{\bA} \asymp \Omega_{\bB} \asymp \sqrt{n} d_1^{-1}$. Thus, asymptotic normality in Theorem \ref{thm: asymptotic normality} requires $n \ll d_1^2$.
This condition reflects a fundamental phase transition in high-dimensional factor models, sharing the same spirit as Theorem 1 of \cite{bai2003inferential} and Theorem 2 of  \cite{chen2023statistical}. When $n \ll d_1^2$, the first-order signal-by-noise cross-product dominates the perturbation $\widehat{\bU}_{\bA} - \bU_{\bA}\bR_{\bA}$, yielding the asymptotic normality stated in the theorem. Otherwise, the first-order noise-by-noise cross-product becomes the dominant contributor, introducing non-negligible bias that requires more sophisticated distributional characterization involving the error covariance structure. The latter case parallels traditional PCA, where debiasing is often necessary to achieve asymptotic normality for linear combinations of principal components \citep{koltchinskii2020efficient}. Notably, if the entries of noise process $\bE_t$ are i.i.d., the condition $n \ll d_1^2$ can be removed, as the population noise covariance does not affect the singular space of the covariance of the projected data.

\end{remark}

\subsection{Data-Driven Statistical Inference}

Although statistical inference for loading matrices has been studied in vector \citep{bai2003inferential}, Tucker \citep{chen2023statistical}, and CP \citep{chen2026estimation} factor models, practical procedures for constructing confidence intervals remain underexplored, where covariance matrix estimation is crucial, particularly under weak factor settings. In this subsection, we develop a data-driven inference approach based on plug-in estimators. %We derive consistent estimators of the variance-covariance matrices.
We first estimate the projected factor covariance matrices, which capture the row-wise covariances of the projected factors $\bU_{\bB\perp} \bU_{\bB\perp}^{\top}\cano \bF_t$ and $\bU_{\bA\perp} \bU_{\bA\perp}^{\top}\cano \bG_t$. Using the estimated factors in \eqref{eq:factors}, the plug-in estimators for the projected factor covariance matrices in \eqref{eqn: sample projected factor covariance} are
\begin{equation} \label{eqn: sample projected factor covariance estimate}
\widehat{\bSigma}_{\bF} = \frac{1}{n} \sum_{t=1}^n \widehat{\bU}_{\bA}^{\top} \bX_t^{\top} \widehat{\bU}_{\bB\perp} \widehat{\bU}_{\bB\perp}^{\top} \bX_t \widehat{\bU}_{\bA}, \quad 
\widehat{\bSigma}_{\bG} = \frac{1}{n} \sum_{t=1}^n \widehat{\bU}_{\bB}^{\top} \bX_t \widehat{\bU}_{\bA\perp} \widehat{\bU}_{\bA\perp}^{\top} \bX_t^{\top} \widehat{\bU}_{\bB} .
\end{equation}
A direct error metric $\|\widehat{\bSigma}_{\bF} - \bSigma_{\bF}\|$ and $\|\widehat{\bSigma}_{\bG} - \bSigma_{\bG}\|$ is uninformative,  since the covariance matrices scale with dimensions, $\|\bSigma_{\bF}\| = O_{\PP}(d_1d_2^{1-\delta_0})$ and $\|\bSigma_{\bG}\| = O_{\PP}(d_2d_1^{1-\delta_0})$, and are subject to rotational ambiguity. Instead, we adopt standardized error metrics $\|\bR_{\bA}^{\top}\bSigma_{\bF}^{-1/2}\bR_{\bA}\widehat{\bSigma}_{\bF}\bR_{\bA}^{\top}\bSigma_{\bF}^{-1/2}\bR_{\bA} - \bI_{r_1}\|$ and $\|\bR_{\bB}^{\top}\bSigma_{\bG}^{-1/2}\bR_{\bB}\widehat{\bSigma}_{\bG}\bR_{\bB}^{\top}\bSigma_{\bG}^{-1/2}\bR_{\bB} - \bI_{r_2}\|$, which accounts for both the scale of $\cano \bF_t,\cano \bG_t$ and rotation. 

\begin{theorem}\label{thm: loading scale estimation}
Suppose the conditions of Theorem~\ref{thm: iteration} hold. Let $\bR_{\bA}, \bR_{\bB}$ be the rotation matrices from Theorem~\ref{thm: asymptotic normality}. Then, with probability at least $1 - \exp(-cn^{\varphi/2})- \exp(-c(d_1 \wedge d_2))$, we have 
\begin{align*}
\left\|\bR_{\bA}^{\top}\bSigma_{\bF}^{-1/2}\bR_{\bA}\widehat{\bSigma}_{\bF} \bR_{\bA}^{\top}\bSigma_{\bF}^{-1/2}\bR_{\bA}- \bI_{r_1}\right\| 
\le & C_1\overline{\varepsilon}_{\bA} + C_1  d_2^{(\delta_1-\delta_0)/2} \overline{\varepsilon}_{\bA}^2 + C_1 d_2^{\delta_1-\delta_0} \overline{\varepsilon}_{\bA}^2 \left(\overline{\varepsilon}_{\bB}+\overline{\varepsilon}_{\bA}^2\right) ,  \\
\left\|\bR_{\bB}^{\top}\bSigma_{\bG}^{-1/2}\bR_{\bB}\widehat{\bSigma}_{\bG} \bR_{\bB}^{\top}\bSigma_{\bG}^{-1/2}\bR_{\bB}- \bI_{r_2}\right\| 
\le & C_2\overline{\varepsilon}_{\bB} + C_2 d_1^{(\delta_1-\delta_0)/2} \overline{\varepsilon}_{\bB}^2 + C_2 d_1^{\delta_1-\delta_0}  \overline{\varepsilon}_{\bB}^2 \left(\overline{\varepsilon}_{\bA}+\overline{\varepsilon}_{\bB}^2\right) ,
\end{align*}
where $C_1,C_2>0$ are constants depending on $\overline{\sigma}_{F}, \underline{\sigma}_{ F}, \overline{\sigma}_{G}, \underline{\sigma}_{G}, \overline{\sigma}$, $\underline{\sigma}$, and $\overline{\varepsilon}_{\bA}$, $\overline{\varepsilon}_{\bB}$ are defined in \eqref{eqn: epsilon bar A bar B}. 
%Here, $C_1, C_2$ depend on the variance components, $\overline{\sigma}_{F}, \underline{\sigma}_{ F}, \overline{\sigma}_{G}, \underline{\sigma}_{G}, \overline{\sigma}$ and $\underline{\sigma}$.
\end{theorem}

Theorem~\ref{thm: loading scale estimation} shows that the convergence rates are controlled by the loading matrix estimation errors $\overline{\varepsilon}_{\bA},\overline{\varepsilon}_{\bB}$. It shows that accurate loading matrix estimation is sufficient for accurate estimation of the factor covariance matrices. The factors $d_1^{\delta_1-\delta_0}$ and $d_2^{\delta_1-\delta_0}$ indicate that heterogeneity in factor loading strengths (larger gap $\delta_1-\delta_0$) degrades the precision of the covariance estimate.

Next, we construct plug-in estimators for the covariance matrices $\bSigma_{\bA,i}$ and $\bSigma_{\bB,j}$. %This requires first estimating the noise covariance matrices $\bSigma_{\bE}$ and $\bSigma_{\bE^\top}$. 
Define the noise covariance estimators
\begin{align}
\widehat{\bSigma}_{\bE} &= \frac{1}{n} \sum_{t=1}^n \Vec (\bX_t - \widehat{\bF}_t \widehat{\bU}_{\bA}^{\top} - \widehat{\bU}_{\bB} \widehat{\bG}_t^{\top}) \Vec (\bX_t - \widehat{\bF}_t \widehat{\bU}_{\bA}^{\top} - \widehat{\bU}_{\bB} \widehat{\bG}_t^{\top})^{\top}  ,\label{eqn: Sigma_hat_E}\\
\widehat{\bSigma}_{\bE^{\top}} &= \frac{1}{n} \sum_{t=1}^n \Vec (\bX_t^{\top} - \widehat{\bU}_{\bA}\widehat{\bF}_t^{\top} - \widehat{\bG}_t\widehat{\bU}_{\bB}^{\top}) \Vec (\bX_t^{\top} - \widehat{\bU}_{\bA}\widehat{\bF}_t ^{\top} - \widehat{\bG}_t\widehat{\bU}_{\bB}^{\top})^{\top},  \label{eqn: Sigma_hat_Et}
\end{align}
for $\bSigma_{\bE}$ and $\bSigma_{\bE^{\top}}$ (defined in Assumption~\ref{assum: error term}). The plug-in estimators for $\bSigma_{\bA,i}$ and $\bSigma_{\bB,j}$ are
\begin{align}
\widehat{\bSigma}_{\bA,i} &= \frac{1}{n}\sum_{t=1}^n\left[e_i^{\top}\widehat{\bU}_{\bA\perp}\widehat{\bU}_{\bA\perp}^{\top} \otimes \widehat{\bF}_t^{\top} \widehat{\bU}_{\bB \perp}\widehat{\bU}_{\bB \perp}^{\top}\right] \widehat{\bSigma}_{\bE} \left[\widehat{\bU}_{\bA\perp}\widehat{\bU}_{\bA\perp}^{\top} e_i \otimes \widehat{\bU}_{\bB\perp} \widehat{\bU}_{\bB\perp}^{\top} \widehat{\bF}_t \right], \label{eqn: Sigma_hat_A}\\ 
\widehat{\bSigma}_{\bB,j} &= \frac{1}{n}\sum_{t=1}^n\left[e_j^{\top}\widehat{\bU}_{\bB\perp}\widehat{\bU}_{\bB\perp}^{\top} \otimes \widehat{\bG}_t^{\top} \widehat{\bU}_{\bA \perp}\widehat{\bU}_{\bA \perp}^{\top}\right] \widehat \bSigma_{\bE^{\top}} \left[\widehat{\bU}_{\bB\perp}\widehat{\bU}_{\bB\perp}^{\top}e_j \otimes \widehat{\bU}_{\bA \perp} \widehat{\bU}_{\bA \perp}^{\top} \widehat{\bG}_t\right], \label{eqn: Sigma_hat_B}
\end{align}
respectively. As in Theorem~\ref{thm: loading scale estimation}, since $\|\bSigma_{\bA,i}\| = O_{\PP}(d_2^{1-\delta_0})$ and $\|\bSigma_{\bB,j}\| = O_{\PP}(d_1^{1-\delta_0})$ with rotational ambiguity, we adopt standardized error metrics $\|\bR_{\bA}^{\top}\bSigma_{\bF}^{-1/2}\bR_{\bA}\widehat{\bSigma}_{\bA,i}\bR_{\bA}^{\top}\bSigma_{\bF}^{-1/2}\bR_{\bA} - \bI_{r_1}\|$ and $\|\bR_{\bB}^{\top}\bSigma_{\bG}^{-1/2}\bR_{\bB}\widehat{\bSigma}_{\bB,j}\bR_{\bB}^{\top}\bSigma_{\bG}^{-1/2}\bR_{\bB} - \bI_{r_2}\|$.

\begin{theorem}\label{thm: covariance matrix}
Suppose the conditions of Theorem~\ref{thm: iteration} hold. Let $\bR_{\bA}, \bR_{\bB}$ be the rotation matrices from Theorem~\ref{thm: asymptotic normality}. Then, with probability at least $1 - \exp(-cn^{\varphi/2})- \exp(-c(d_1 \wedge d_2))$, we have
\begin{align*}
\left\|\bR_{\bA}^{\top}\bSigma_{\bA,i}^{-1/2}\bR_{\bA}\widehat{\bSigma}_{\bA,i}\bR_{\bA}^{\top}\bSigma_{\bA,i}^{-1/2}\bR_{\bA} - \bI_{r_1}\right\|
\le &  C_1 d_2^{\delta_1-\delta_0} \left( d_2^{1-\delta_0} \overline{\varepsilon}_{\bA}^2 +  d_1^{1-\delta_0} \overline{\varepsilon}_{\bB}^2\right)  \left(1 + \frac{(d_1+d_2)^{1/\varphi}}{n}\right)  , \\
\left\|\bR_{\bB}^{\top}\bSigma_{\bB,j}^{-1/2}\bR_{\bB}\widehat{\bSigma}_{\bB,j}\bR_{\bB}^{\top}\bSigma_{\bB,j}^{-1/2}\bR_{\bB} - \bI_{r_1}\right\| 
\le & C_2 d_1^{\delta_1-\delta_0} \left( d_1^{1-\delta_0} \overline{\varepsilon}_{\bB}^2 +  d_2^{1-\delta_0} \overline{\varepsilon}_{\bA}^2 \right)  \left(1 + \frac{(d_1+d_2)^{1/\varphi}}{n}\right)  ,
\end{align*}
where $C_1,C_2>0$ are constants depending on $\overline{\sigma}_{F}, \underline{\sigma}_{ F}, \overline{\sigma}_{G}, \underline{\sigma}_{G}, \overline{\sigma}$, $\underline{\sigma}$, and $\overline{\varepsilon}_{\bA}$, $\overline{\varepsilon}_{\bB}$ are defined in \eqref{eqn: epsilon bar A bar B}. 
%Here, $C_1, C_2$ depend on the variance components, $\overline{\sigma}_{F}, \underline{\sigma}_{ F}, \overline{\sigma}_{G}, \underline{\sigma}_{G}, \overline{\sigma}$ and $\underline{\sigma}$.
\end{theorem}

Theorem~\ref{thm: covariance matrix} shows that the convergence rates are governed by the loading matrix estimation errors and a factor $(d_1+d_2)^{1/\varphi}/n$ reflecting temporal dependence. The terms $d_1^{\delta_1-\delta_0}$ and $d_2^{\delta_1-\delta_0}$ capture the effect of heterogeneity in factor loading strengths.

Combining the covariance matrices estimators from Theorems~\ref{thm: loading scale estimation} and \ref{thm: covariance matrix}, we establish a feasible data-driven inference procedure.

\begin{theorem}\label{cor: data driven normality}
Suppose the conditions of Theorem~\ref{thm: iteration} hold. Let $\calF_{r_k}$ denote the Borel $\sigma$-algebra on $\RR^{r_k}$ for $k=1,2$, and let $e_i$ denote the $i$-th standard basis vector. Then, for all $i \in [d_2]$ and $j \in [d_1]$,
\begin{align*}
\sup_{\calA \in \calF_{r_1}}\left|\PP\left(\sqrt{n} \widehat{\bSigma}_{\bA,i}^{-\frac{1}{2}}\widehat{\bSigma}_{\bF}\left(\widehat{\bU}_{\bA}^{\top}- \bR_{\bA}^{\top}\bU_{\bA}^{\top}\right)e_i \in \calA\right) - \PP\left(\mathcal{N}(0, \bI_{r_1}) \in \calA \right)\right| \le & C_1 \widehat{\Omega}_{\bA}, \\
\sup_{\calB \in \calF_{r_2}}\left|\PP\left(\sqrt{n} \widehat{\bSigma}_{\bB,j}^{-\frac{1}{2}}\widehat{\bSigma}_{\bG}\left(\widehat{\bU}_{\bB}^{\top} -\bR_{\bB}^{\top}\bU_{\bB}\right)e_j \in \calB\right) - \PP\left(\mathcal{N}(0, \bI_{r_2}) \in \calB\right)\right| \le & C_2 \widehat{\Omega}_{\bB},
\end{align*}
where $C_1,C_2>0$ are constants depending on $\overline{\sigma}_{F}, \underline{\sigma}_{ F}, \overline{\sigma}_{G}, \underline{\sigma}_{G}, \overline{\sigma}$, $\underline{\sigma}$, and the error bounds are
\begin{align*}
\widehat{\Omega}_{\bA}
= & d_2^{3(\delta_1-\delta_0)/2} \sqrt{r_1\log(nd_1)}\left( d_2^{1-\delta_0} \overline{\varepsilon}_{\bA}^2 + d_1^{1-\delta_0} \overline{\varepsilon}_{\bB}^2 \right)  \left(1 + \frac{(d_1+d_2)^{1/\varphi}}{n}\right) \\
&\quad + d_2^{(\delta_1-\delta_0)/2} \cdot \sqrt{n} d_2^{(1-\delta_0)/2}d_1^{1/2} \overline{\varepsilon}_{\bA}^2 , \\
\widehat{\Omega}_{\bB}
= & d_1^{3(\delta_1-\delta_0)/2} \sqrt{r_2\log(nd_2)} \left( d_1^{1-\delta_0} \overline{\varepsilon}_{\bB}^2 + d_2^{1-\delta_0} \overline{\varepsilon}_{\bA}^2 \right)  \left(1 + \frac{(d_1+d_2)^{1/\varphi}}{n}\right) \\
&\quad + d_1^{(\delta_1-\delta_0)/2} \cdot \sqrt{n} d_1^{(1-\delta_0)/2}d_2^{1/2} \overline{\varepsilon}_{\bB}^2.
\end{align*}
%Here, $C_1, C_2$ depend on the variance components, $\overline{\sigma}_{F}, \underline{\sigma}_{ F}, \overline{\sigma}_{G}, \underline{\sigma}_{G}, \overline{\sigma}$ and $\underline{\sigma}$.
\end{theorem}

In Theorem \ref{cor: data driven normality}, the rotation matrices $\bR_{\bA}$ and $\bR_{\bB}$ naturally cancel in the plug-in procedure. Compared with Theorem~\ref{thm: asymptotic normality}, the error bounds are inflated by factors $d_1^{(\delta_1-\delta_0)/2}$ and $d_2^{(\delta_1-\delta_0)/2}$ due to repeated use of the estimated factors $\widehat{\bF}_t$ and $\widehat{\bG}_t$, respectively. In addition, accurate loading matrix estimation $\widehat{\bU}_{\bA}$ and $\widehat{\bU}_{\bB}$ is required to counteract the factor strength contributions $d_2^{1-\delta_0}$ from $\bA$ and $d_1^{1-\delta_0}$ from $\bB$ when estimating $\bSigma_{\bE}$ and $\bSigma_{\bE^{\top}}$. The term $(d_1+d_2)^{1/\varphi}/n$ accounts for dependence between loading matrix estimation and factors/noises.

\begin{remark}
Assume $d_1 \asymp d_2 $ with fixed ranks $r_1,r_2$. Up to logarithmic factors, the error bounds $\widehat{\Omega}_{\bA} $ and $ \widehat{\Omega}_{\bB} $ have the rate
$$
\sqrt{n} d_1^{\frac{5}{2} \delta_1-\delta_0-1} + d_1^{\frac{5}{2} \delta_1-\frac{3}{2} \delta_0-\frac{1}{2}} + \frac{d_1^{\frac{5}{2} \delta_1-2 \delta_0}}{\sqrt{n}} 
+  \Big(d_1^{\frac{7}{2} \delta_1-\frac{5}{2} \delta_0-1}+\frac{d_1^{\frac{7}{2} \delta_1-3 \delta_0-\frac{1}{2}}}{\sqrt{n}}+\frac{d_1^{\frac{7}{2} (\delta_1-\delta_0 )}}{n} \Big) \max \Big\{1, \frac{ d_1^{1 / \varphi}}{n} \Big\}.
$$
For equal factor strength $\delta_1=\delta_0$, this simplifies to
$n^{1/2} d_1^{3\delta_0/2-1} + d_1^{\delta_0- 1/2} + n ^{-1/2} d_1^{\delta_0/2}
+  (d_1^{ \delta_0-1}+n^{-1/2} d_1^{\delta_0/2-1/2}  ) \max \{1, n^{-1}d_1^{1 / \varphi} \} .$
For strong factors ($\delta_0=\delta_1 = 0$) with independent factor processes ($\theta_1 = \infty$) and sub-Gaussian tails ($\theta_2 = 2$), we have $\varphi = 1$, and valid data-driven inference requires $d_1^{1/3} \lesssim n \lesssim d_1^{2}$. For strong factors ($\delta_0=\delta_1 = 0$) with standard exponential mixing ($\theta_1 = 1$) and sub-Gaussian tails ($\theta_2 = 2$), we have $\varphi = 1/2$, requiring $d_1^{1/2} \lesssim n \lesssim d_1^{2}$ for valid inference. 
\end{remark}

\section{Matrix Bernstein Inequality for Quadratic Forms of Matrix Time Series}\label{sec:ineq}

Tail probability inequalities play an important role in statistical estimation and inference. In this section, we derive powerful tail probability inequalities for quadratic forms of matrix time series under exponential-type tail conditions. The following lemma, proved in Appendix~C.3, extends classical matrix Bernstein inequalities for i.i.d. random matrices to quadratic forms of dependent matrix time series and is of independent interest. We refer readers to Appendices~C.2 and C.4 for tail probability inequalities involving Gaussian random matrices and matrix time series with exponential-type tails.

\begin{lemma}\label{lem: concentration GtWG FtWF main text} %[cf. Lemma~\ref{lem: concentration GtWG FtWF}]
Suppose Assumptions~\ref{assum: alpha-mixing}, \ref{assum: independent factor rows}, and \ref{assum: factor tail bound} hold. Let $1/\varphi = 1/\theta_1 + 2/\theta_2$. Then, for all $x\gtrsim \sqrt{n r_1} + r_1^{1/\phi}$,
\begin{align}
\label{eqn: concentration FtWF main text}
\PP\left(\left\|\sum_{t=1}^n (\bF_t^\top \bW \bF_t - \EE[\bF_t^\top \bW \bF_t]) \right\| \ge \overline{\sigma}_{F}^2 \sqrt{s}  x \right) 
\le &  n\exp \left\{-c x^\varphi \right\} + \exp \left\{- \frac{c x^2}{n} \right\} ,
\end{align}
where $\bW \in \calT_{d_1, s} =  \left\{ \bW^* \in \RR^{d_1 \times d_1} : \| \bW^* \| \le 1,\ \bW^* = \bW^{*\top},\ {\rm rank}(\bW^*) \le s \right\}$, and $c>0$ is a constant.
\end{lemma}

\section{Simulation}\label{sec:simul}
In this section, we evaluate the empirical performance of the two-stage estimation procedure, MINE and COMPAS, and
compare it with a computationally efficient variant of COMPAS that uses partial (subset) orthogonal complement projections, validating the empirical efficiency loss of this practical alternative. We also examine the asymptotic normality of rows of the estimated loading matrices and verify that data-driven statistical inference procedures achieve nominal coverage rates, demonstrating accurate uncertainty quantification in finite samples.

\noindent
\textbf{Data generation.} The matrix time series $\{\bX_t\}_{t=1}^n \in \RR^{d_1 \times d_2}$ are generated according to MAFM model \eqref{eqn: factor model}. Each row of the latent row-factor matrix $\bfv_{t,i} \in \RR^{r_1}$ follows an independent VAR(1) process $\bfv_{t,i} = \bPhi_{F,i} \bfv_{t-1,i} + \boldsymbol{\varepsilon}_{t,i}$ with $\boldsymbol{\varepsilon}_{t,i} \stackrel{\text{i.i.d.}}{\sim} \cN(0, \bI_{r_1})$, and similarly for each row of the latent column-factor matrix $\bg_{t,j} \in \RR^{r_2}$. To induce heterogeneous temporal dynamics, each coefficient matrix $\bPhi_{F,i}$ (or $\bPhi_{G,j}$) is constructed as $\bQ_i \bD_i(\phi_1, \phi_2) \bQ_i^\top$, where $\bD_i(\phi_1, \phi_2) = {\rm diag}(\underbrace{\phi_1, \ldots, \phi_1}_{\lceil r/2\rceil}, \underbrace{\phi_2, \ldots, \phi_2}_{r-\lceil r/2\rceil})$ with $r = r_1$ for row factors and $r = r_2$ for column factors, and $\bQ_i$ is an orthogonal matrix from QR decomposition of a standard Gaussian random matrix. Each coordinate is uniformly assigned to one of three representative eigenvalue pairs $(\phi_1, \phi_2) \in \{(0.90, 0.70),(0.50, -0.50),(-0.90, -0.70)\}$, representing highly persistent, moderately persistent, and oscillatory dynamics, respectively. 

The loading matrices $\bA \in \RR^{d_2 \times r_1}$ and $\bB \in \RR^{d_1 \times r_2}$ are constructed via SVD to satisfy the loading strength conditions in Assumption~\ref{assum: loading matrices scaling}. Specifically, singular values are geometrically spaced between $d_k^{(1-\delta_0)/2}$ and $d_k^{(1-\delta_1)/2}$ for $k \in \{1,2\}$. This ensures $\|\bA\|_2 \asymp d_2^{(1-\delta_0)/2} $ and $\sigma_{r_1}(\bA) \asymp d_2^{(1-\delta_1)/2}$, with analogous conditions for $\bB$. We consider two factor loading strength regimes. In the strong factor loading regime, $(\delta_0,\delta_1)=(0,0)$, so that all row and column factors are pervasive and the loading matrix singular values grow at the same rate; in the weak factor loading regime, $(\delta_0,\delta_1)=(0.3,0.5)$, which increases the spread between the largest and smallest singular values for the loading matrices. 
The noise matrix $\bE_t$ has i.i.d. $N(0,\sigma^2_\varepsilon)$ entries with $\sigma^2_\varepsilon=1$, independent of the row and column factor processes.

\begin{figure}[htbp!]
    \centering
    \includegraphics[width=0.8\textwidth]{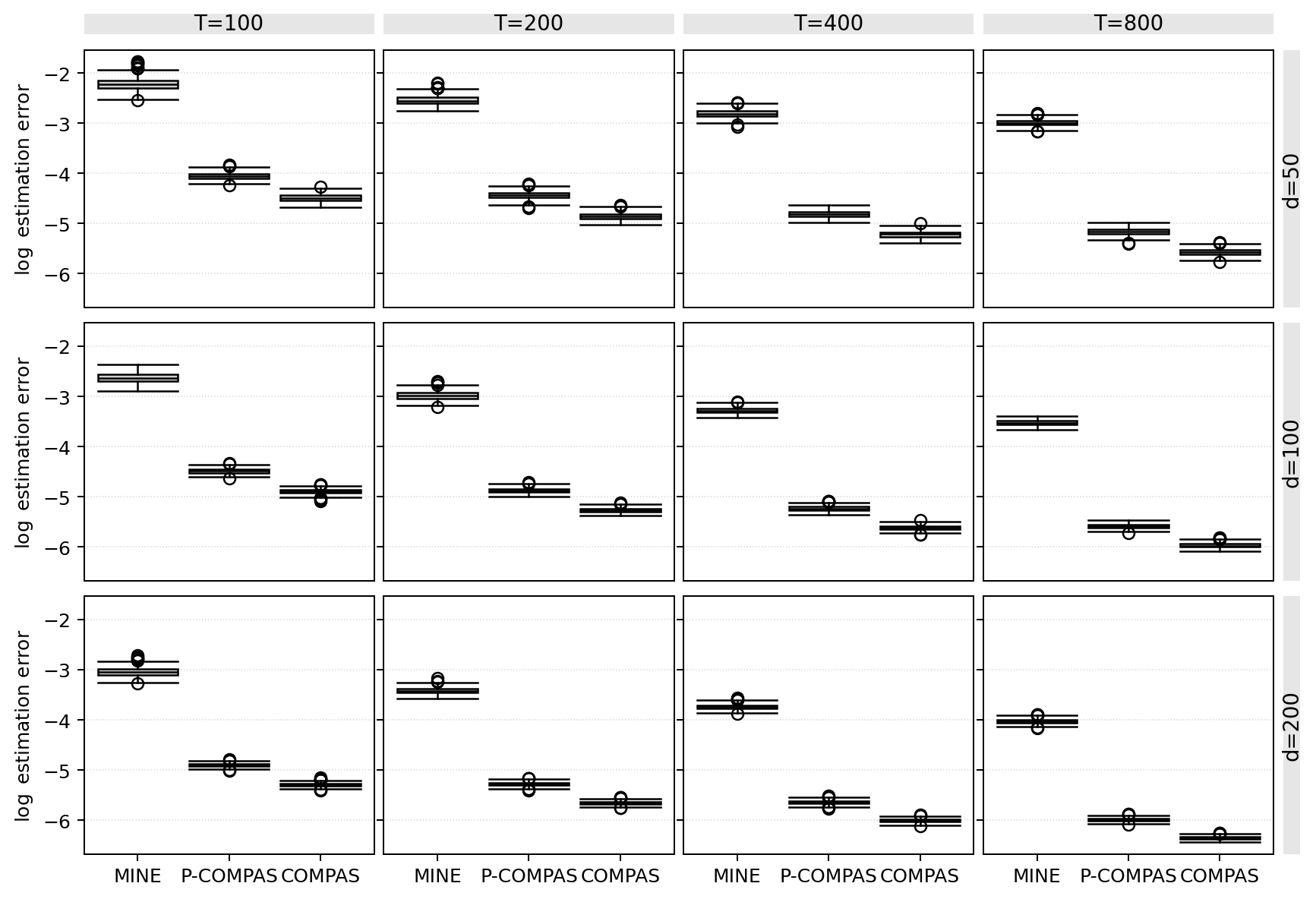}
    \caption{\small Estimation error $\log(\cD(\bU_{\bA}, \widehat\bU_{\bA}))$ under strong factor loading strength regime with $(\delta_0, \delta_1) = (0, 0)$ across dimensions $d$ and sample sizes $T$ for three estimation methods: MINE, P-COMPAS, and COMPAS.}
    \label{fig:subspace_A_strong}
\end{figure}

\begin{figure}[htbp!]
    \centering
    \includegraphics[width=0.8\textwidth]{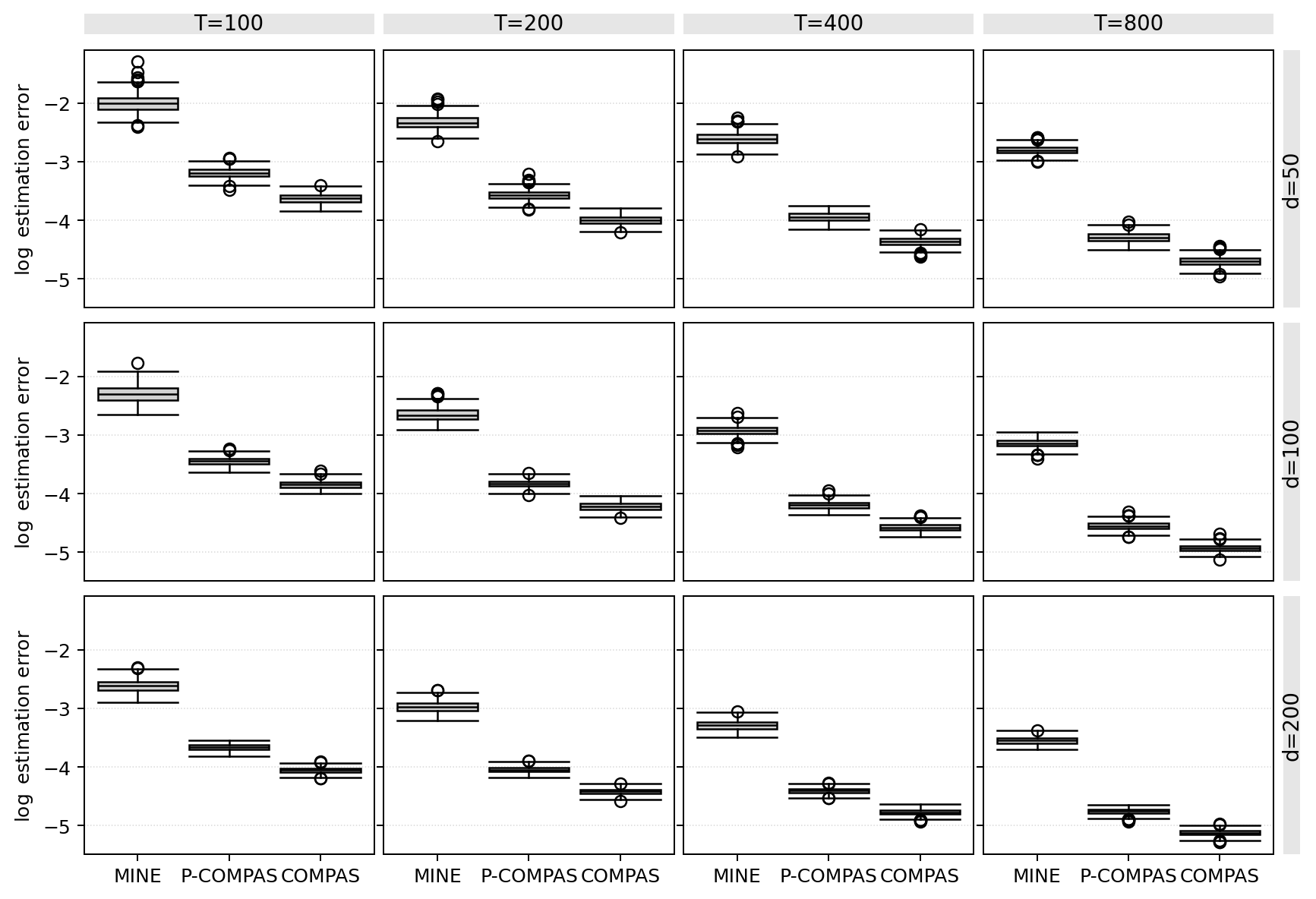}
    \caption{\small Estimation error $\log(\cD(\bU_{\bA}, \widehat\bU_{\bA}))$ under weak factor loading strength regime with  $(\delta_0, \delta_1) = (0.3, 0.5)$ across dimensions $d$ and sample sizes $T$ for three estimation methods: MINE, P-COMPAS, and COMPAS.}
    \label{fig:subspace_A_weak}
\end{figure}

\noindent
\textbf{Finite-sample validation of estimation error bounds.} Figures~\ref{fig:subspace_A_strong} and \ref{fig:subspace_A_weak} display the loading space estimation accuracy across varying dimensions $d_1 = d_2 \in \{50, 100, 200\}$ and sample sizes $T \in \{100, 200, 400, 800\}$ under both strong and weak factor loading strength regimes. For each configuration $(d, T)$, we fix one realization of $(\bA, \bB)$ and generate $500$ independent Monte Carlo replicates with new factor and noise processes. The figures display the average logarithm of estimation error $\log(\cD(\bU_{\bA}, \widehat \bU_{\bA}))$, where $\cD(\bU_{\bA}, \widehat \bU_{\bA})=\|\widehat \bU_{\bA} \widehat \bU_{\bA}^\top - \bU_{\bA}\bU_{\bA}^\top\|_2$, for three methods: MINE (Algorithm~\ref{alg: MINE}, initialization only), P-COMPAS (Algorithm~\ref{alg: COMPAS} with partial complement projections and MINE initialization), and COMPAS (Algorithm~\ref{alg: COMPAS} with full complement projections and MINE initialization).

The results reveal several key findings. First, the proposed COMPAS consistently achieves the lowest loading space estimation error across all configurations, substantially improving over the MINE initialization. Second, P-COMPAS exhibits degraded performance compared to COMPAS, confirming that partial (subset) orthogonal complement projections sacrifice statistical accuracy for computational efficiency, a tradeoff formalized in Theorem~\ref{thm: partial_iteration}. Third, a striking pattern emerges when comparing the two loading strength regimes: despite similarly poor MINE initialization in both cases, the strong factor loading regime (Figure~\ref{fig:subspace_A_strong}) enables substantially better final recovery than the weak loading regime (Figure~\ref{fig:subspace_A_weak}). This occurs because stronger loadings provide larger signal strength, allowing iterative refinement to effectively leverage the modewise additive structure and achieve sharper error bounds, whereas weaker factors fundamentally limit recovery accuracy. Analogous results for $\hat \bU_{\bB}$ appear in Appendix \ref{sec:simul_appendix}.

\begin{figure}[htbp!]
  \centering
  \resizebox{0.9\textwidth}{!}{%
  \begin{subfigure}{0.32\textwidth}
    \centering
    \includegraphics[width=\linewidth]{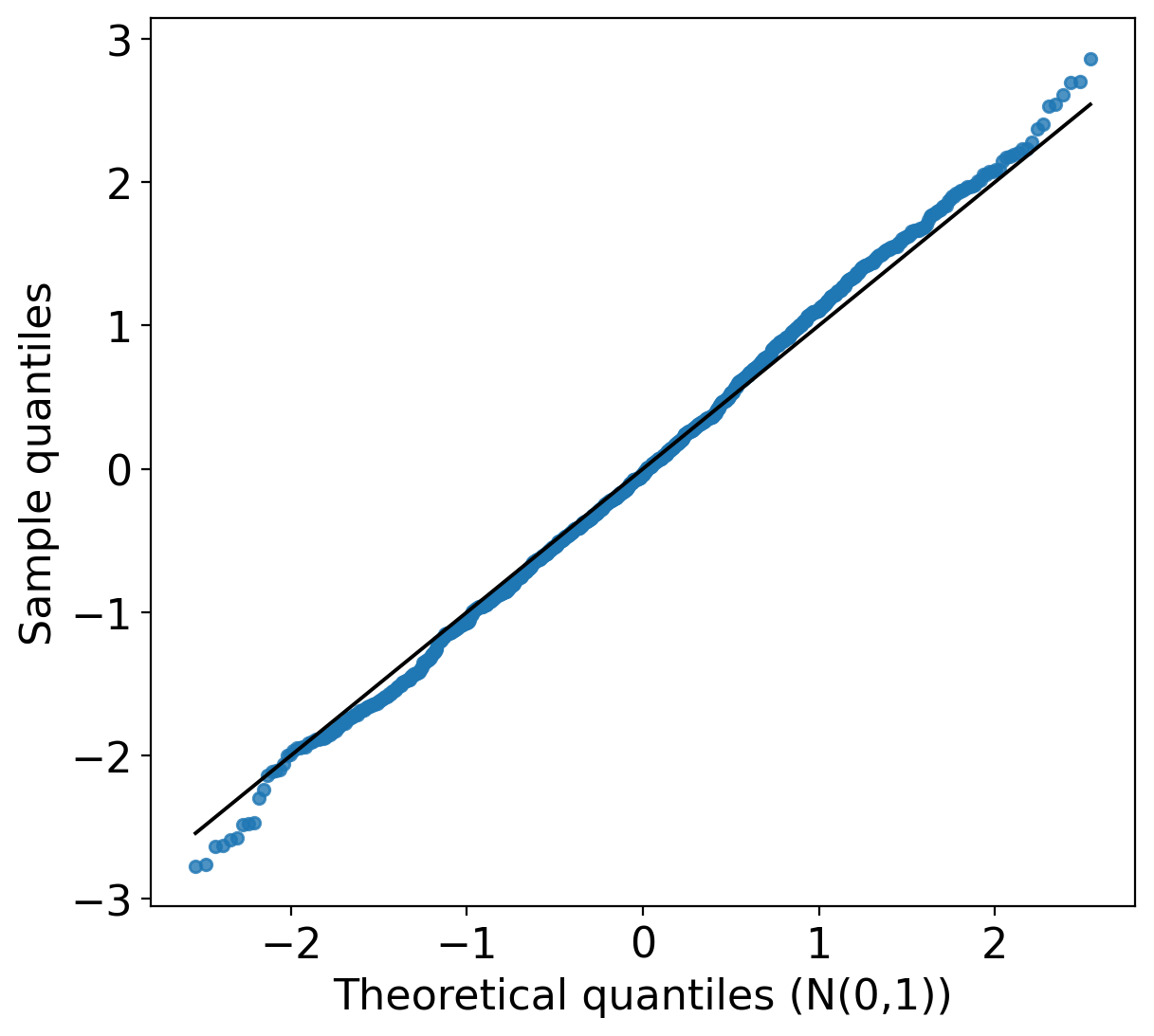}
    \caption{$d=20$, oracle}
  \end{subfigure}\hfill
  \begin{subfigure}{0.32\textwidth}
    \centering
    \includegraphics[width=\linewidth]{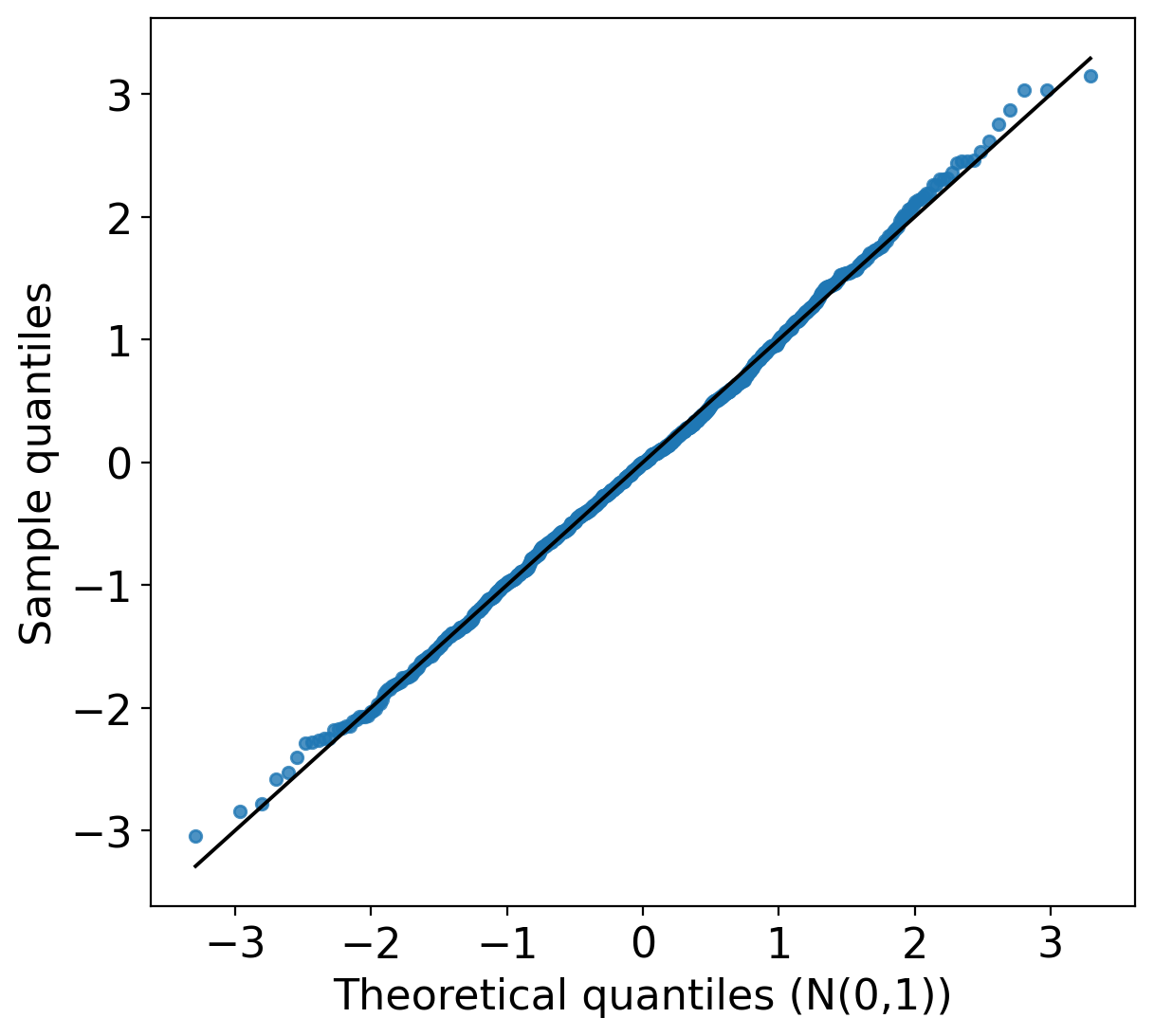}
    \caption{$d=50$, oracle}
  \end{subfigure}\hfill
  \begin{subfigure}{0.32\textwidth}
    \centering
    \includegraphics[width=\linewidth]{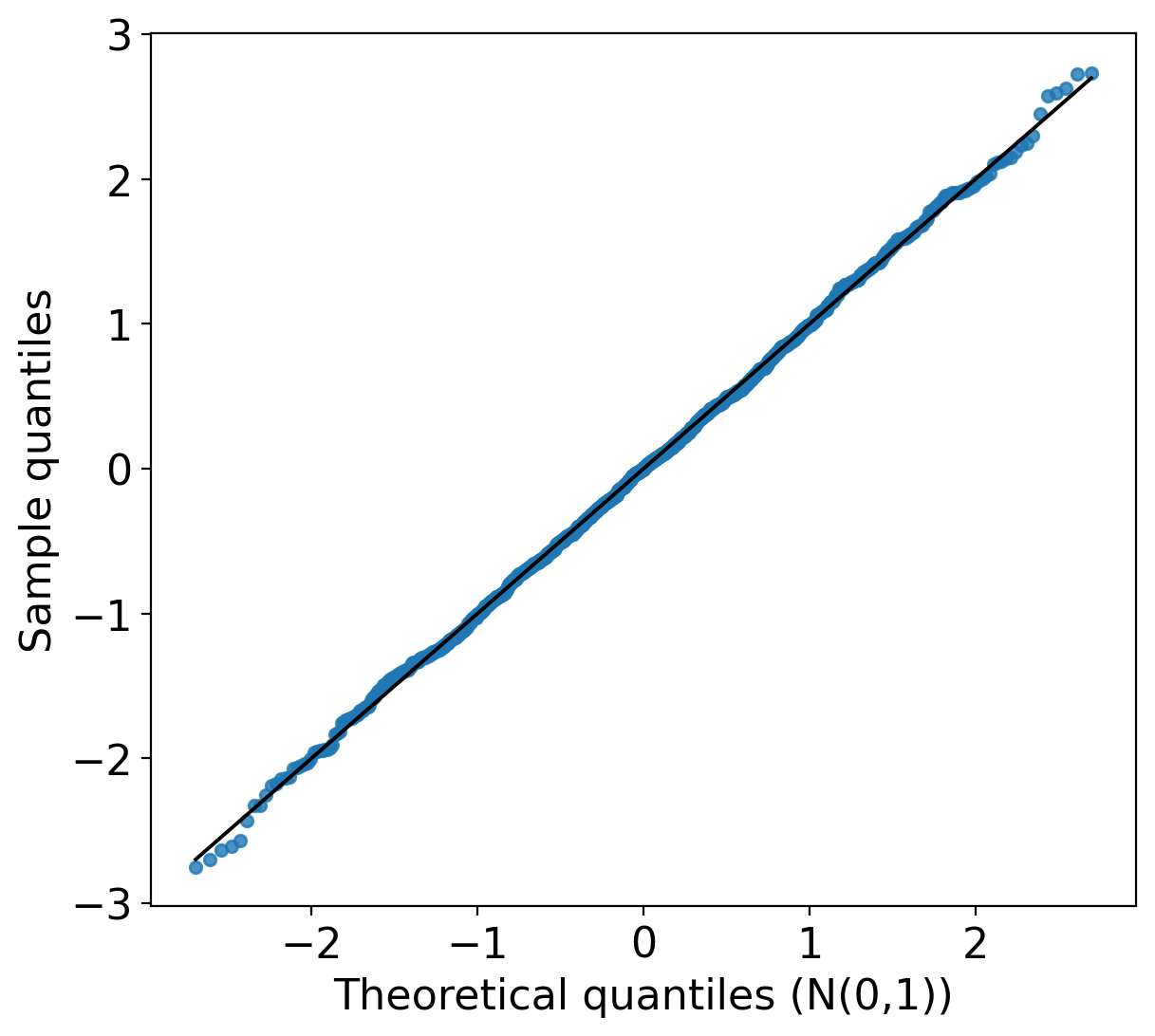}
    \caption{$d=100$, oracle}
  \end{subfigure}}
  \vspace{0.35em}
  \resizebox{0.9\textwidth}{!}{%
  \begin{subfigure}{0.32\textwidth}
    \centering
    \includegraphics[width=\linewidth]{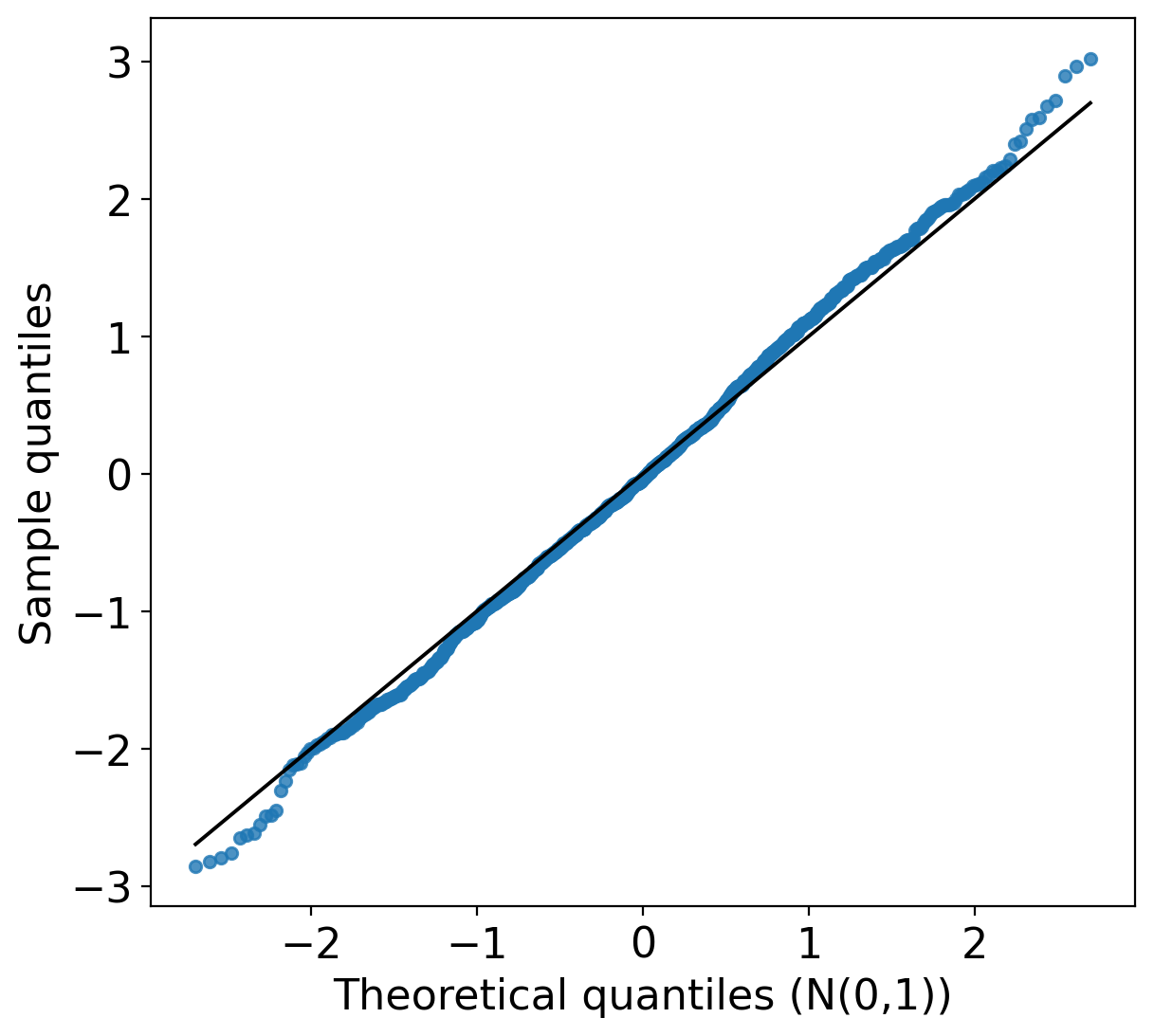}
    \caption{$d=20$, data-driven}
  \end{subfigure}\hfill
  \begin{subfigure}{0.32\textwidth}
    \centering
    \includegraphics[width=\linewidth]{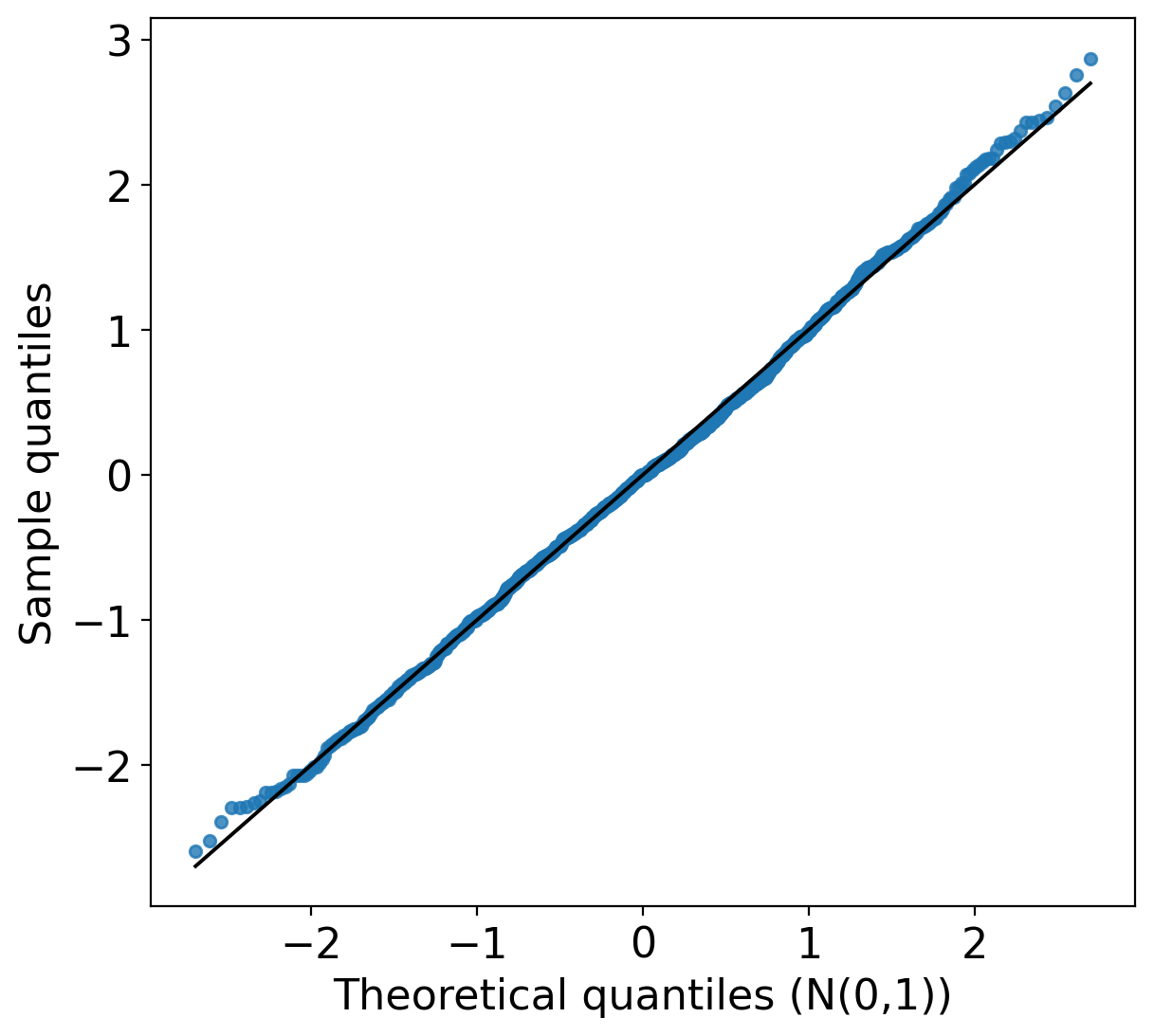}
    \caption{$d=50$, data-driven}
  \end{subfigure}\hfill
  \begin{subfigure}{0.32\textwidth}
    \centering
    \includegraphics[width=\linewidth]{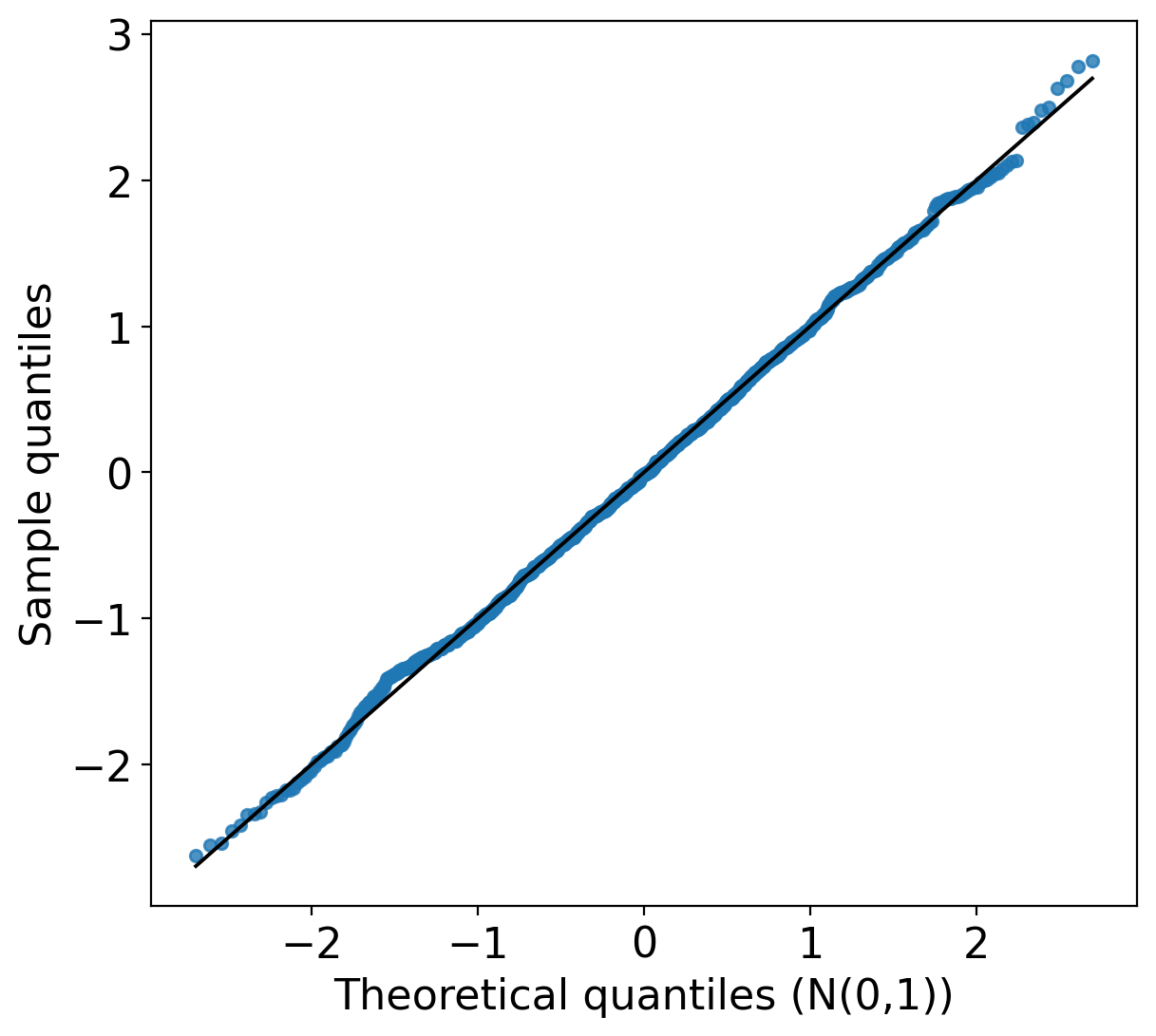}
    \caption{$d=100$, data-driven}
  \end{subfigure}}
  \caption{\small Normal QQ plots for the first coordinate of the standardized first row of $\widehat\bU_{\bA}-\bU_{\bA}\bR_{\bA}$ with $T=200$ under the strong loading strength regime.
  The top row uses the oracle inference procedure with population covariance matrices, and the bottom row uses the feasible data-driven inference procedure with plug-in covariance estimators.
  Columns correspond to $d\in\{20,50,100\}$.}
  \label{fig:qq_A_oracle_dd}
\end{figure}

\begin{figure}[t]
  \centering
  \resizebox{0.75\textwidth}{!}{%
  \begin{subfigure}{0.32\textwidth}
    \centering
    \includegraphics[width=\linewidth]{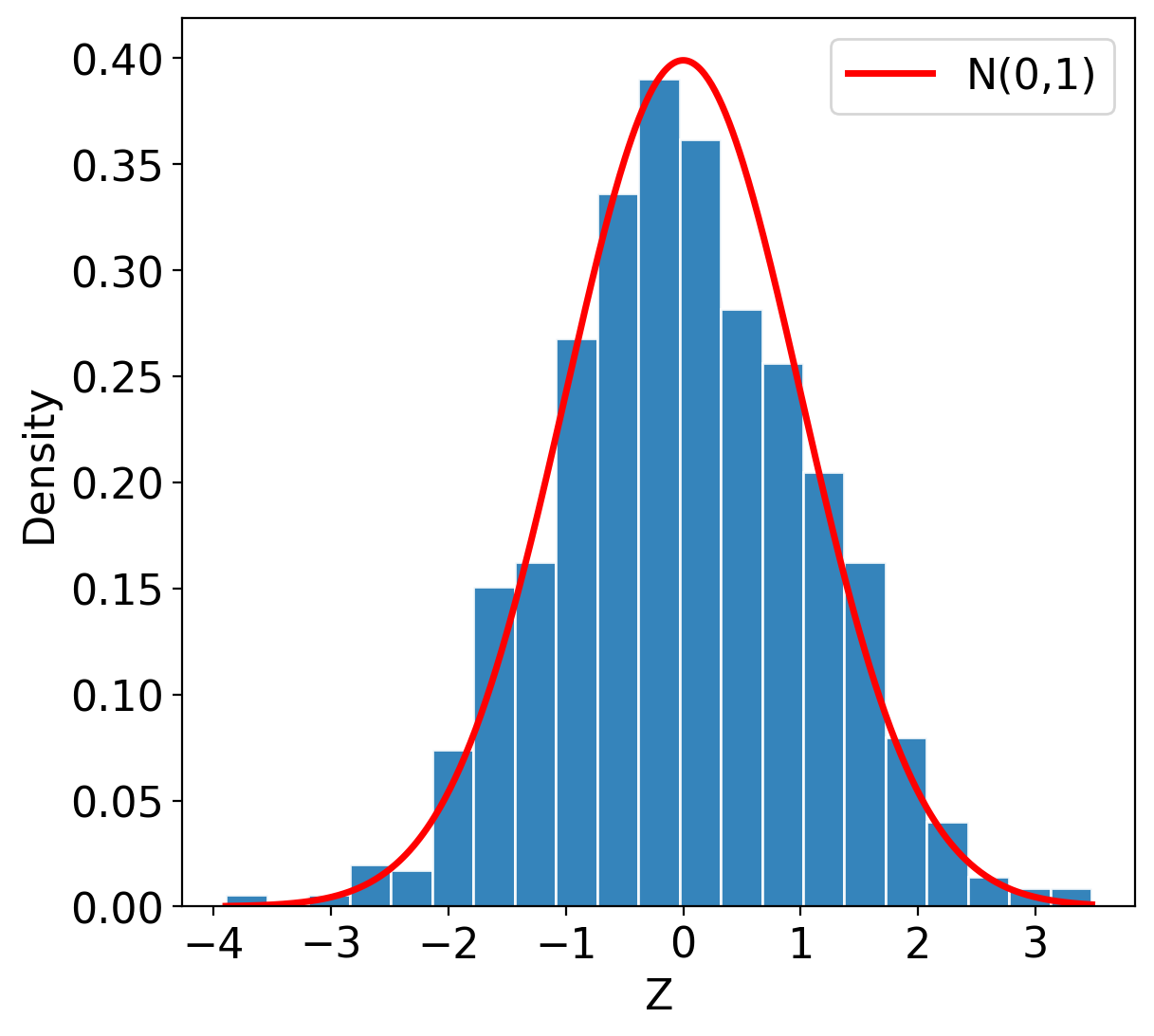}
    \caption{$d=20$, oracle}
  \end{subfigure}\hfill
  \begin{subfigure}{0.32\textwidth}
    \centering
    \includegraphics[width=\linewidth]{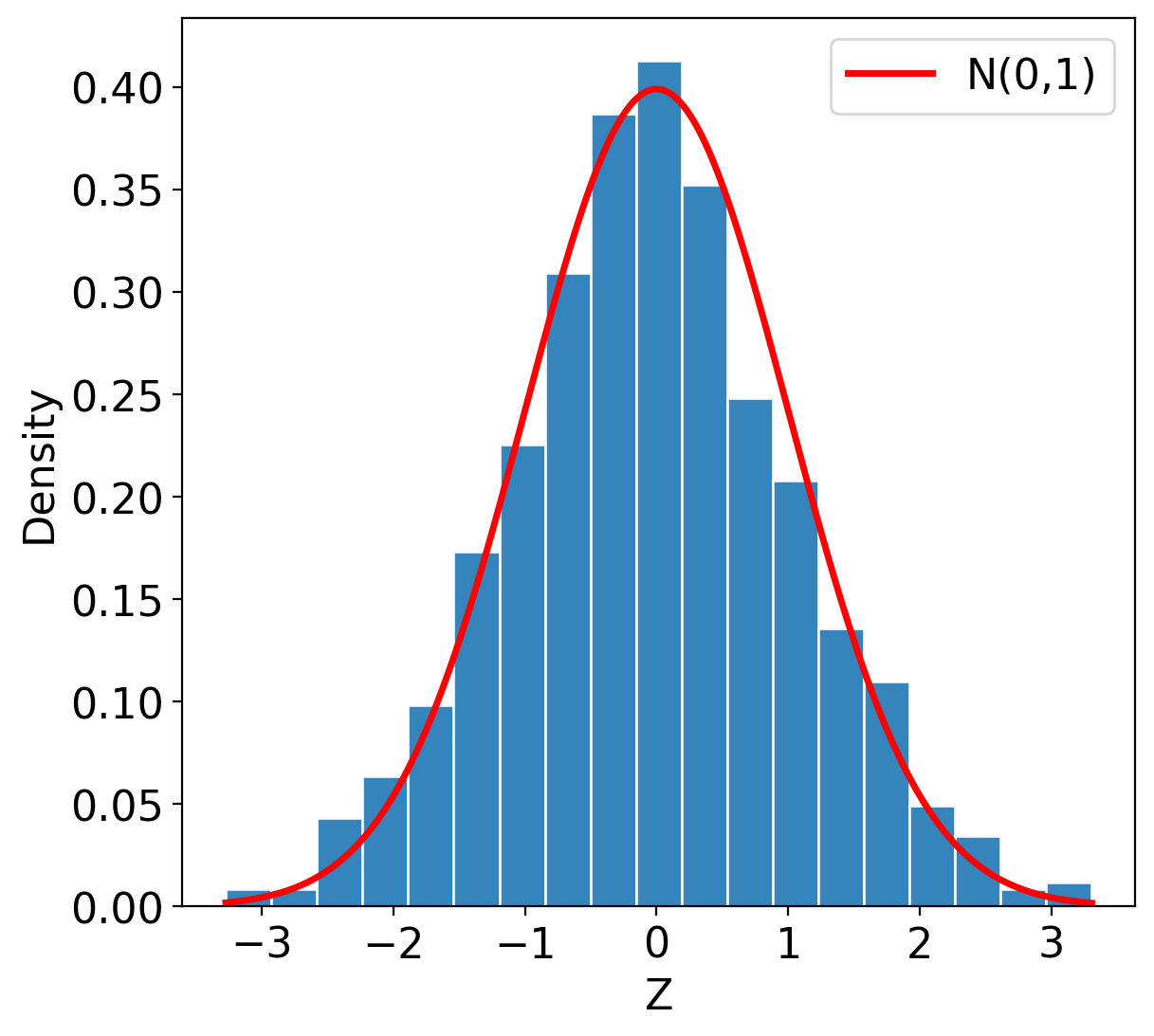}
    \caption{$d=50$, oracle}
  \end{subfigure}\hfill
  \begin{subfigure}{0.32\textwidth}
    \centering
    \includegraphics[width=\linewidth]{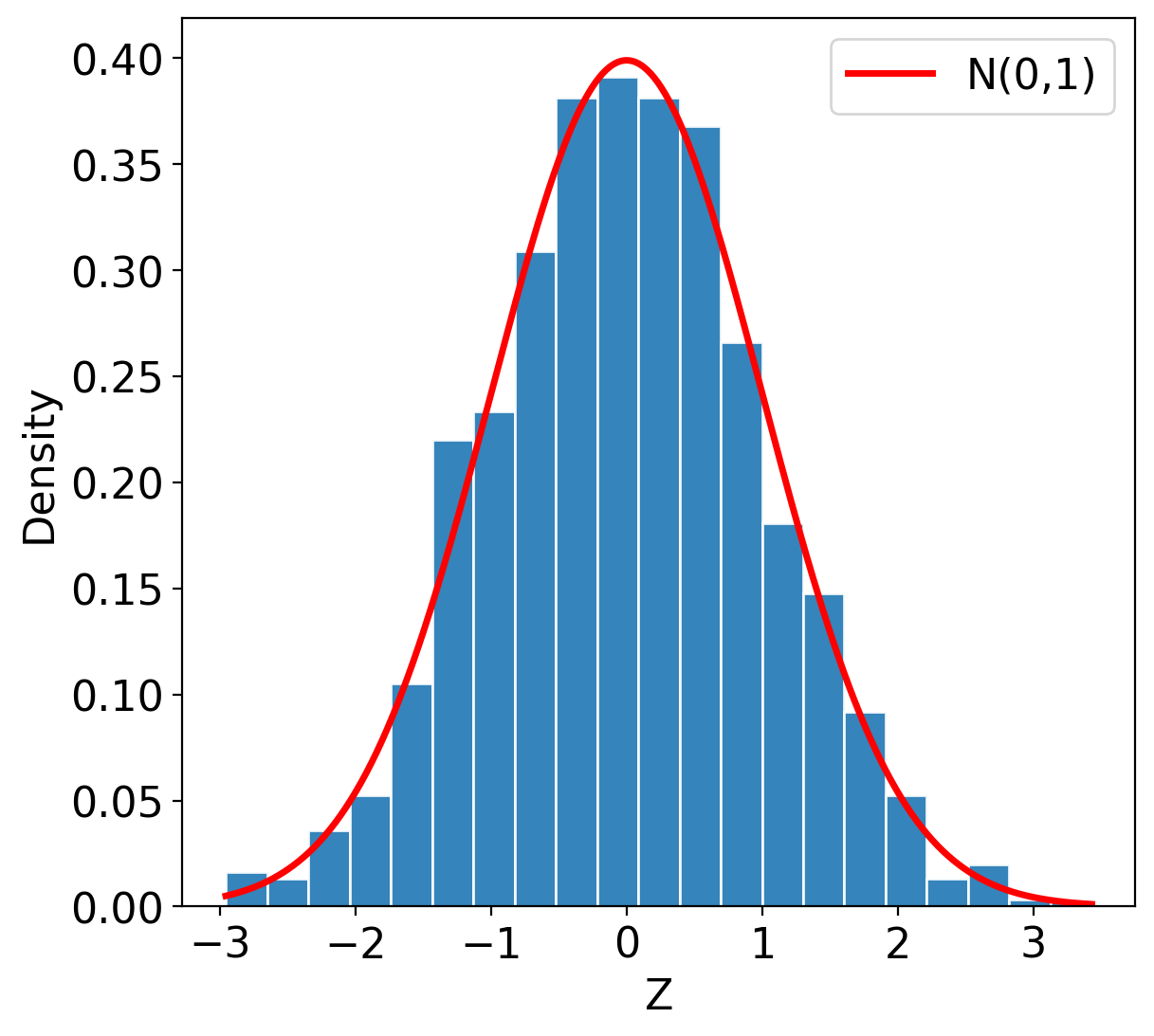}
    \caption{$d=100$, oracle}
  \end{subfigure}}
  \vspace{0.35em}
  \resizebox{0.75\textwidth}{!}{%
  \begin{subfigure}{0.32\textwidth}
    \centering
    \includegraphics[width=\linewidth]{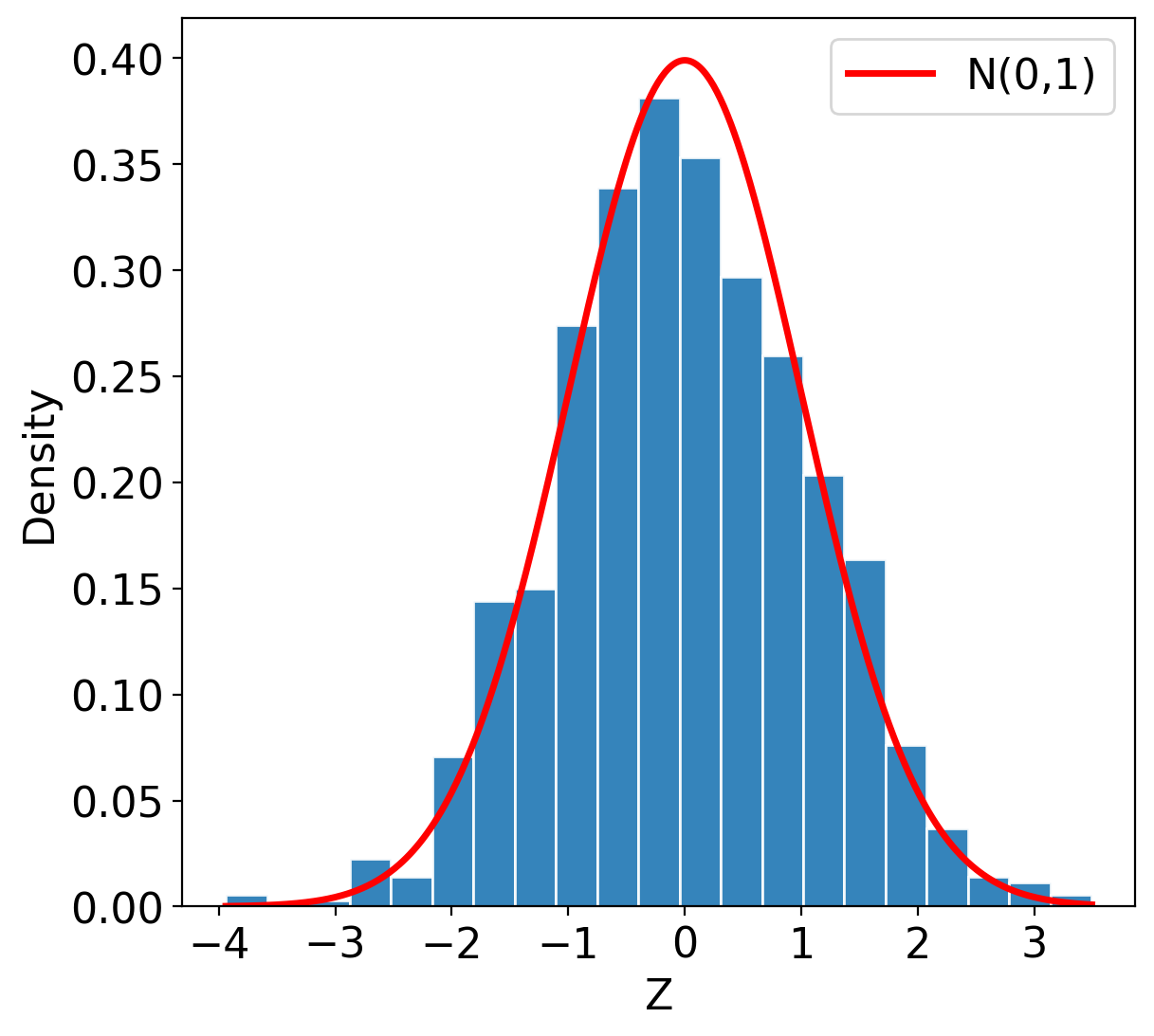}
    \caption{$d=20$, data-driven}
  \end{subfigure}\hfill
  \begin{subfigure}{0.32\textwidth}
    \centering
    \includegraphics[width=\linewidth]{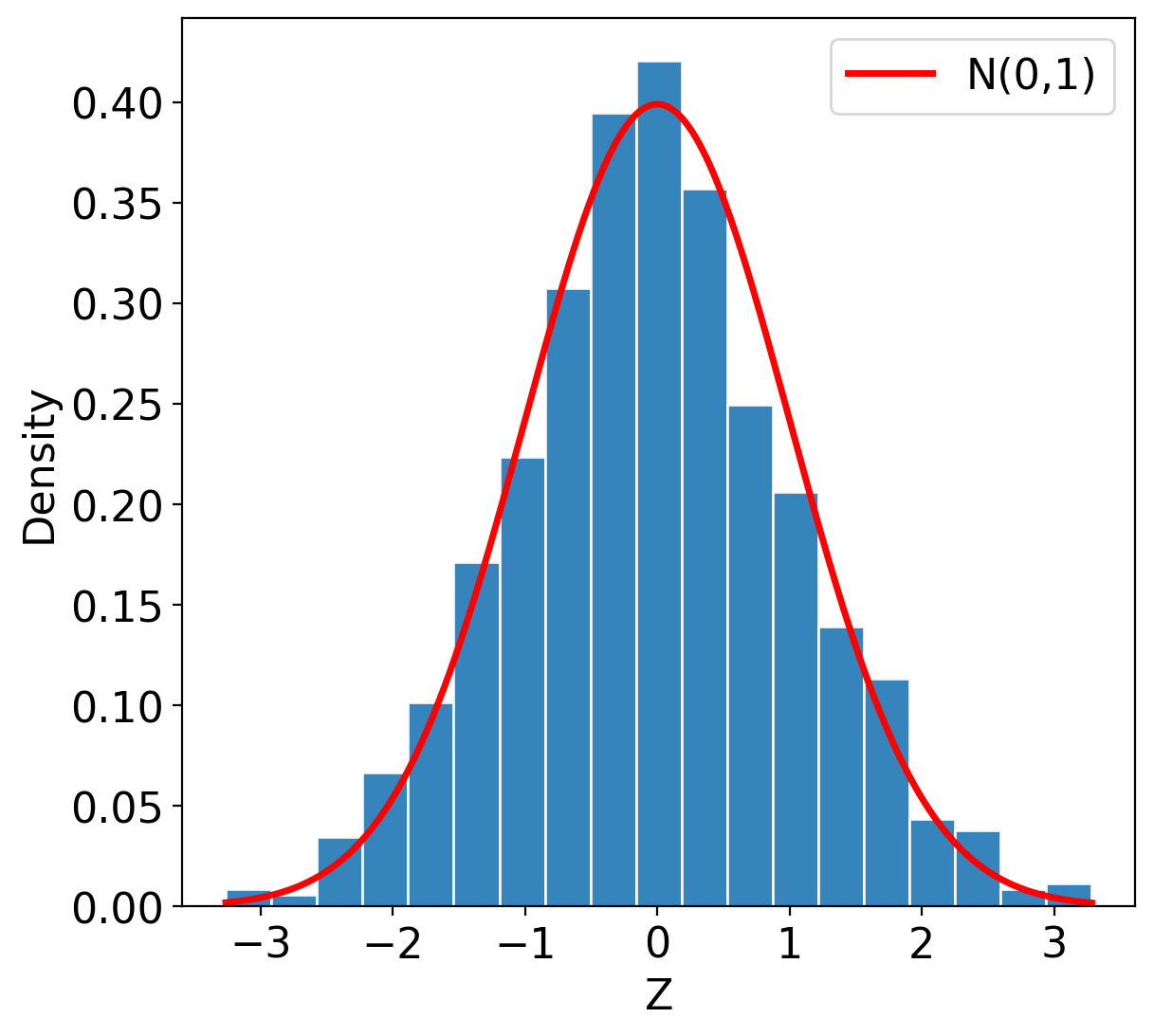}
    \caption{$d=50$, data-driven}
  \end{subfigure}\hfill
  \begin{subfigure}{0.32\textwidth}
    \centering
    \includegraphics[width=\linewidth]{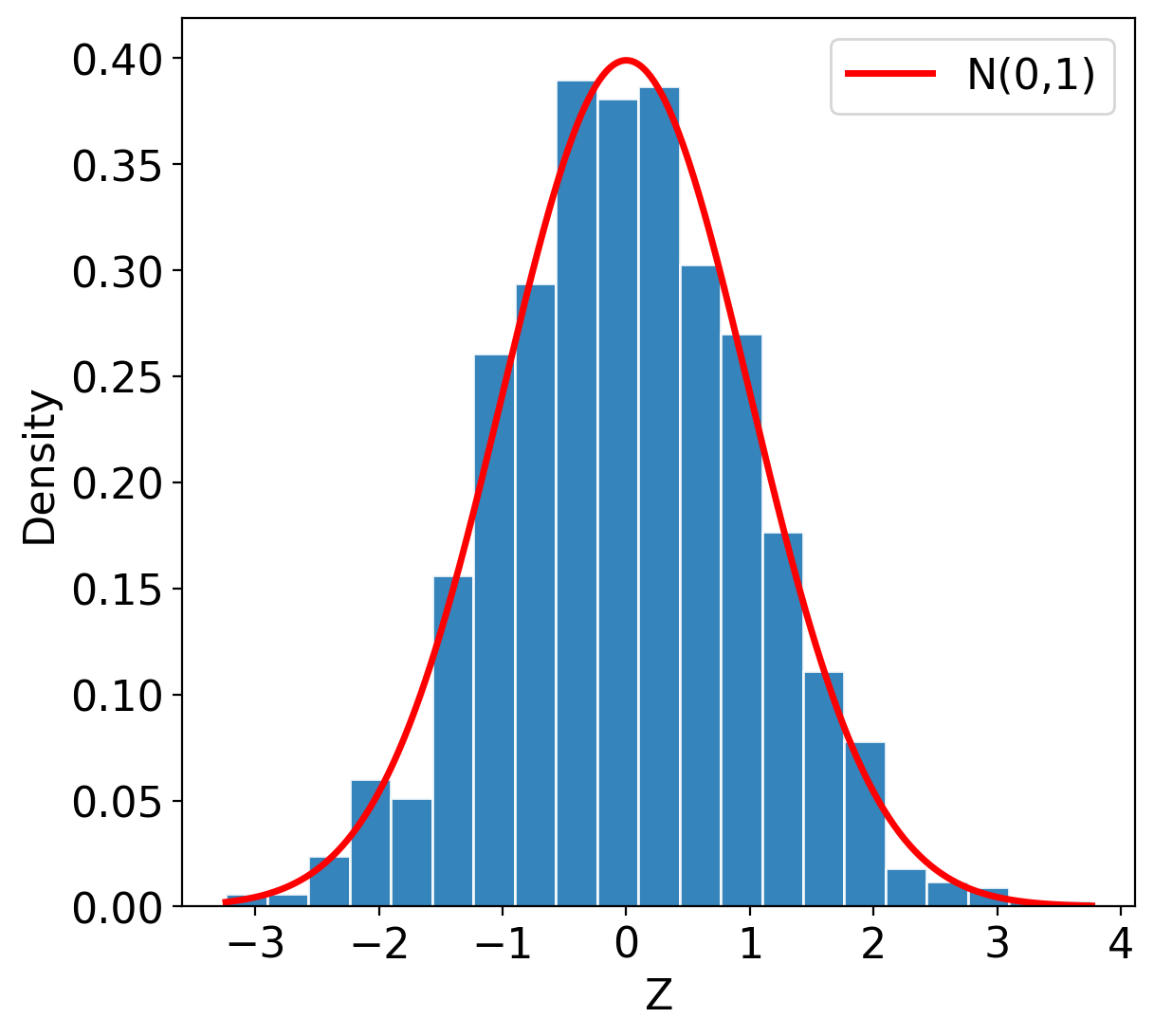}
    \caption{$d=100$, data-driven}
  \end{subfigure}}
  \caption{\small Histograms for the first coordinate of the standardized first row of $\widehat\bU_{\bA}-\bU_{\bA}\bR_{\bA}$ with $T=200$ under the strong loading strength regime, overlaid with the $\cN(0,1)$ density (red curve).
  The top row uses the oracle inference procedure with population covariance matrices, and the bottom row uses the feasible data-driven inference procedure with plug-in covariance estimators.
  Columns correspond to $d\in\{20,50,100\}$.}
  \label{fig:hist_A_oracle_dd}
\end{figure}

\noindent
\textbf{Asymptotic normality for rows of the loading matrices.} Figures~\ref{fig:qq_A_oracle_dd} and~\ref{fig:hist_A_oracle_dd} assess the asymptotic normality for the estimated loading matrix $\widehat \bU_{\bA}$, evaluating the oracle asymptotic distribution from Theorem~\ref{thm: asymptotic normality} and the feasible data-driven inference procedure from Theorem~\ref{cor: data driven normality}. We focus on the first row of the loading matrix $\widehat \bU_{\bA}$ under configurations $d_1=d_2\in\{20,50,100\} $ and $T = 200$ in the strong factor loading regime across $500$ Monte Carlo replicates. Corresponding results for $\widehat \bU_{\bB}$ appear in Appendix \ref{sec:simul_appendix}. For each replicate, we construct both oracle standardized rows of the loading matrix using true population covariance matrices and feasible standardized rows of the loading matrix using plug-in covariance estimators from the estimated factors and residual-based noise estimates. We assess the quality of normal approximation using the first coordinate of the standardized first row of $\widehat\bU_{\bA}-\bU_{\bA}\bR_{\bA}$ through Q-Q plots (Figure~\ref{fig:qq_A_oracle_dd}) and histograms overlaid with the standard normal density (Figure~\ref{fig:hist_A_oracle_dd}). The plots for other components are similar.

Both diagnostics demonstrate strong agreement with $\cN(0,1)$: Q-Q plots align closely with the reference line, and histograms match the theoretical density well. These results validate the asymptotic normality theory and confirm that the feasible data-driven inference procedure accurately captures the asymptotic distribution of the estimated loading matrices without requiring knowledge of population covariance matrices.

\section{Real Data Analysis}\label{sec:real}

We evaluate the proposed MAFM on a quarterly OECD macroeconomic panel in which each observation is naturally represented as a country-by-indicator matrix observed over time.

\noindent \textbf{Data and preprocessing.} 
The data are compiled from OECD releases that provide consistent quarterly coverage for a common set of advanced economies and macroeconomic indicators. After intersecting country availability across sources, we obtain a balanced panel of $d_1 = 18$ countries observed over $n=131$ quarters (1991Q2–2023Q4) with $d_2 = 9$ indicators capturing prices, real activity, financial conditions, and labor markets. Following standard macroeconometric practice, price indices, production indices, and GDP are transformed via log first differences, while interest rates and unemployment are first differenced and scaled by $1/100$. At each quarter $t$, the transformed indicators form the raw matrix $\bY_t \in \mathbb{R}^{d_1 \times d_2}$, with rows corresponding to countries and columns to indicators.

To remove scale differences across indicators, we apply pooled standardization: for each indicator $j\in[d_2]$, we compute the pooled mean $\mu_j$ and standard deviation $\sigma_j$ across all country-quarter observations and set $X_{t,ij} = (Y_{t,ij}-\mu_j)/\sigma_j$. This follows standard practice in macroeconomic panel studies \citep{ottinger2025history}. All estimation and forecasting use the standardized series $\{\bX_t\}_{t=1}^n$.

\noindent \textbf{Competing methods.} 
We compare MAFM using MINE initialization and COMPAS refinement against four alternative approaches: (1) TwDFM \citep{yuan2023twoway}, which decomposes matrix time series into separate row and column components with autoregressive dynamics via two-step quasi-maximum likelihood estimation; (2) Tucker factor model \citep{han2024tensor} with cross-covariance based estimation; (3) CP factor model \citep{chen2026estimation} with cross-covariance based estimation; and (4) VecFactor \citep{bai2003inferential}, a vectorized baseline that flattens matrix observations into vectors before applying standard factor analysis with cross-covariance estimation. For fair comparison, all methods use AR models to forecast factors, which are then used to reconstruct predicted observations.

\begin{figure}[ht]
    \centering
    \resizebox{0.7\textwidth}{!}{
    \begin{minipage}{\textwidth}
        \begin{subfigure}[t]{\textwidth}
            \centering
            \includegraphics[width=\textwidth]{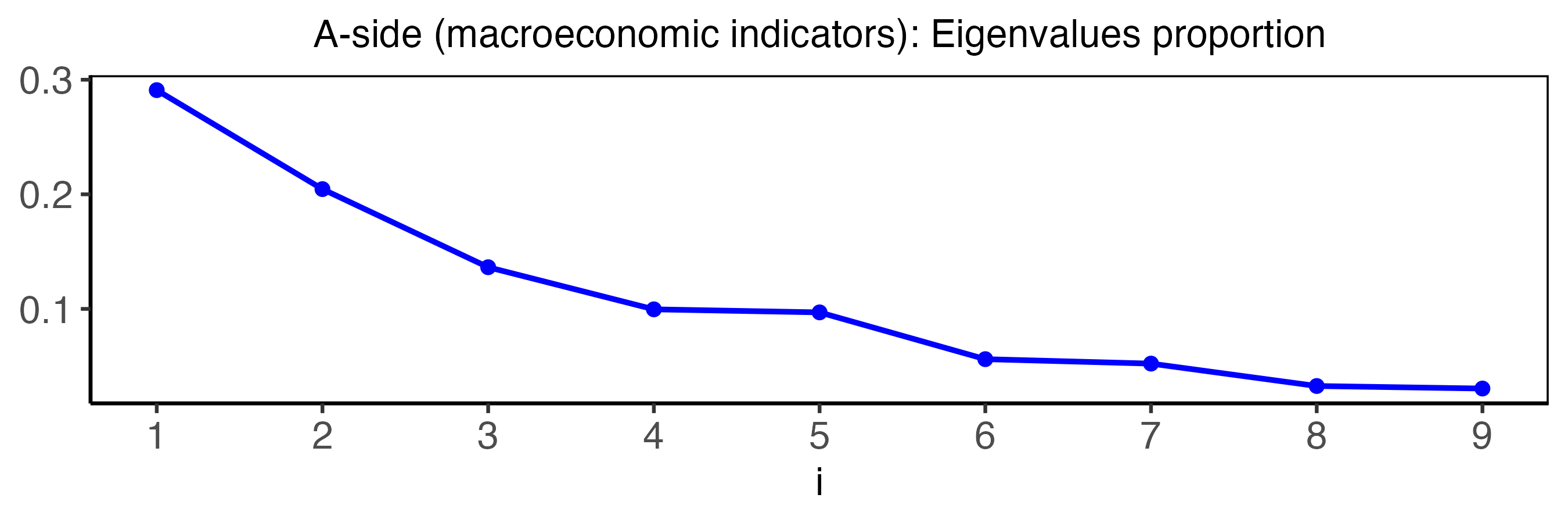}
            \caption{Row-factor component}
            \label{fig:eigprop_A}
        \end{subfigure}
        
        \begin{subfigure}[t]{\textwidth}
            \centering
            \includegraphics[width=\textwidth]{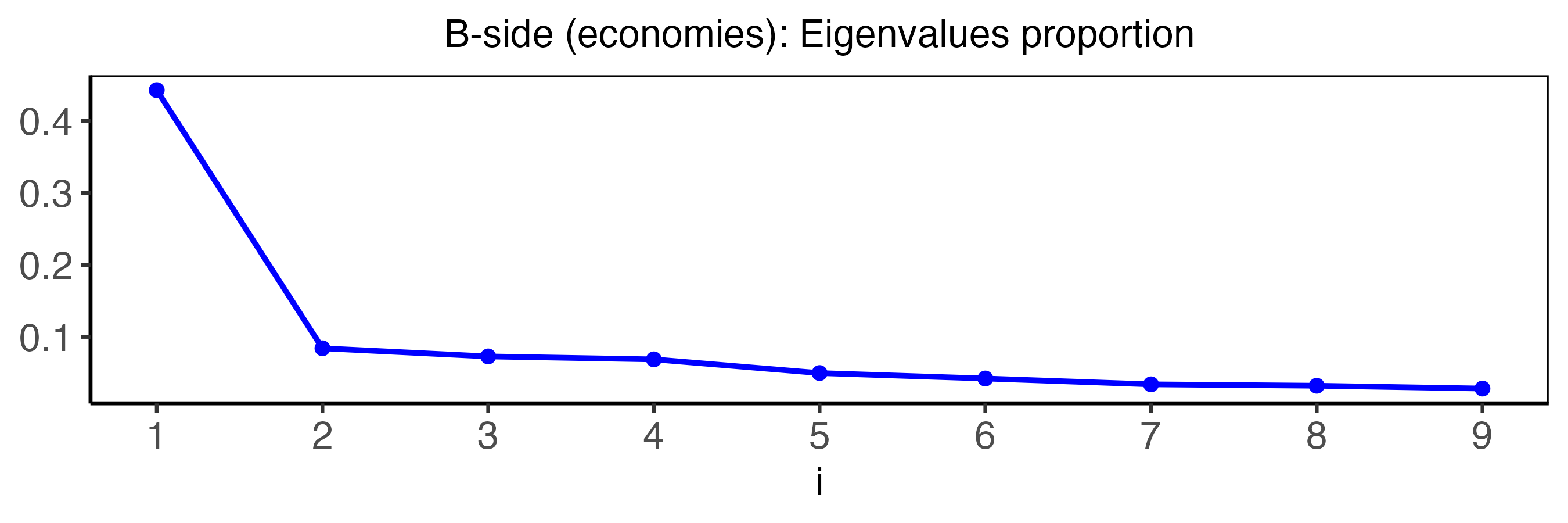}
            \caption{Column-factor component}
            \label{fig:eigprop_B}
        \end{subfigure}
    \end{minipage}
    }
    \caption{\small Eigenvalue proportion diagnostics for rank selection. (a) Eigenvalue proportions $\lambda_i / \sum_j \lambda_j$ of the row-factor component matrix $n^{-1}\sum_{t=1}^n \bX_t^\top \bX_t$; (b) Eigenvalue proportions of the column-factor component matrix $n^{-1}\sum_{t=1}^n \bX_t \bX_t^\top$, where $\lambda_i$ denotes the $i$th largest eigenvalue.}
%    The upper panel reports the eigenvalue proportions of the A-side (row-factor component) matrix $n^{-1}\sum_{t=1}^n \bX_t^\top \bX_t$, and the lower panel reports the eigenvalue proportions of the B-side matrix (column-factor component) $n^{-1}\sum_{t=1}^n \bX_t \bX_t^\top$. In each panel, the $y$-axis is $\lambda_i/\sum_j \lambda_j$, where $\lambda_i$ denotes the $i$th largest eigenvalue.}
    \label{fig:eigprop_AB}
\end{figure}

\noindent \textbf{Factor rank selection.} Figure~\ref{fig:eigprop_AB} displays eigenvalue proportion diagnostics for factor rank selection. For MAFM, the A-side (row-factor component) scree plot suggests $r_1=4$ factors (clear elbow at component 4), while the B-side (column-factor component) plot suggests $r_2 = 2$ (sharp decay after component 2), yielding $(r_1, r_2) = (4, 2)$. For robustness, we also examine $(4, 1)$, $(5, 1)$, and $(5, 2)$. Competing methods are evaluated across multiple factor ranks as shown in Table~\ref{tab:oecd_insample}.

\noindent \textbf{In-sample fit.} We assess in-sample performance using all observations for estimation. For a given method with fitted values $\widehat{\bX}_t$, define the fitting error as $\text{Fit-err} =  (d_1d_2n)^{-1} \sum_{t=1}^{n} \| \bX_t - \widehat\bX_t \|_{\rm F}^2$, and the coefficient of determination as $R^2 = 1 - \sum_{t=1}^{n} \| \bX_t - \widehat \bX_t \|_{\rm F}^2/\sum_{t=1}^{n} \| \bX_t - \bar{x} \mathbf{1}_{d_1} \mathbf{1}_{d_2}^\top \|_{\rm F}^2$, where $\bar{x} = (d_1d_2n)^{-1} \sum_{t=1}^{n} \mathbf{1}_{d_1}^\top \bX_t \mathbf{1}_{d_2} $ is the grand mean.
Table~\ref{tab:oecd_insample} reports Fit-err and in-sample $R^2$ for MAFM and the four competing methods across factor rank specifications. Table~\ref{tab:oecd_runtime} reports the total computation times for all configurations. MAFM consistently outperforms alternatives across all specifications while maintaining computational efficiency. Notably, MAFM is substantially faster than TwDFM, which achieves only slightly lower in-sample $R^2$ values. This demonstrates the computational advantage of COMPAS over quasi-maximum likelihood estimation.

\begin{table}[htbp!]
\centering
\setlength{\tabcolsep}{4pt}
\renewcommand{\arraystretch}{1.12}
\resizebox{\textwidth}{!}{%
\begin{tabular}{ccc ccc ccc ccc ccc}
\toprule
\multicolumn{3}{c}{MAFM} &
\multicolumn{3}{c}{TwDFM} &
\multicolumn{3}{c}{Tucker Factor} &
\multicolumn{3}{c}{VecFactor} &
\multicolumn{3}{c}{CP Factor} \\
\cmidrule(lr){1-3}\cmidrule(lr){4-6}\cmidrule(lr){7-9}\cmidrule(lr){10-12}\cmidrule(lr){13-15}
\# factor & $R^2$ & Fit-err &
\# factor & $R^2$ & Fit-err &
\# factor & $R^2$ & Fit-err &
\# factor & $R^2$ & Fit-err &
\# factor & $R^2$ & Fit-err \\
\midrule
(4,1) & {\bf 0.8209} & {\bf 0.1777} & (4,1) & 0.8153 & 0.1833 & (4,1) & 0.2407 & 0.7535 & 4 & 0.4836 & 0.5124 & 2 & 0.2306 & 0.7635 \\
(4,2) & {\bf 0.8539} & {\bf 0.1449} & (4,2) & 0.8450 & 0.1538 & (4,2) & 0.3923 & 0.6030 & 5 & 0.5303 & 0.4661 & 3 & 0.2600 & 0.7343 \\
(5,1) & {\bf 0.8867} & {\bf 0.1125} & (5,1) & 0.8746 & 0.1245 & (5,1) & 0.2494 & 0.7449 & 6 & 0.5712 & 0.4255 & 4 & 0.2630 & 0.7313 \\
(5,2) & {\bf 0.9119} & {\bf 0.0875} & (5,2) & 0.8997 & 0.0995 & (5,2) & 0.4079 & 0.5875 & 7 & 0.6055 & 0.3914 & 5 & 0.2340 & 0.7602 \\
\bottomrule
\end{tabular}%
}
\caption{\small In-sample performance ($R^2$ and Fit-err) on the OECD data.}
\label{tab:oecd_insample}
\end{table}

\begin{table}[htbp!]
\centering
\begin{tabular}{lccccc}
\toprule
Method & MAFM & TwDFM & Tucker Factor & VecFactor & CP Factor \\
\midrule
Time (sec) & 0.760 & 13.217 & 0.374 & 0.092 & 1.726 \\
\bottomrule
\end{tabular}
\caption{\small Computation time (seconds) for all factor rank combinations in Table~\ref{tab:oecd_insample}.}
\label{tab:oecd_runtime}
\end{table}

\noindent \textbf{Out-of-sample forecasting.}
We evaluate predictive performance using an expanding-window design with one-step-ahead forecasts. At each forecast origin $w$, we re-estimate the model using $\{\bX_t\}_{t=1}^{w}$ through the following steps: (i) estimate loading matrices $\widehat{\bU}_{\bA}, \widehat{\bU}_{\bB}$ and factor sequences $\{\widehat{\bF}_t\}_{t=1}^{w}, \{\widehat{\bG}_t\}_{t=1}^{w}$; (ii) fit independent $AR(p)$ models to each factor component with $p$ selected by AIC; and (iii) generate factor forecasts $\widehat{\bF}_{w+1}, \widehat{\bG}_{w+1}$ and reconstruct $\widehat{\bX}_{w+1|w} = \widehat{\bF}_{w+1}\widehat{\bU}_{\bA}^{\top} + \widehat{\bU}_{\bB}\widehat{\bG}_{w+1}^{\top}$. The one-step-ahead forecast error is ${\rm FE}(w+1) = (d_1d_2)^{-1} \|\bX_{w+1}-\widehat \bX_{w+1|w}\|_{\rm F}^2 $. To assess recent forecasting performance, we report the averaged forecast errors over the last $h$ forecasts: $\overline{{\rm FE}}_{h} = h^{-1}\sum_{w=n-h}^{n-1} {\rm FE}(w+1)$ for $h \in \{5, 10, 15, 20,25,30\}$. Table~\ref{tab:oecd_oos} reports results for MAFM and competing methods. MAFM consistently achieves the lowest forecast errors across all horizons, confirming that its modewise additive structure yields genuine predictive gains beyond in-sample fit.

\begin{table}[htbp!]
\centering
\setlength{\tabcolsep}{8pt}
\renewcommand{\arraystretch}{1.15}
\begin{tabular}{lcccccc}
\toprule
Method & $\overline{{\rm FE}}_5$ & $\overline{{\rm FE}}_{10}$ & $\overline{{\rm FE}}_{15}$ & $\overline{{\rm FE}}_{20}$ & $\overline{{\rm FE}}_{25}$ & $\overline{{\rm FE}}_{30}$ \\
\midrule
MAFM         & {\bf 1.267} & {\bf 1.433} & {\bf 1.170} & {\bf 0.988} & {\bf 0.855} & {\bf 0.773} \\
TwDFM        & 1.351 & 1.476 & 1.191 & 1.012 & 0.877 & 0.796 \\
Tucker Factor & 1.450 & 1.621 & 1.300 & 1.098 & 0.949 & 0.858 \\
VecFactor    & 1.317 & 1.498 & 1.211 & 1.023 & 0.883 & 0.804 \\
CP Factor    & 1.477 & 1.562 & 1.301 & 1.100 & 0.957 & 0.863 \\
\bottomrule
\end{tabular}
\caption{\small Out-of-sample rolling one-step-ahead forecast error $\overline{{\rm FE}}_h$ on OECD data. Factor ranks: MAFM (4,2), TwDFM (4,2), Tucker (4,2), VecFactor 7, CP 4.}
\label{tab:oecd_oos}
\end{table}

\begin{figure}[htbp!]
    \centering
    \includegraphics[width=0.75\textwidth]{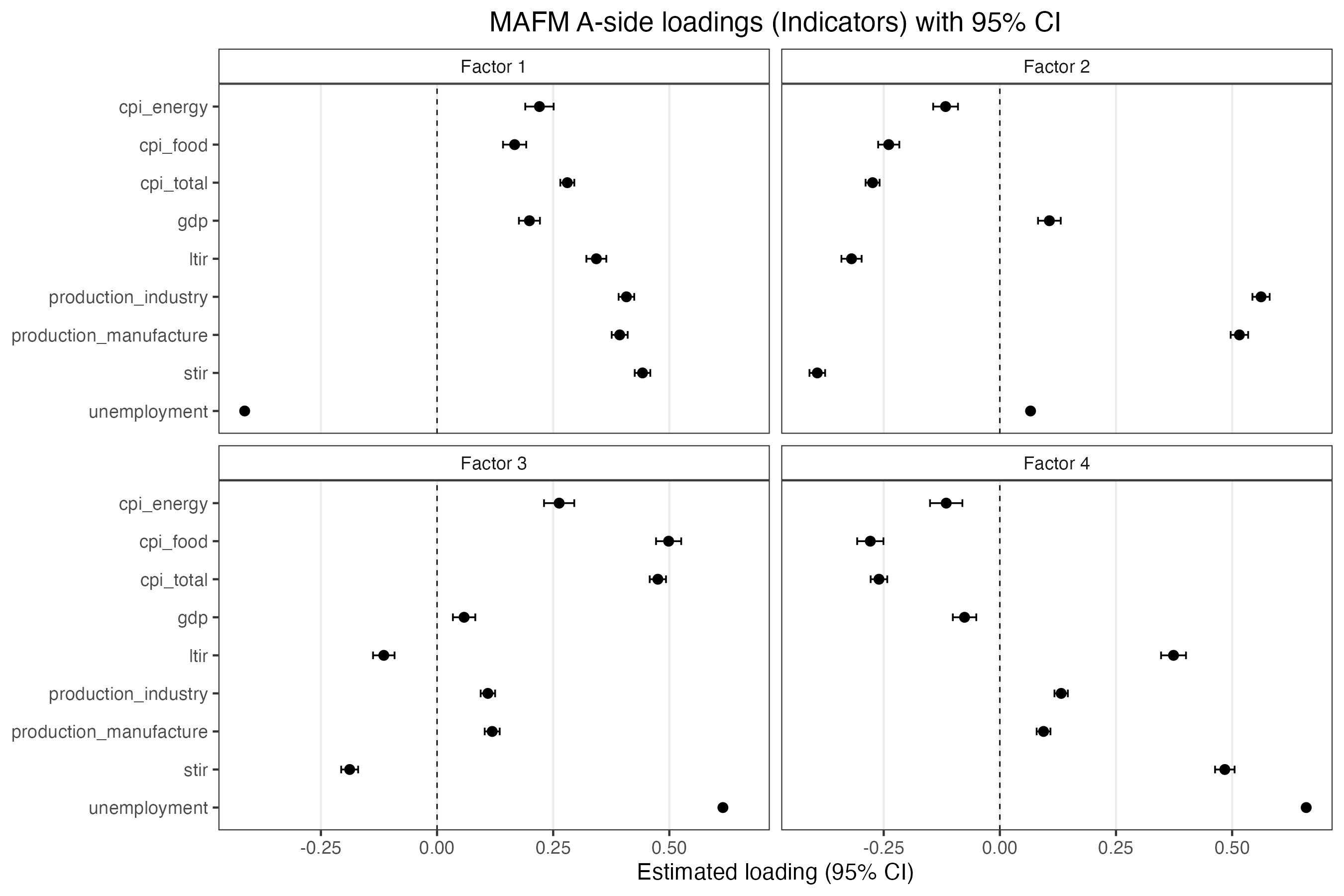}
    \caption{\small Row-factor loadings (indicators) with 95\% confidence intervals under $(r_1, r_2) = (4,2)$.}
    \label{fig:loadings_indicators}
\end{figure}

\begin{figure}[htbp!]
    \centering
    \includegraphics[width=0.75\textwidth]{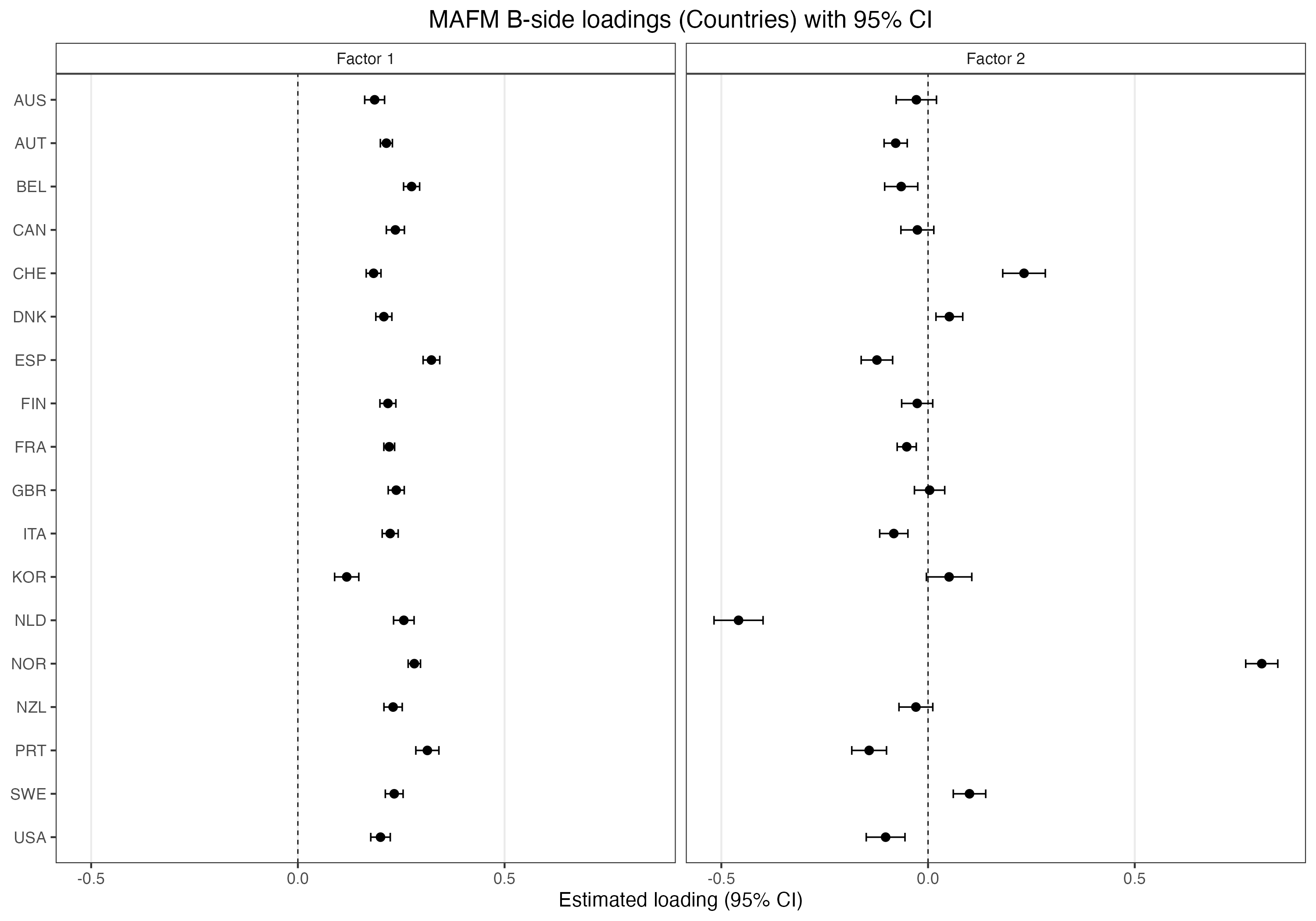}
    \caption{\small Column-factor loadings (countries) with 95\% confidence intervals under $(r_1, r_2) = (4,2)$.}
    \label{fig:loadings_countries}
\end{figure}

\noindent\textbf{Interpretation of estimated loadings.} Using the feasible data-driven inference procedure in our MAFM framework (Theorem \ref{cor: data driven normality}), we construct 95\% confidence intervals for the factor loadings, displayed in Figures \ref{fig:loadings_indicators} and \ref{fig:loadings_countries}. The uniformly tight confidence intervals indicate that the main patterns are well-identified and precisely estimated.

For the row-factor loadings (indicators), Factor 1 exhibits a clear global macro comovement pattern: broadly positive loadings on prices, output, production, and interest rates, paired with a strongly negative loading on unemployment. The remaining factors capture more specialized contrasts among specific indicator subgroups. For the column-factor loadings (countries), Factor 1 loads positively and significantly on essentially every country, indicating a pervasive common component affecting all economies. In contrast, Factor 2 is geographically concentrated, Norway and the Netherlands exhibit large opposite-signed loadings while most other countries remain near zero. Overall, the decomposition reveals one dominant common factor alongside higher-order factors that capture selective cross-sectional variation.

\section{Conclusion}
\label{sec:discussion}

In this paper, we propose a modewise additive factor model for matrix time series that extends Tucker and CP factor models by separately capturing row-specific and column-specific effects, rather than attributing all variation to global factors. To address the cross-modal interference inherent in this structure, we develop a two-stage estimation procedure, MINE for initialization and COMPAS for iterative refinement via orthogonal complement projections. We establish convergence rates and asymptotic distributions under high-dimensional settings where each mode dimension is comparable to or exceeds the sample size. As a technical contribution of independent interest, we derive sharp Bernstein-type inequalities for quadratic forms of dependent matrix time series. Simulations demonstrate the advantages of our method, and an application to OECD macroeconomic data illustrates its practical relevance.

%-------
%
%  bibliography
%

%\clearpage
%\nocite{*} % include all bib's
% \bibliographystyle{\mybibsty}
% \bibliography{\mybib}
\bibliographystyle{apalike}
\bibliography{main}

%-------
%
%  appendix
%
\clearpage
\setcounter{page}{1}
\begin{appendices}
    \begin{center}
        {\bf\Large Supplementary Material of ``Modewise Additive Factor Model for Matrix Time Series''}

    \end{center}

\section{Additional Simulation Results}\label{sec:simul_appendix}

\begin{figure}[htbp!]
    \centering
    \includegraphics[width=0.75\textwidth]{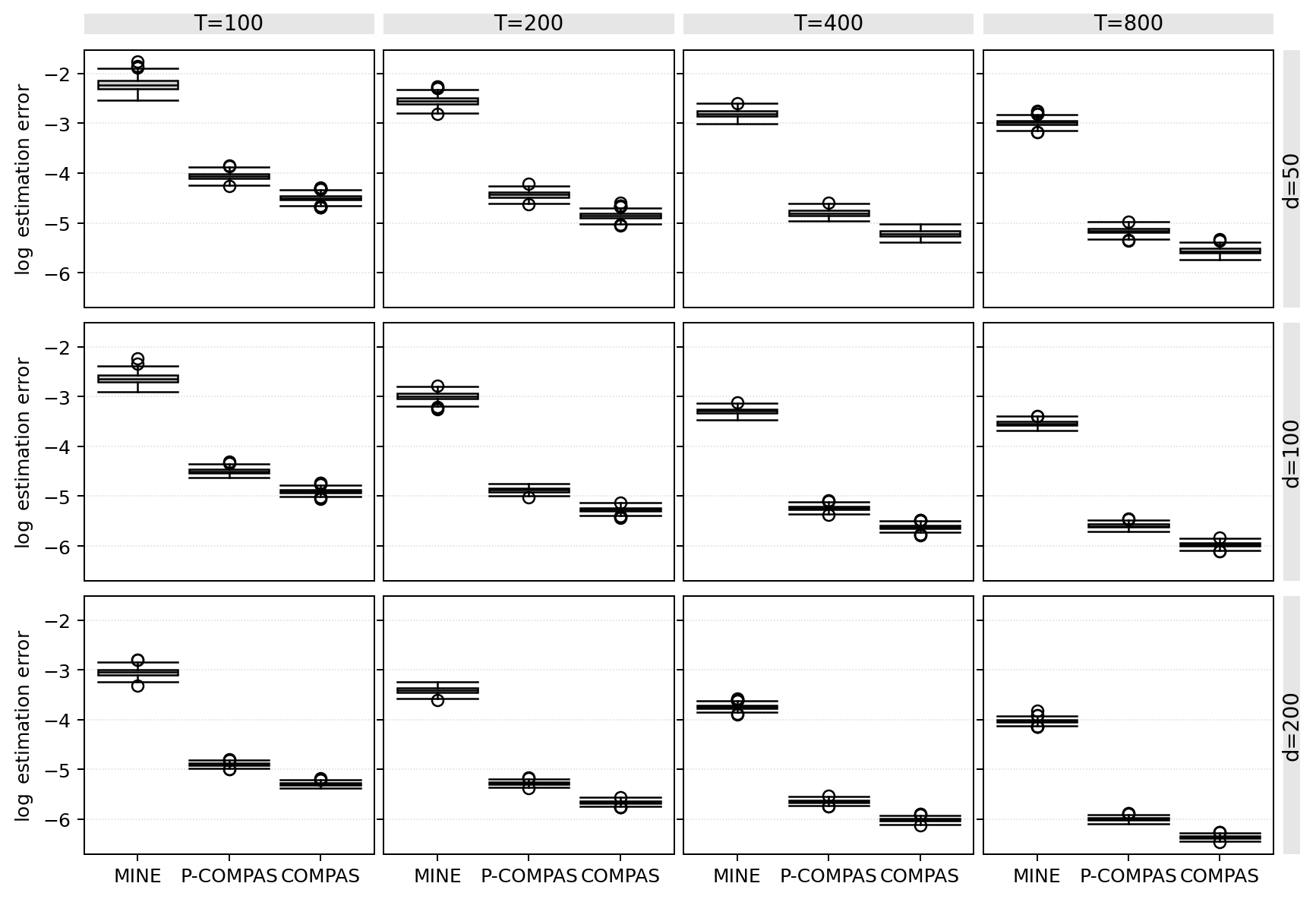}
    \caption{\small Estimation error $\log(\cD(\bU_{\bB}, \widehat\bU_{\bB}))$ under strong factor loading with $(\delta_0, \delta_1) = (0, 0)$ across dimensions $d$ and sample sizes $T$ for three estimation methods: MINE, P-COMPAS, and COMPAS.}
    \label{fig:subspace_B_strong}
\end{figure}

\begin{figure}[htbp!]
    \centering
    \includegraphics[width=0.75\textwidth]{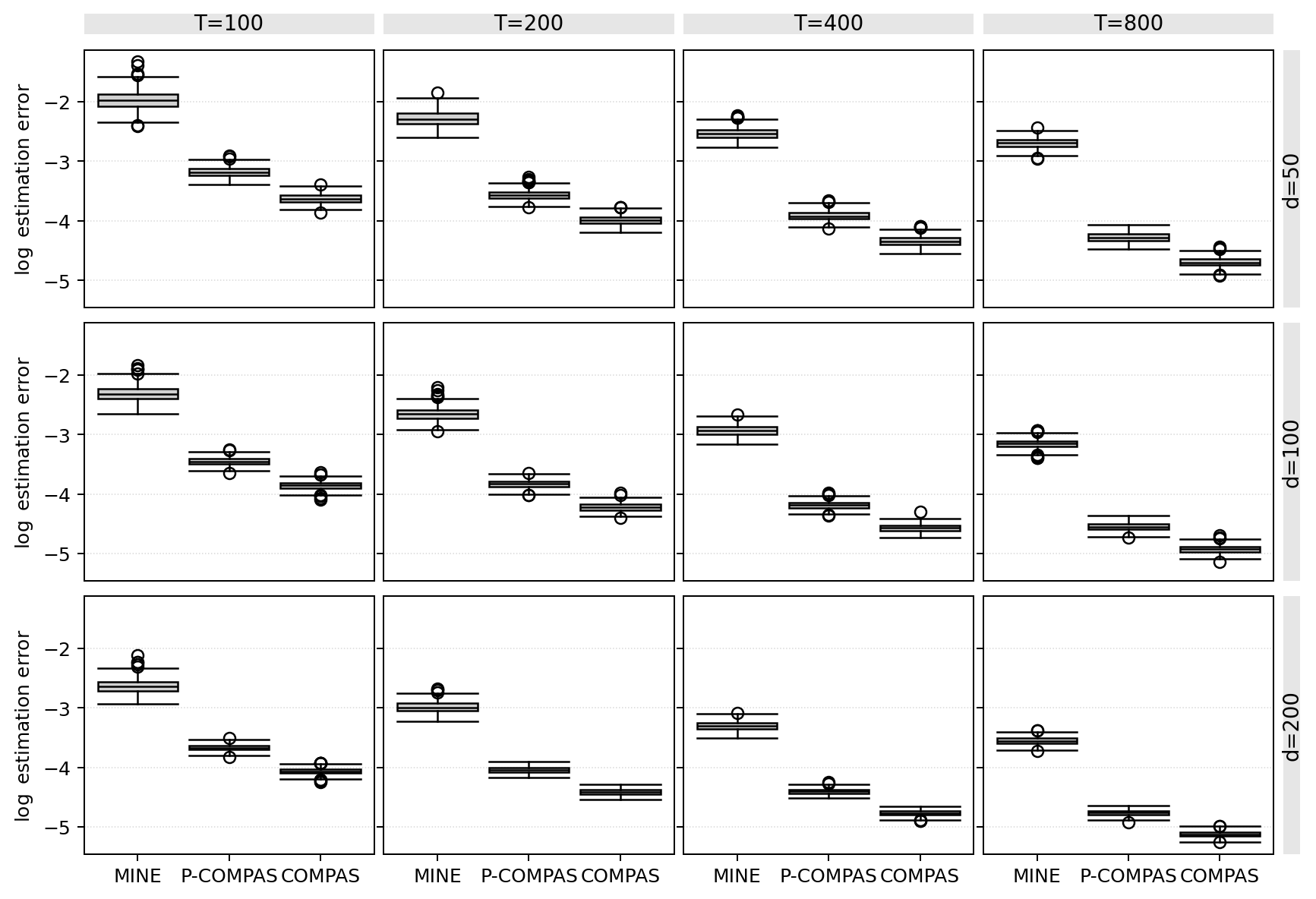}
    \caption{\small Estimation error $\log(\cD(\bU_{\bB}, \widehat\bU_{\bB}))$ under weak factor loading with  $(\delta_0, \delta_1) = (0.3, 0.5)$ across dimensions $d$ and sample sizes $T$ for three estimation methods: MINE, P-COMPAS, and COMPAS.}
    \label{fig:subspace_B_weak}
\end{figure}

\begin{figure}[htbp!]
  \centering
  \resizebox{0.75\textwidth}{!}{%
  \begin{subfigure}{0.32\textwidth}
    \centering
    \includegraphics[width=\linewidth]{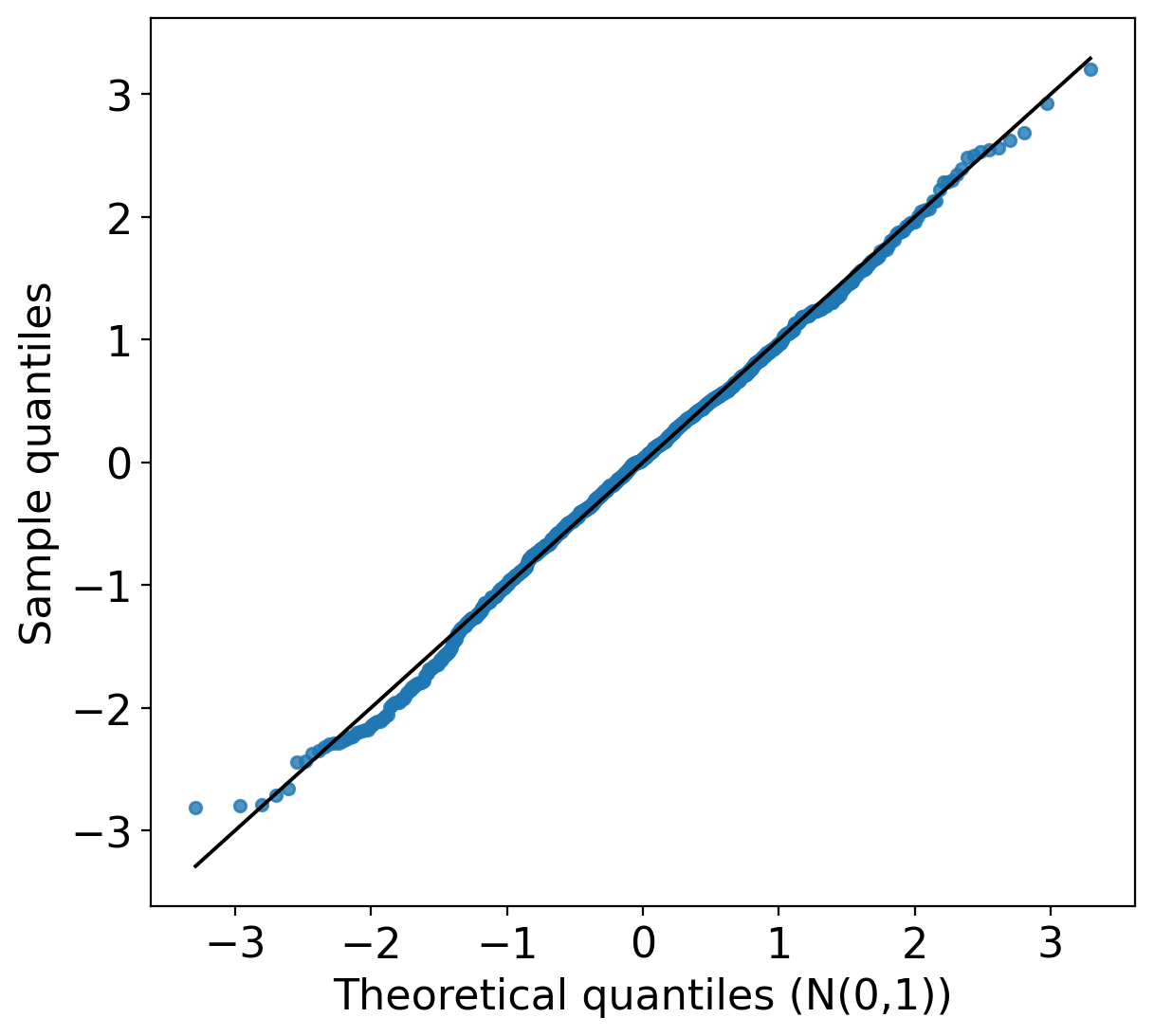}
    \caption{$d=20$, oracle}
  \end{subfigure}\hfill
  \begin{subfigure}{0.32\textwidth}
    \centering
    \includegraphics[width=\linewidth]{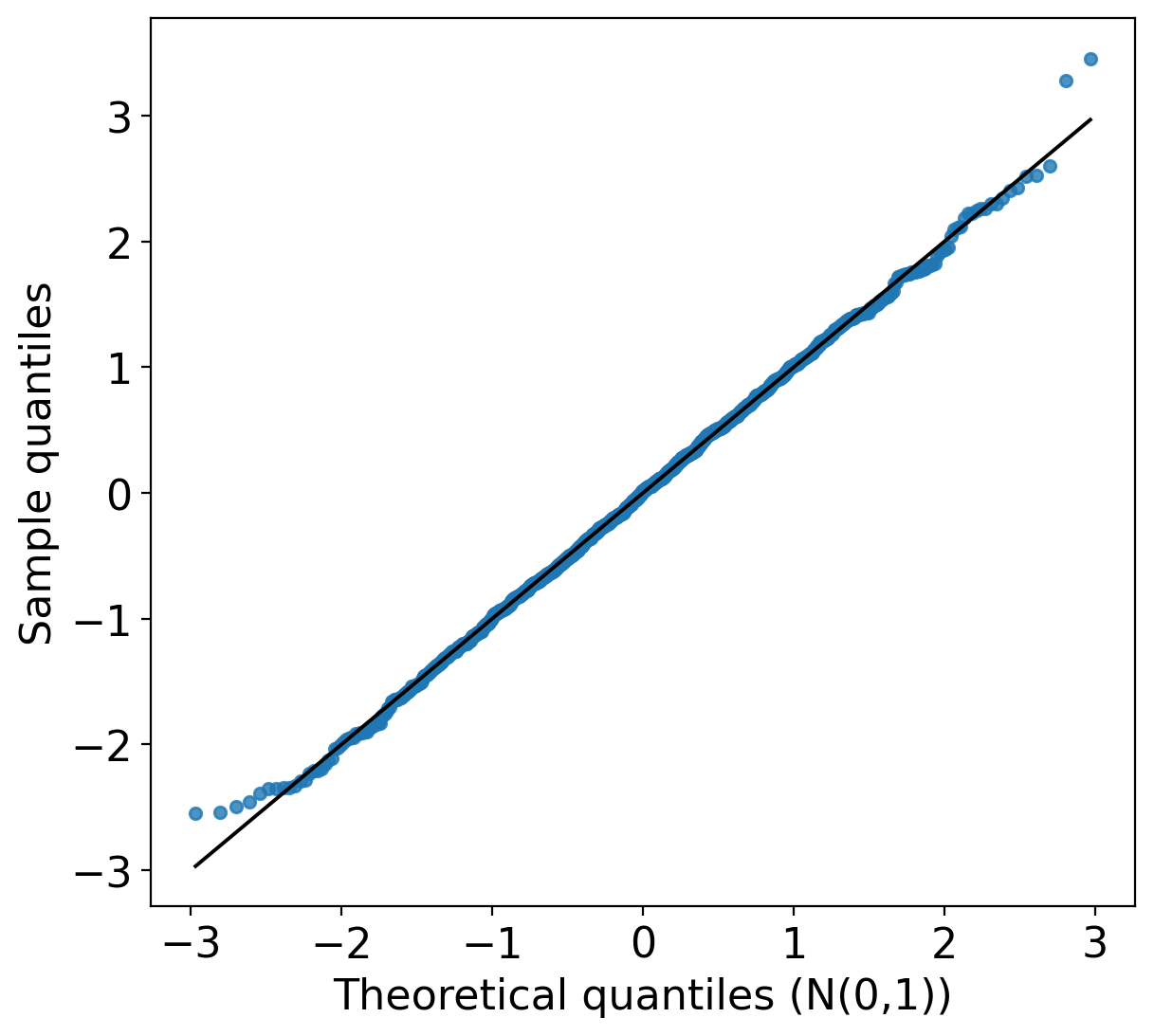}
    \caption{$d=50$, oracle}
  \end{subfigure}\hfill
  \begin{subfigure}{0.32\textwidth}
    \centering
    \includegraphics[width=\linewidth]{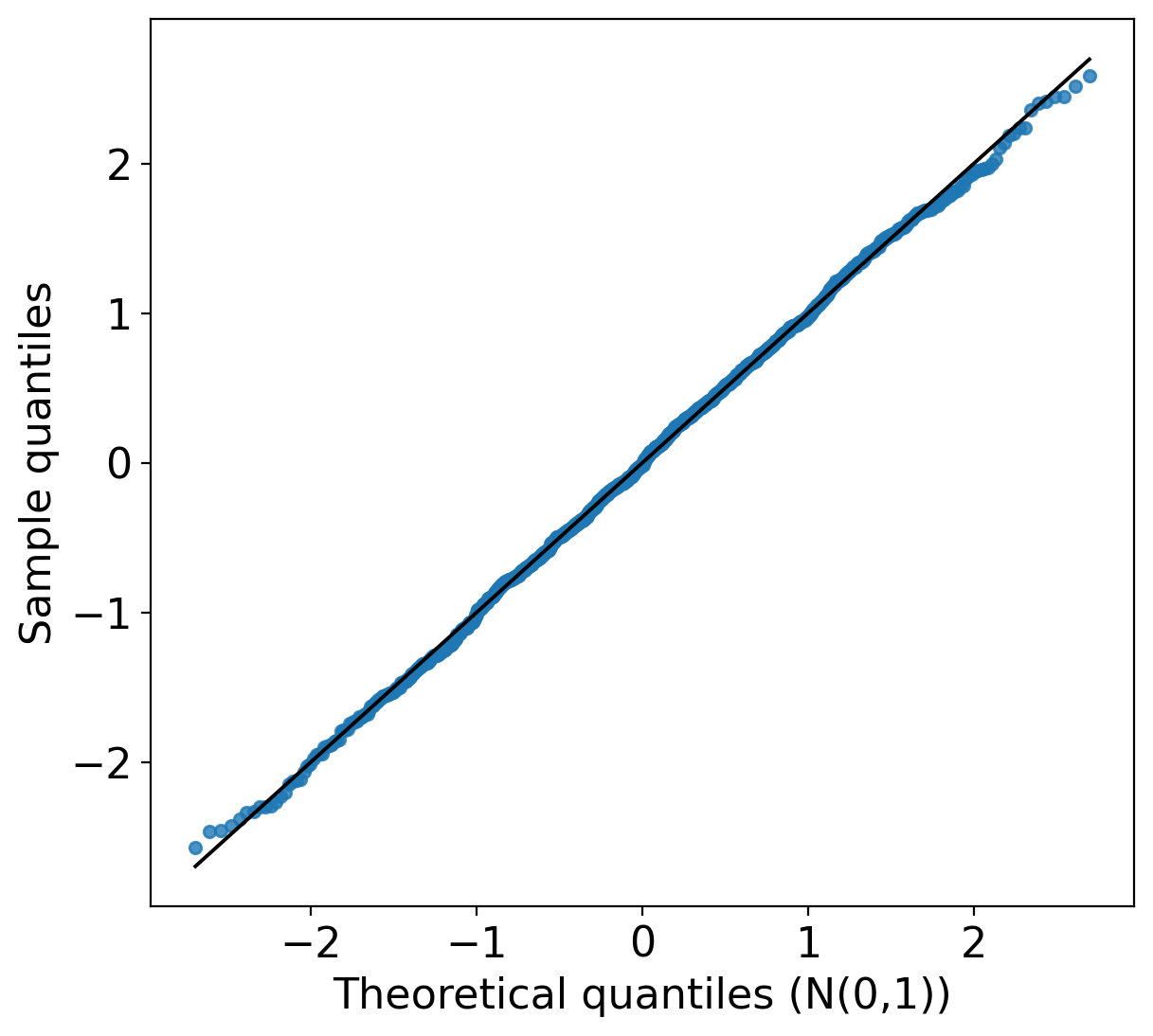}
    \caption{$d=100$, oracle}
  \end{subfigure}}
  \vspace{0.35em}
  \resizebox{0.75\textwidth}{!}{%
  \begin{subfigure}{0.32\textwidth}
    \centering
    \includegraphics[width=\linewidth]{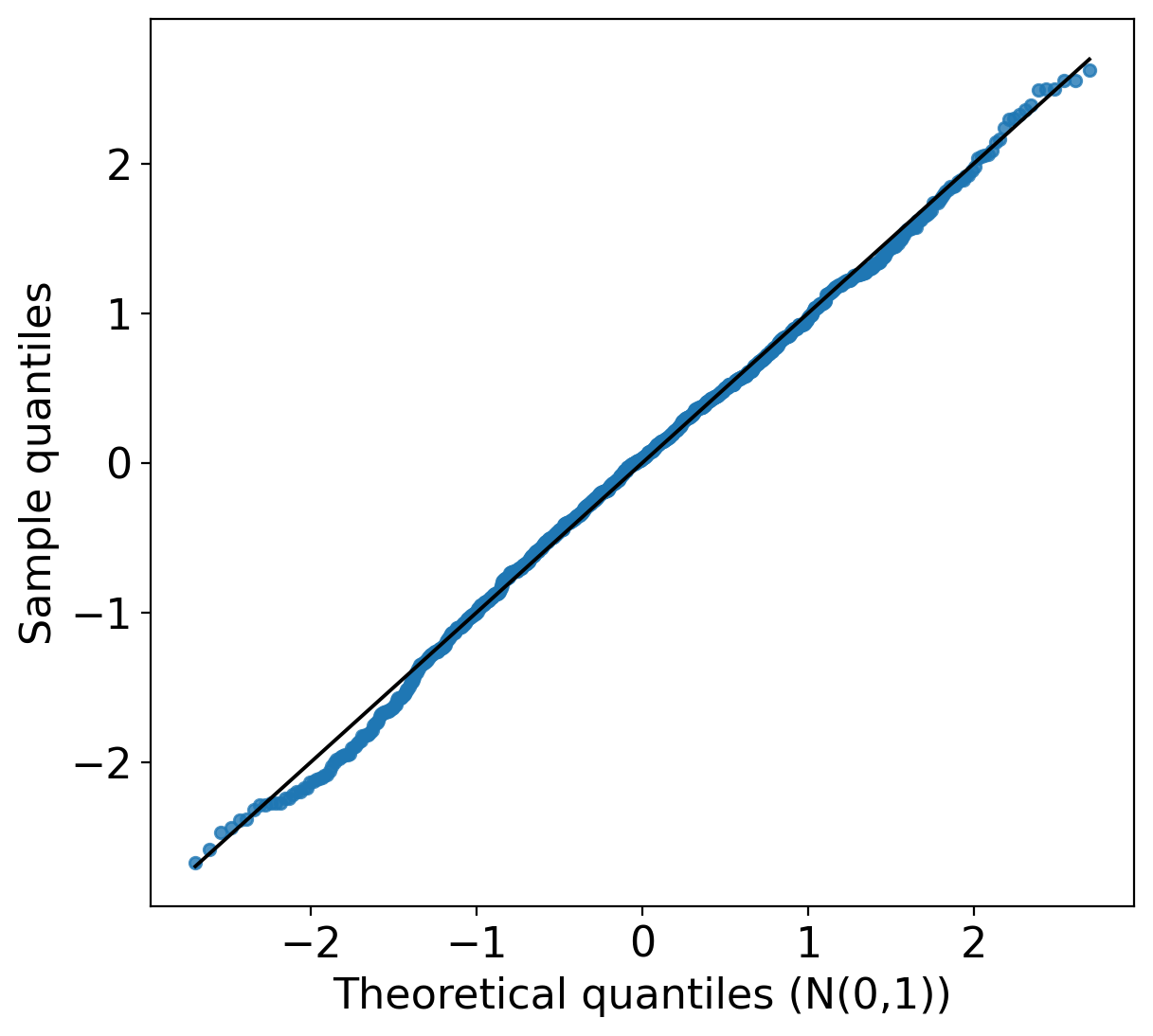}
    \caption{$d=20$, data-driven}
  \end{subfigure}\hfill
  \begin{subfigure}{0.32\textwidth}
    \centering
    \includegraphics[width=\linewidth]{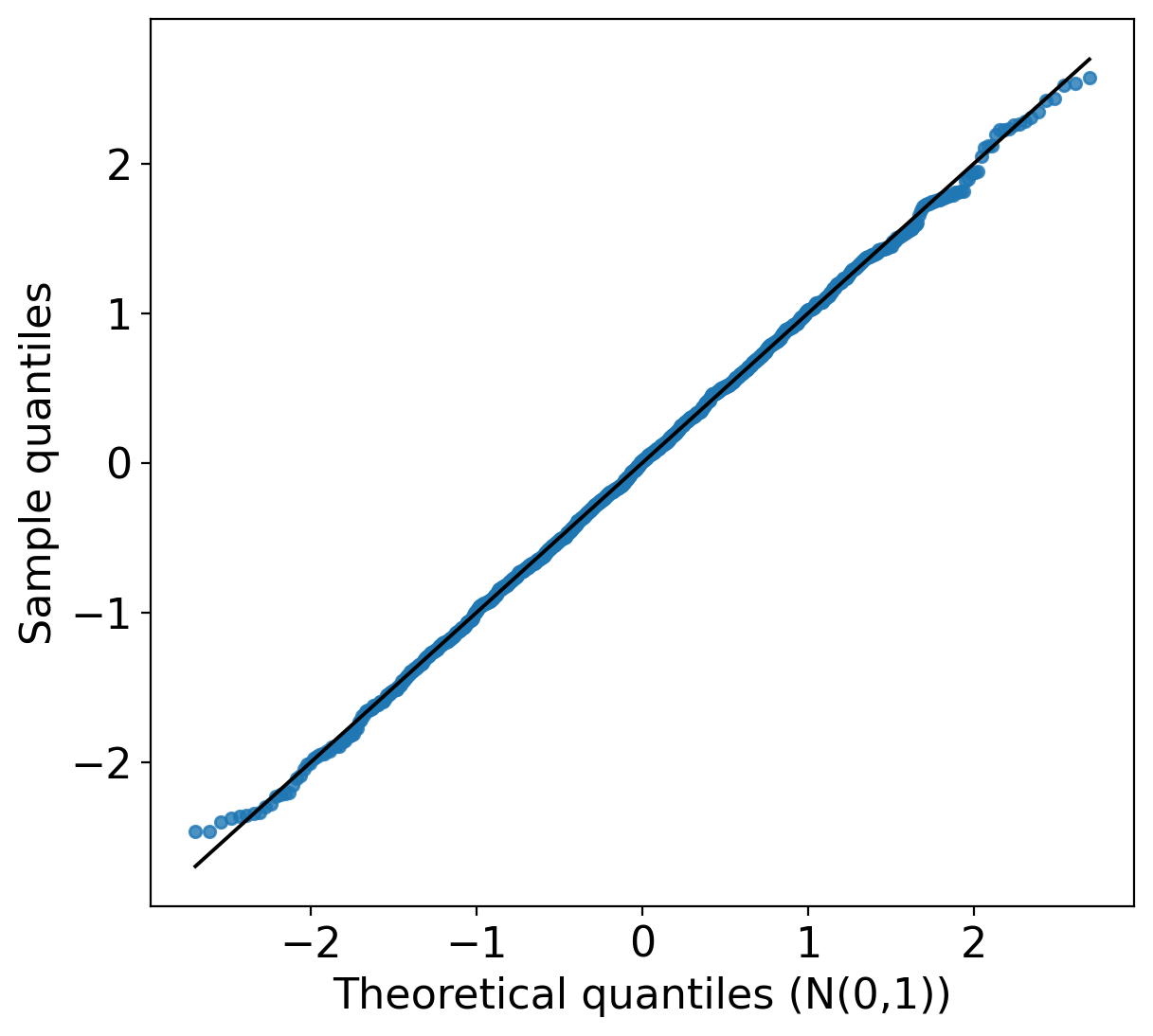}
    \caption{$d=50$, data-driven}
  \end{subfigure}\hfill
  \begin{subfigure}{0.32\textwidth}
    \centering
    \includegraphics[width=\linewidth]{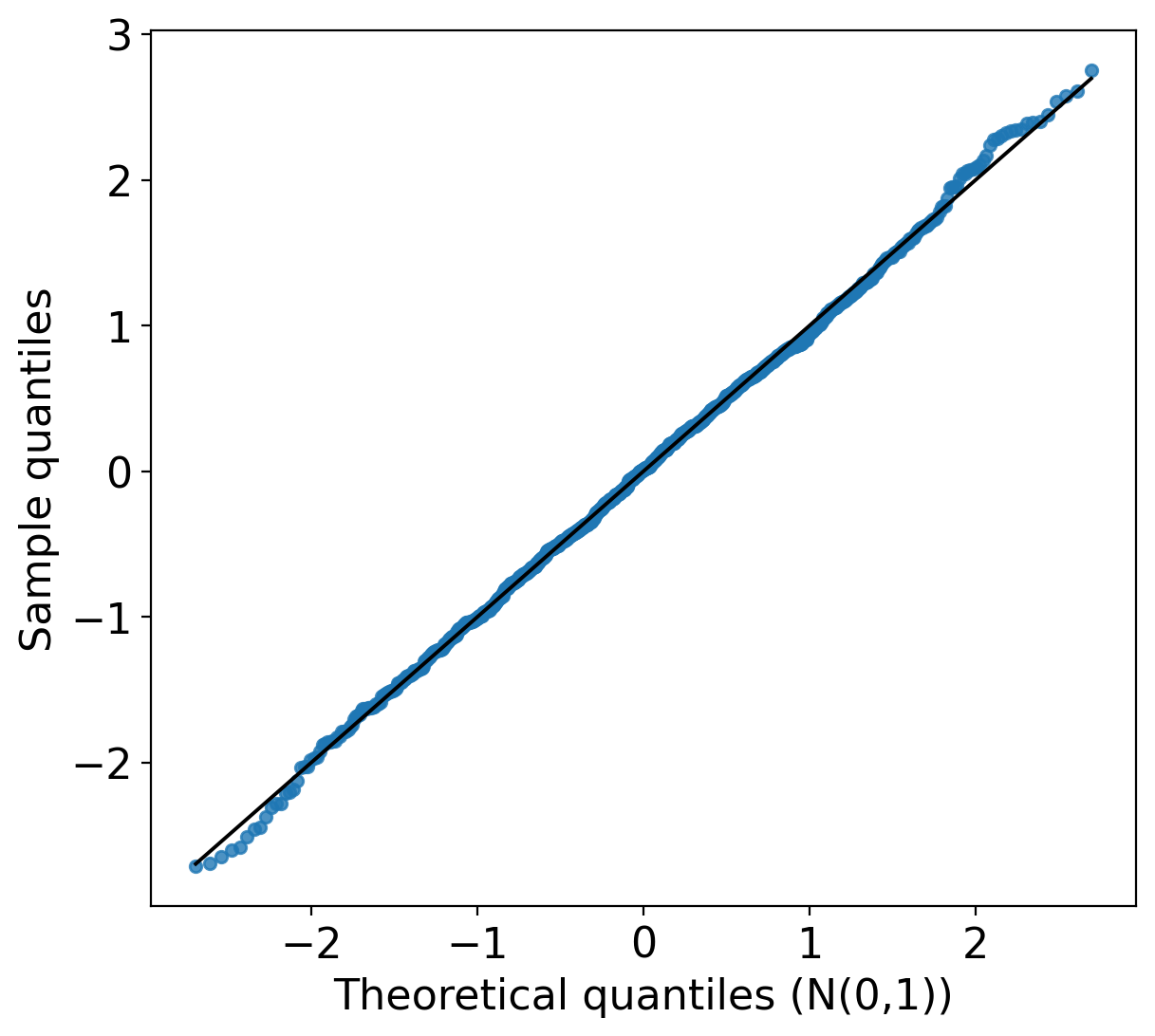}
    \caption{$d=100$, data-driven}
  \end{subfigure}}
  \caption{\small Normal QQ plots for the first coordinate of the coordinate first row of $\widehat\bU_{\bB}-\bU_{\bB}\bR_{\bB}$ with $T=200$ under the strong loading strength regime.
  The top row uses the oracle inference procedure with population covariance matrices, and the bottom row uses the feasible data-driven inference procedure with plug-in covariance estimators.
  Columns correspond to $d\in\{20,50,100\}$.}
  \label{fig:qq_B_oracle_dd}
\end{figure}

\begin{figure}[htbp!]
  \centering
  \resizebox{0.75\textwidth}{!}{%
  \begin{subfigure}{0.32\textwidth}
    \centering
    \includegraphics[width=\linewidth]{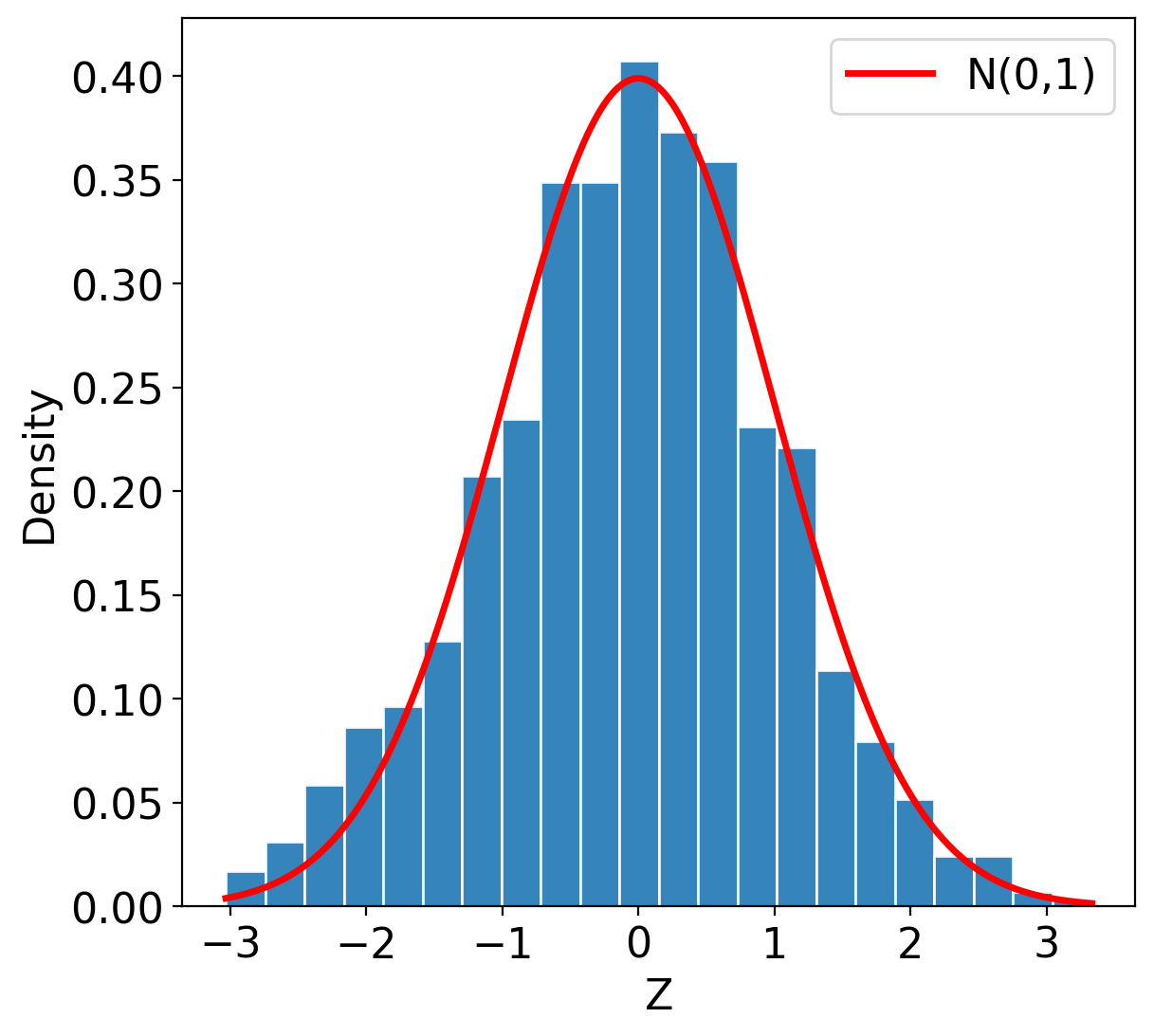}
    \caption{$d=20$, oracle}
  \end{subfigure}\hfill
  \begin{subfigure}{0.32\textwidth}
    \centering
    \includegraphics[width=\linewidth]{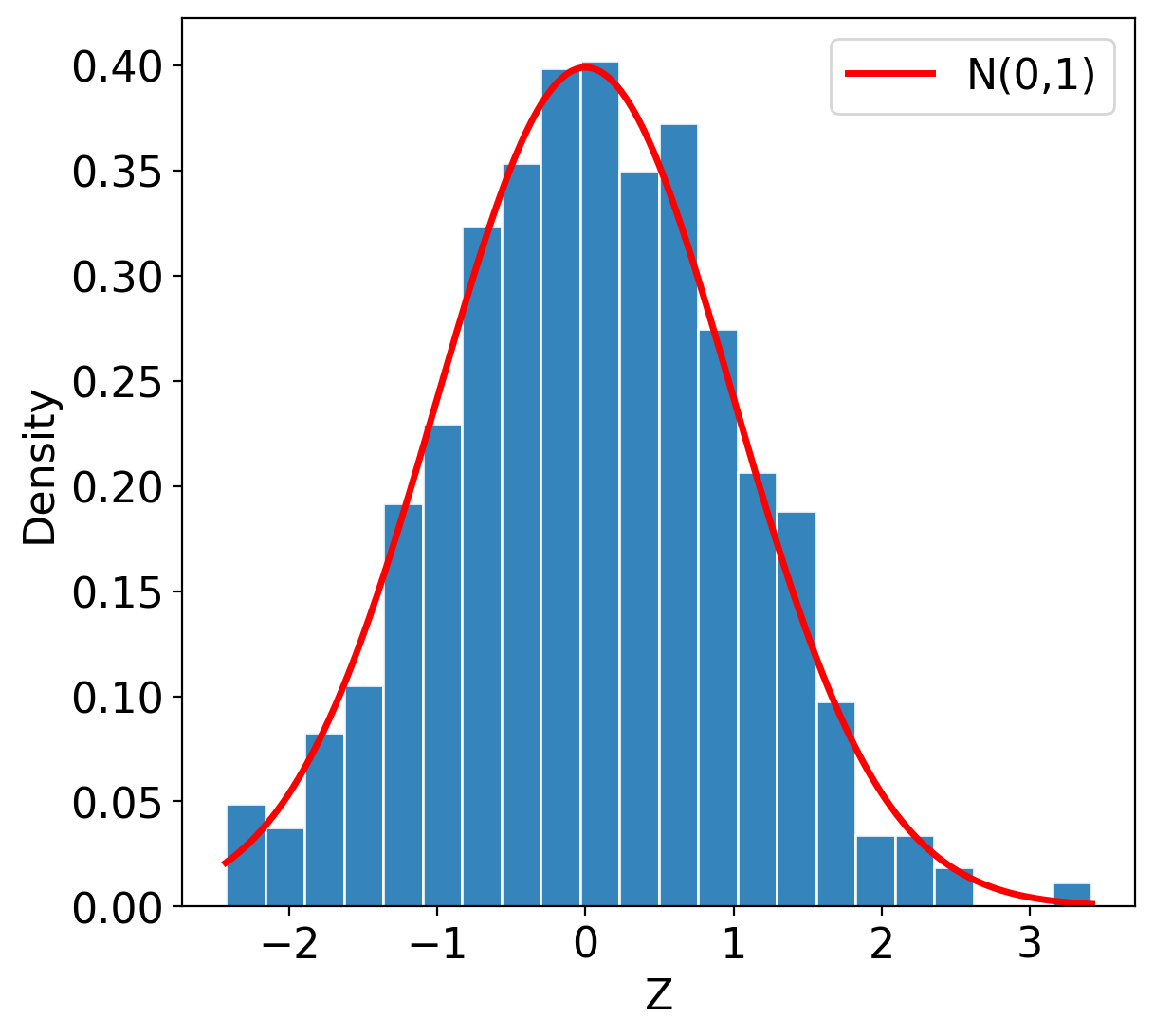}
    \caption{$d=50$, oracle}
  \end{subfigure}\hfill
  \begin{subfigure}{0.32\textwidth}
    \centering
    \includegraphics[width=\linewidth]{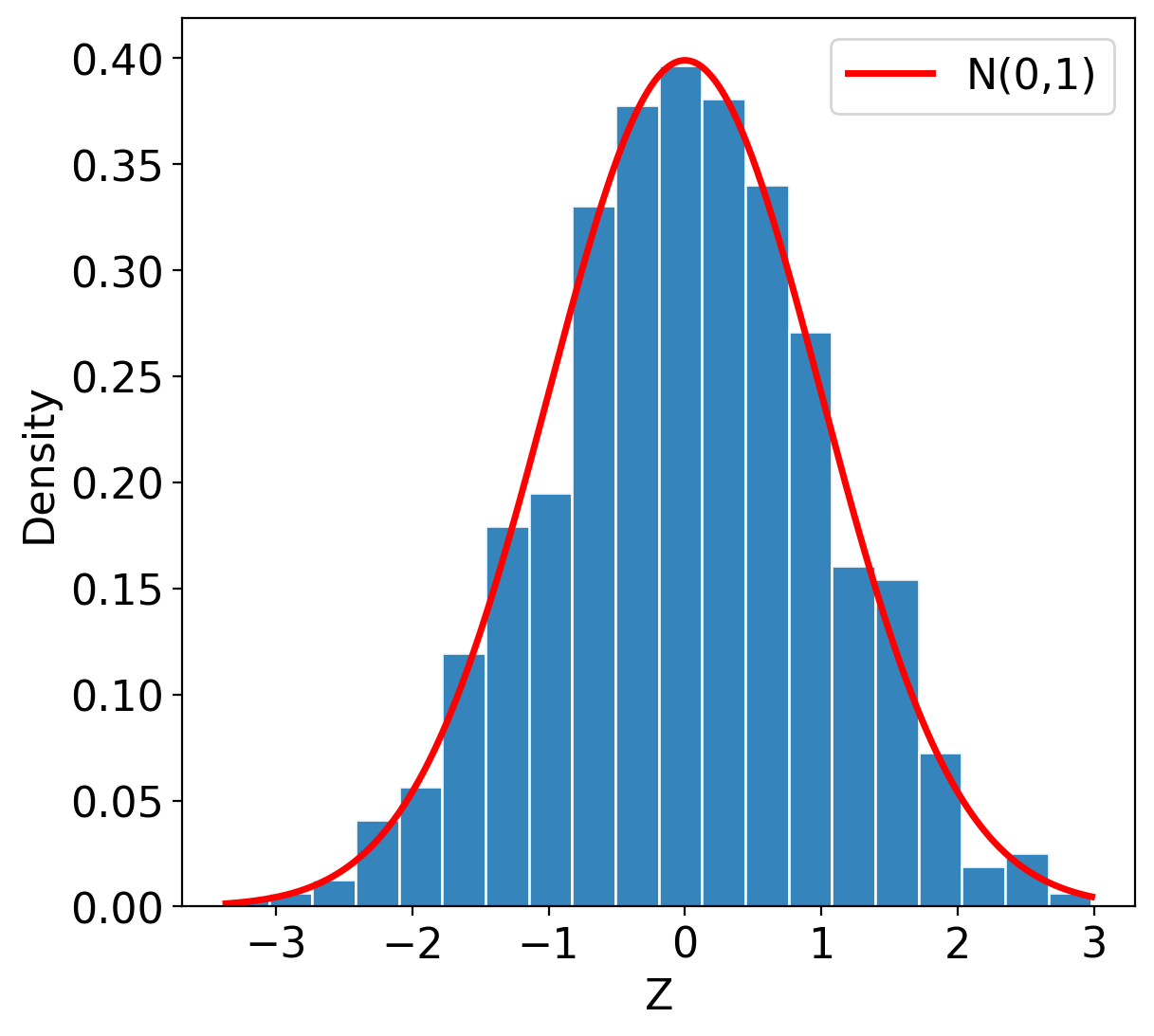}
    \caption{$d=100$, oracle}
  \end{subfigure}}
  \vspace{0.35em}
  \resizebox{0.75\textwidth}{!}{%
  \begin{subfigure}{0.32\textwidth}
    \centering
    \includegraphics[width=\linewidth]{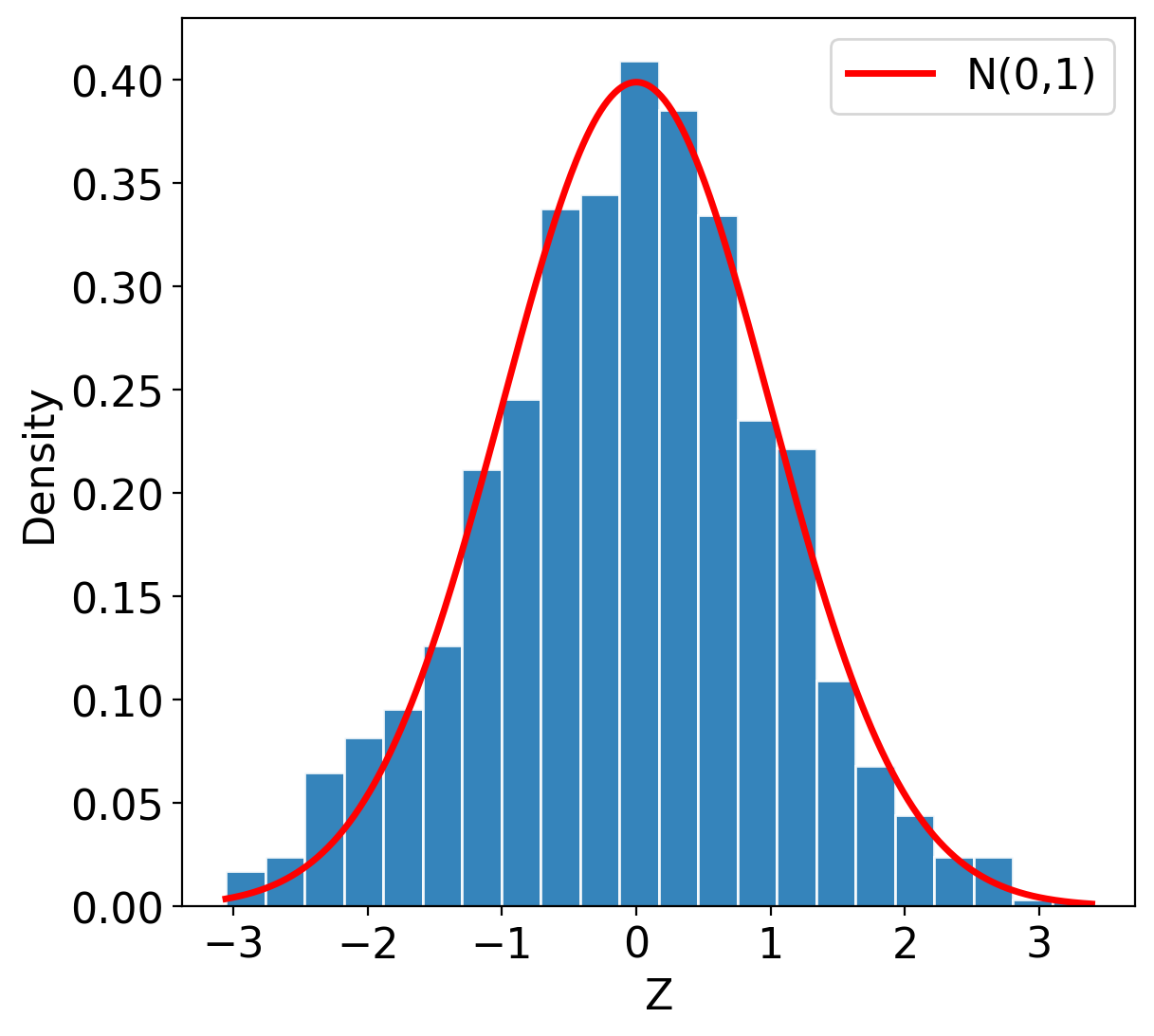}
    \caption{$d=20$, data-driven}
  \end{subfigure}\hfill
  \begin{subfigure}{0.32\textwidth}
    \centering
    \includegraphics[width=\linewidth]{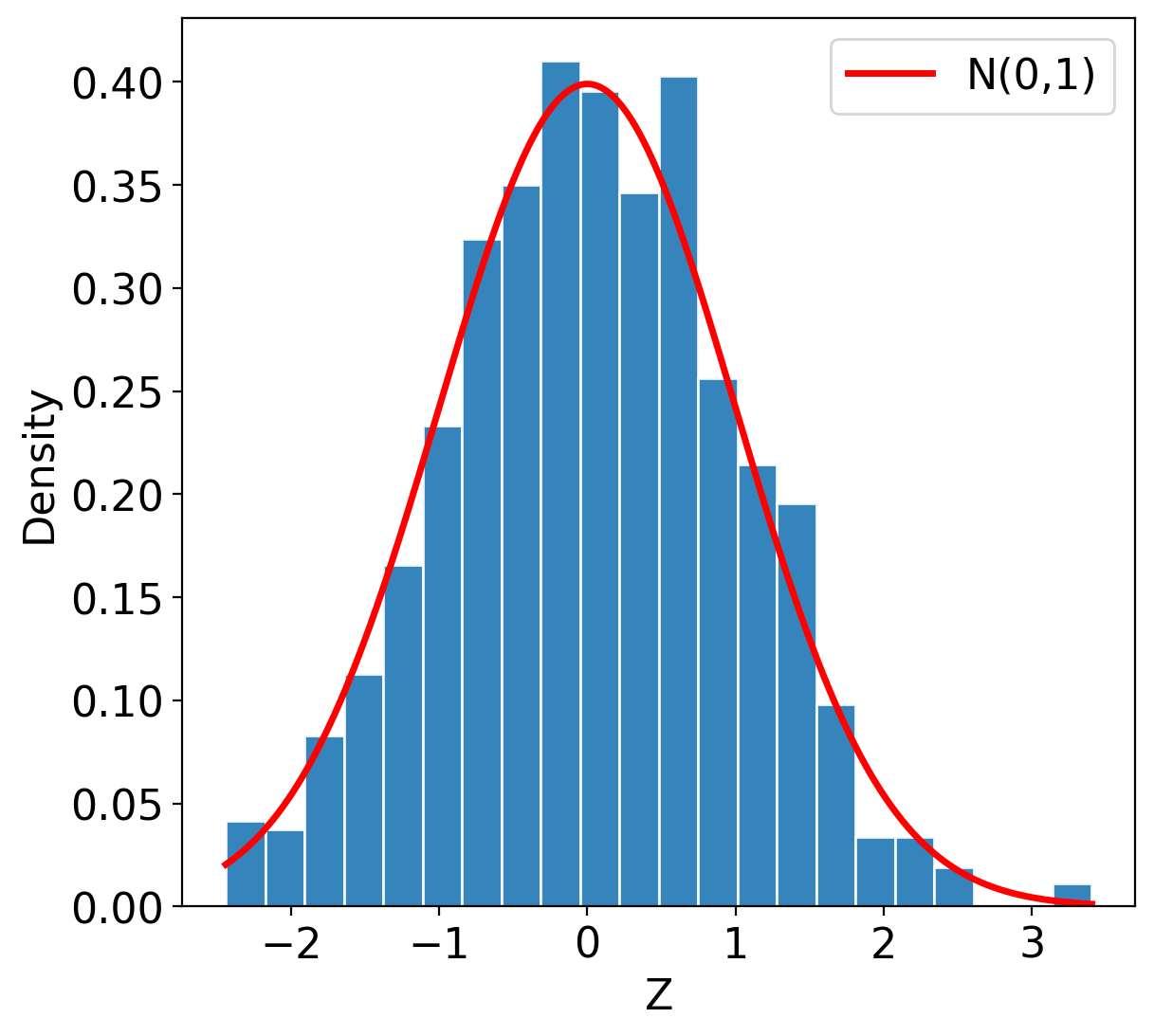}
    \caption{$d=50$, data-driven}
  \end{subfigure}\hfill
  \begin{subfigure}{0.32\textwidth}
    \centering
    \includegraphics[width=\linewidth]{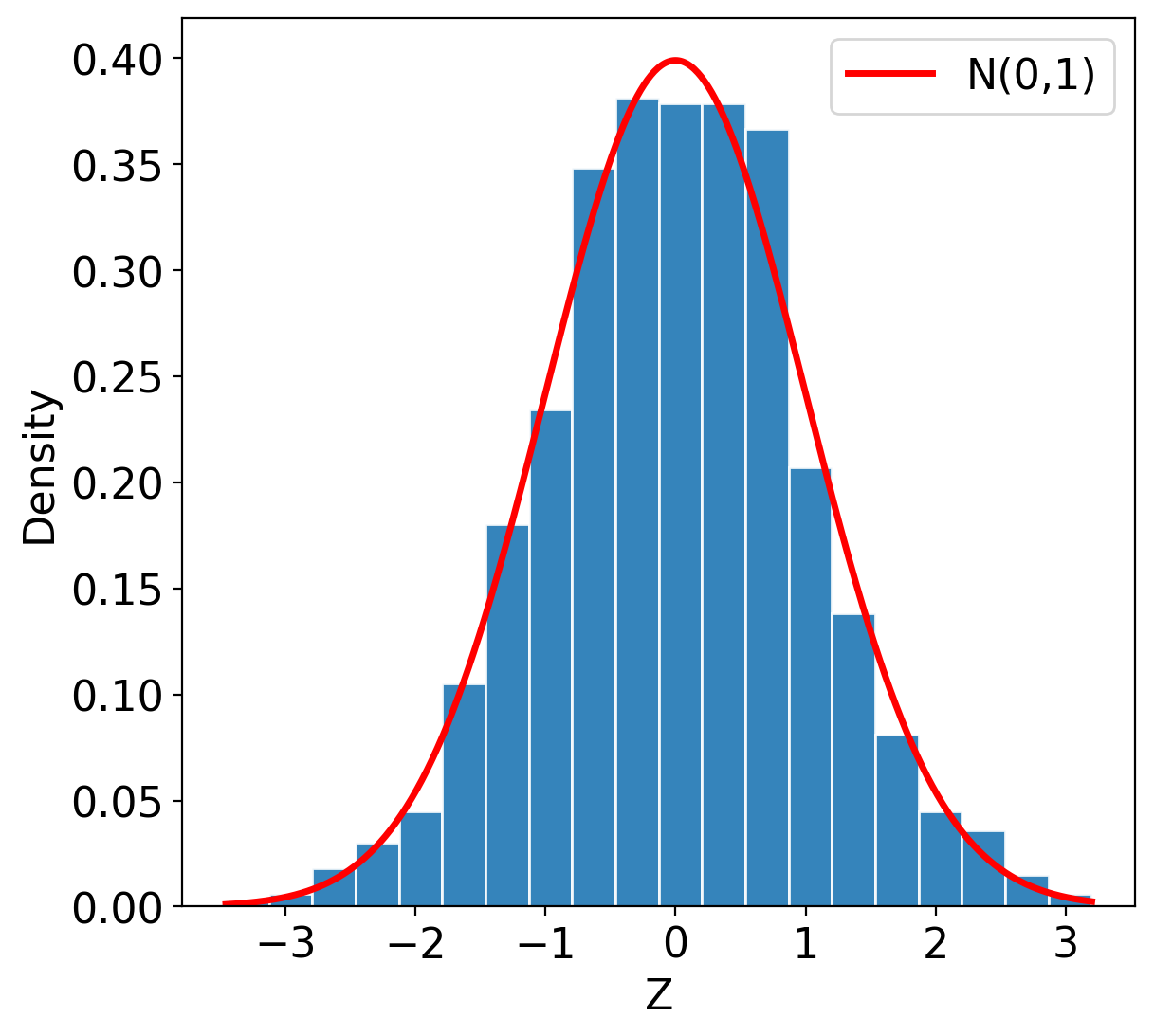}
    \caption{$d=100$, data-driven}
  \end{subfigure}}
  \caption{\small Histograms for the first coordinate of the standardized first row of $\widehat\bU_{\bB}-\bU_{\bB}\bR_{\bB}$ with $T=200$ under the strong loading strength regime, overlaid with the $\cN(0,1)$ density (red curve).
  The top row uses the oracle inference procedure with population covariance matrices, and the bottom row uses the feasible data-driven inference procedure with plug-in covariance estimators.
  Columns correspond to $d\in\{20,50,100\}$.}
  \label{fig:hist_B_oracle_dd}
\end{figure}

This appendix presents complementary simulation results for the loading matrix $\hat \bU_{\bB}$ under identical settings to Section \ref{sec:simul}. Figures \ref{fig:subspace_B_strong} and \ref{fig:subspace_B_weak} display estimation errors under strong and weak factor loading regimes, while Figures \ref{fig:qq_B_oracle_dd} and \ref{fig:hist_B_oracle_dd} assess asymptotic normality via Q-Q plots and histograms. The results mirror those for $\widehat{\bU}_{\bA}$: COMPAS outperforms MINE and P-COMPAS, strong loadings yield superior recovery, and data-driven inference achieves nominal coverage rates.

\end{appendices}

\end{document}